  \documentclass{cambridge6A}
  \usepackage{natbib}

  \usepackage{rotating}
  \usepackage{floatpag}
  \rotfloatpagestyle{empty}

  \usepackage{amsthm}
  \usepackage{graphicx}

  \usepackage{multind}
  \makeindex{authors}
  \makeindex{subject}

\usepackage{natbib}
\usepackage{longtable}

  \theoremstyle{plain}

  \theoremstyle{definition}

  \theoremstyle{remark}

  \usepackage{aas_macros}
  \usepackage[sectionbib]{chapterbib}

\begin{document}

\title[ \it XXVII Canary Islands Winter School of Astrophysics]
{\bf HIGH TIME RESOLUTION ASTROPHYSICS}

\author{\rm TARIQ SHAHBAZ, \\
{\it Instituto de Astrof\'{\i}sica de Canarias, 
Tenerife, Spain \\ 
Departamento de  Astrof\'\i{}sica, Universidad de La Laguna, 
Tenerife, Spain} \\
\vspace{5mm}
JORGE CASARES VEL\'AZQUEZ \\ 
{\it Instituto de Astrof\'{\i}sica de Canarias, 
Tenerife, Spain \\ 
Departamento de  Astrof\'\i{}sica, Universidad de La Laguna, 
Tenerife, Spain} \\
\vspace{5mm}
TEODORO MU\~{N}OZ DARIAS \\ 
{\it Instituto de Astrof\'{\i}sica de Canarias, 
Tenerife, Spain \\ 
Departamento de  Astrof\'\i{}sica, Universidad de La Laguna, 
Tenerife, Spain
}
}

  \frontmatter
  \maketitle
  \tableofcontents

  \pagenumbering{arabic}
 
  \mainmatter

\author[T.M. Belloni]{TOMASO M. BELLONI}
\chapterauthor{TOMASO M. BELLONI}

\chapter[Black-hole and neutron-star binaries]
{X-ray emission from black-hole and neutron-star binaries
\label{Belloni}}

\contributor{Tomaso M. Belloni
\affiliation{Osservatorio Astronomico di Brera, Italy}}

\abstract {In this chapter, I present the main X-ray observational
characteristics of black-hole binaries and low magnetic field neutron-star
binaries, concentrating on what can be considered similarities or differences,
with particular emphasis on  their fast-timing behaviour. }

\section{Introduction}\label{sec:1}

The large amount of information that has become available in the past two
decades on high-energy emission from accreting X-ray binaries, both in terms of
number of observations and high-resolution high-sensitivity data, have given us
a much more complete view of the emission properties from the inner regions of
the accretion flow, close to the compact object. While it is now nearly
impossible to review all observational aspects, in this chapter I will present
an overview of the X-ray characteristics with a special emphasis on the
comparison between Galactic neutron-star (NS) and black hole (BH) accreting
systems. Since the strong dipolar magnetic field (B) in the `standard' X-ray
pulsars dominates the accretion flow and prevents the formation of
nearly-equatorial accretion in the innermost few gravitational radii around the
NS, I will concentrate on low-field NS objects, most of which are low-mass X-ray
binaries (LMXB). As the size of the region inside the innermost stable circular
orbit (ISCO) around a BH and the size of a NS are roughly comparable, one can
expect that many properties of the accretion flow and its emission would be in
common between the two classes of systems. However, the presence of a solid
surface on a NS, which prevents advection of energy,  must necessarily add
another emission component, which can be very complex as a boundary layer is
formed. Moreover, the dipolar magnetic field in a LMXB, estimated to be in the
range $10^8-10^9$\,G, is still not negligible and must alter the properties of
accretion in the innermost regions, where its energy density is higher.

Optical observations of the companion star
have now provided robust dynamical mass measurements 
\citep[see][for a recent review]{casjon2014}. From the early 
days of X-ray astronomy,
after it was realised that the presence of a BH and/or NS was
the origin of the strong high-energy
emission, we have tried to identify specific `signatures' in the high-energy emission, which would
allow us to identify unambiguously the nature of the compact object. While we know a few of these
signatures for NSs, such as coherent pulsations (now detected in more than a dozen accreting LMXBs) and
thermonuclear X-ray bursts, `smoking guns' that reveal directly the presence of a BH are
still elusive \citep[see][for a review of the observational status of the 
comparison between NS and BH
before {\it XMM-Newton}, {\it Chandra} and RossiXTE]{Klis:1994}. This issue is strongly connected to the
detection of effects due to General Relativity (GR) in the strong field regime as can only found in the
proximity of a compact object, where the theory still needs to be tested. The recent direct detection
of gravitational waves from coalescing BHs is of course one 
\citep[][]{abbott}, but the most
precise measurements to date come from binary pulsars, which are still hundreds of thousands kilometres
apart \citep[see][]{burgay}.

On the one hand, we are looking for a common observational scenario that can be interpreted within a
single accretion model, since the accretion flow around BHs and weakly magnetised NSs
is expected to be very similar. On the other hand, it is important to find crucial observables that
allow a direct identification of the presence of a BH, in the way that pulsations or
thermonuclear X-ray bursts indicate the presence of a surface and therefore of a NS. These
observables would also allow us to measure additional parameters predicted by GR such 
as BH
spin, relativistic precessions and the presence of an ISCO \citep[see][]{psaltis2008}. There is
considerable observational effort aimed at measuring  BH spins through X-ray spectra
\citep[see][]{middleton}, although these methods all need to assume a BH mass, which until now
can only be measured through dynamical methods, i.e. indirectly. Although it is true that optical
measures have become much more accurate in the past decades, the issue of absence of direct evidences
does not depend on measurement accuracy and cannot be solved in that way. Along the way, we moved from
the expression `black hole candidate' to `black hole binary' (which admittedly sounds much better)
for dynamically-confirmed cases, relegating the former to systems without mass measurement, but the
situation has not changed and the associated caveats remain.

This cannot be considered a full review. Here I outline the main X-ray emission
properties of accreting binaries containing a BH or a low-B neutron
star, limiting myself to those that can be used to establish similarities or
differences between classes. The focus is on timing properties, although
spectral information is also discussed as discussing timing only would not allow
to present a complete picture, necessary to understand these systems.
Moreover, I will limit myself to the X-ray band. This means I will ignore huge
parts of the field of the astrophysics of accretion onto stellar-mass compact
objects, which would also require much more space than available. This text
should be sufficient as an introduction and a companion to extricate oneself
through the extensive literature in the field.

\section{Neutron-Star binaries}\label{sec:2}
\subsection{Evolution and subclasses}

Since the 1980s, NS LMXBs have been divided into classes, based on the timing
and spectral characteristics. Since the spectral decomposition for these systems
is complex (see below) and significant variability takes place on a time-scale
of minutes to hours, the spectral variations are best followed through the use
of colour-colour diagrams (CCDM), where two spectral `colours' (ratios between
detected counts in two energy bands) integrated over short time-scales (minutes)
are plotted against each other. NS LMXBs trace very clear patterns on their CCDM,
moving in a smooth way without major jumps in time \citep[see figures\,2.4
in][]{Klis:2006}. Depending on the shape of the CCDM track and the associated
timing properties, classes and subclasses were identified. 

Six bright sources, all persistently at high X-ray flux, were called 'Z sources'
as they trace a path roughly shaped as a letter Z. Of course the precise shape
of the path depends on the instrument and on the chosen energy bands; moreover,
at different times the path can be observed to shift in the CCDM. The six sources
are: Sco\,X-1, Cyg\,X-2, GX\,340+0, GS\,349+2, GX\,5-1 and GX\,17+2. A second subclass
is that of `atoll' sources, who owe their bizarre name to the fact that their
CCDM looks like a tropical atoll. This class is much more populated, with dozens
of systems known to date. In addition, there is a class of sources at low
luminosity, which typically show type-I X-ray bursts, commonly referred to as
`low-luminosity bursters.' A unique system such as Cir\,X-1 has long been
considered not to be classifiable in this scheme, although now it is added to
the Z-source class.

\begin{figure}
\begin{tabular}{cc}
   \includegraphics[scale=0.35]{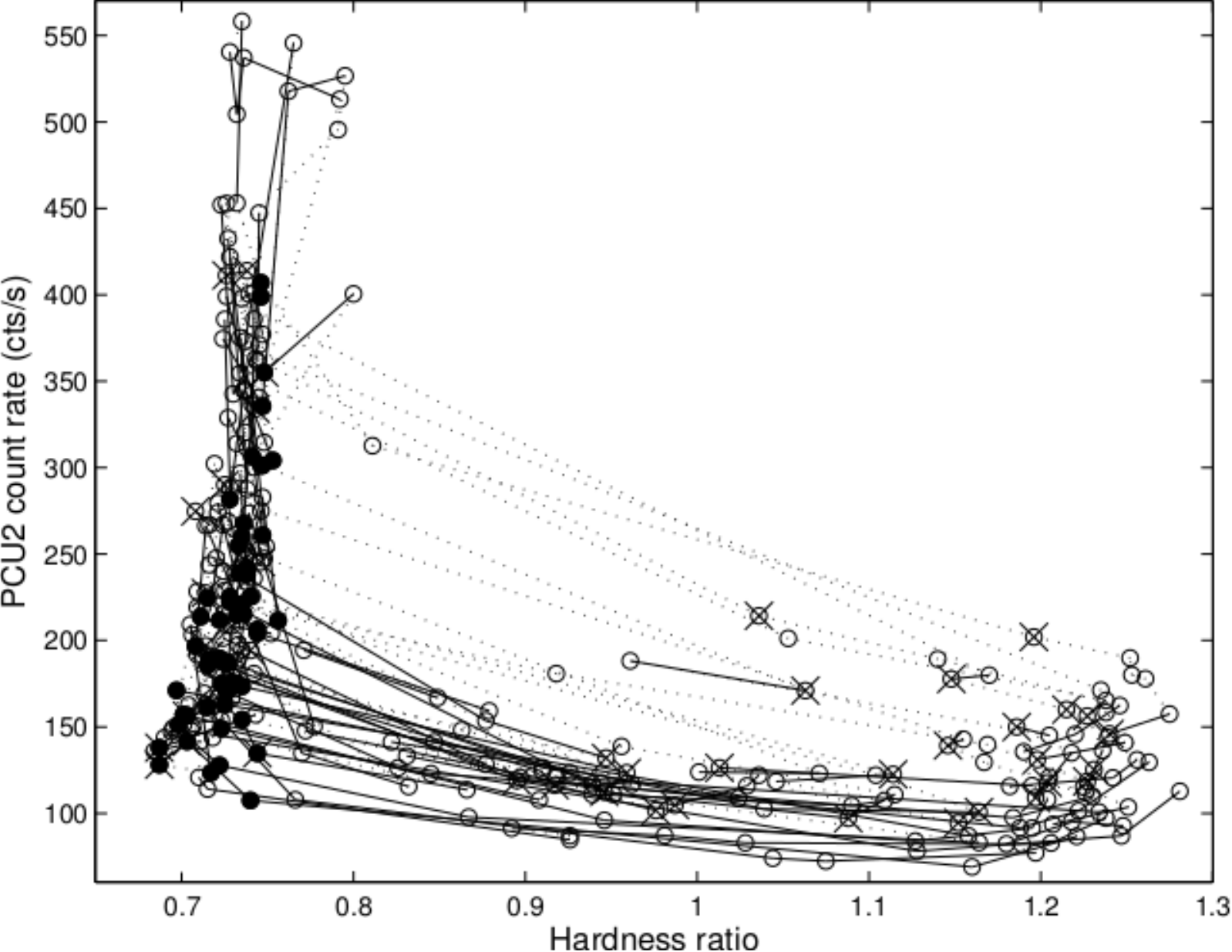} & 
   \includegraphics[scale=0.58,viewport= 0 0 271 250,clip]{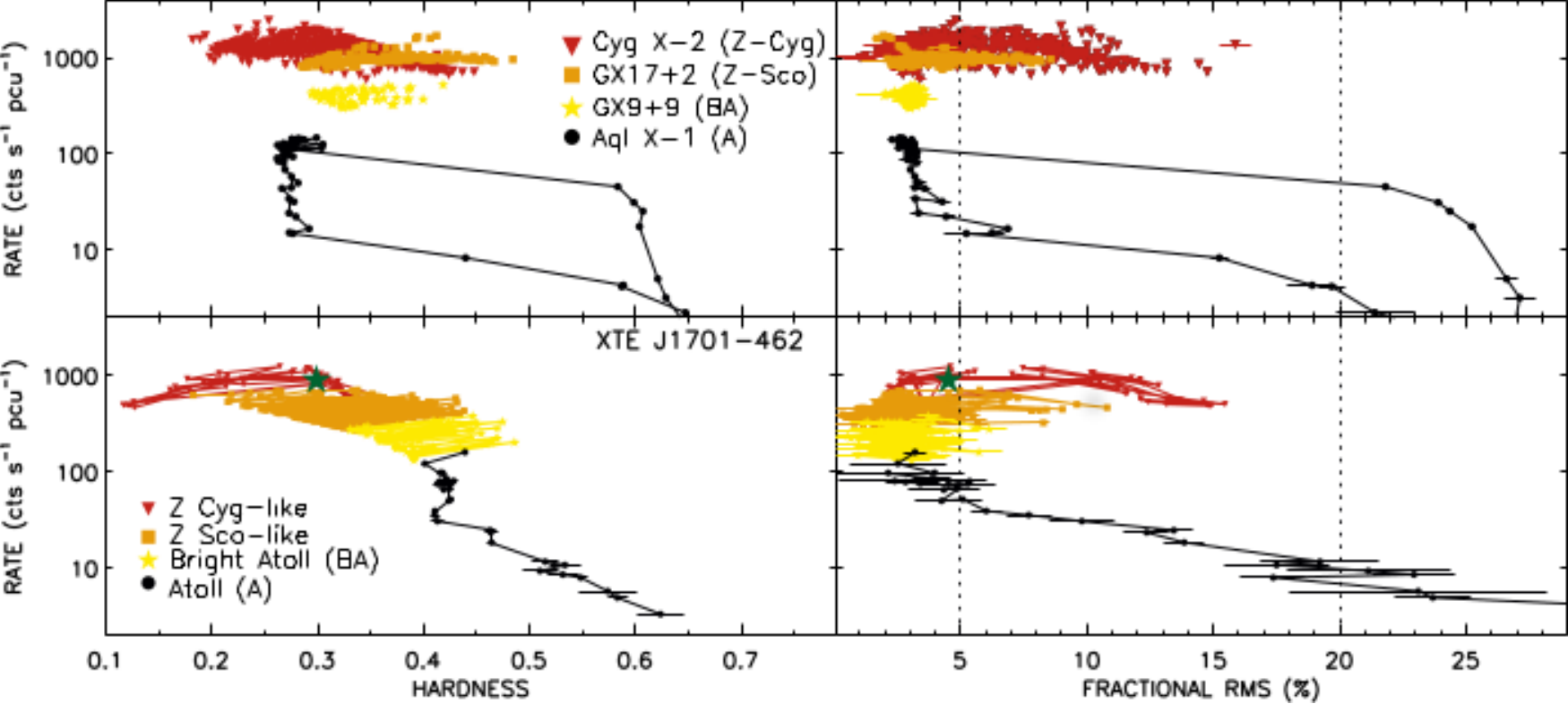} \\ 
\end{tabular}
\caption{Left panel: HID of 4U\,1636-53 from {\it RXTE} data. Solid lines connect indicate movement to the 
right (hardening), dotted lines movement to the left (softening). 
\citep[From][]{Belloni2007}. 
Right panel: HID for the Z sources Cyg\,X-2 and GX\,17+2, the atoll source source 
GX\,9+9 and the 
transient Aql\,X-1 (top), HID for the Z/atoll transient XTE\,J1701-462. 
\citep[From][]{Teo2014}.
}
\label{fig:ns_hid}
\end{figure}

The idea of following systems in the CCDM, which led to this colourful
terminology, comes from the difficulty of disentangling spectral components in
these systems. The presence of a number of spectral components that emit in a
comparable energy band make it difficult to follow changes on short time-scales.
The use of a purely observational tool such as the CCDM, although
instrument-dependent, allows to track these changes very effectively. In
addition, it was found that the properties of fast timing correlate rather
precisely with the position on the CCDM. In analogy with BH binaries
(BHBs), it is interesting to produce also a hardness-intensity diagram (HID),
where a single hardness is plotted versus intensity, for Z sources and for
transient and persistent atoll sources (see Figure\,\ref{fig:ns_hid}). One can see
that atoll sources, both persistent (such as 4U\,1636-53) and transient (such as
Aql\,X-1), follow a hysteresis cycle in the HID. In the case of 4U\,1636-53,
during RossiXTE observations the flux oscillated with a quasi-period of 30-40\,d, 
following a loop in the HID. Other systems are not so regular, but their
behaviour is consistent \citep[see figure\,2 in][]{Teo2014}.

Until a decade ago, there was no agreement as to what the fundamental difference
between Z and atoll sources was. Although it was suspected that accretion rate
was responsible, the effect of a different value of the dipolar magnetic field
of the NS could not be excluded \citep[see][for a review]{HK:1989}. Since the
distances to these systems are in most cases not determined, it is difficult to
measure directly their luminosity. Even for single Z sources, accretion rate was
found not to be tracked by source intensity \cite{HK:1989}. In January 2006, a
bright transient system appeared in the sky, XTE\,J1701-462
\citep[][]{Remillard2006}. It was soon realised that this was a transient
Z-source \citep[][]{Homan2007}. As the outburst progressed, the system passed
through phases where it displayed CCDM shapes typical of all subclasses of Z
sources, until after about 1.5\,yr the accretion rate dropped rapidly and
before reaching quiescence the system went through a brief `atoll' phase,
showing clearly that the difference between classes can be ascribed solely to
the effect of accretion rate \citep[see][see also
Figure\,\ref{fig:ns_hid}]{Homan2010}.

Working with a large sample of systems from the RossiXTE archive, \cite{Teo2014}
have compared the HID properties of NS LMXBs and proposed a general scheme to
put together different classes (and to compare them to BHBs, see below). Their
unified diagram can be seen in Figure\,\ref{fig:teo}, where a RMS-intensity
diagram (RID) is shown [here hardness is replaced by fractional root-mean-squared (RMS) variability,
which however correlates well with hardness, and luminosity rather than count
rate is on the Y axis, necessary to compare different systems]. In this sketch,
the connection between Z and atoll sources is clear and only associated to a
difference in luminosity. In this diagram, the low-luminosity bursters live on
bottom-right branch, always in the hard state and at low luminosity.

One state is missing in this description, namely the quiescent state of
transients, when accretion rate is low and instruments such as the {\it RXTE}/PCA
cannot detect X-ray emission. Most of the time, transients are in quiescence, at
an accretion luminosity below $10^{33}$\,erg/s. While in the case of BHBs (see
below), any X-ray emission observed in quiescence can only come from the
accretion flow (with a possible contribution of the X-ray emission of the
companion late-type star), NS LMXBs feature a hard surface. This has two
practical effects: the first is that the energy from any residual low-level
accretion must be released, even in case of inefficient accretion \citep[see
e.g.][]{menou1999,kong2002}. The second is that if accretion stops completely,
after the effects of accretion have disappeared, the cooling curve of the
NS can be measured \citep[see e.g.][]{BrownCumming2009}. The presence
of the surface emission makes the positioning on the HID not very relevant.

\begin{figure}
\includegraphics[scale=0.22]{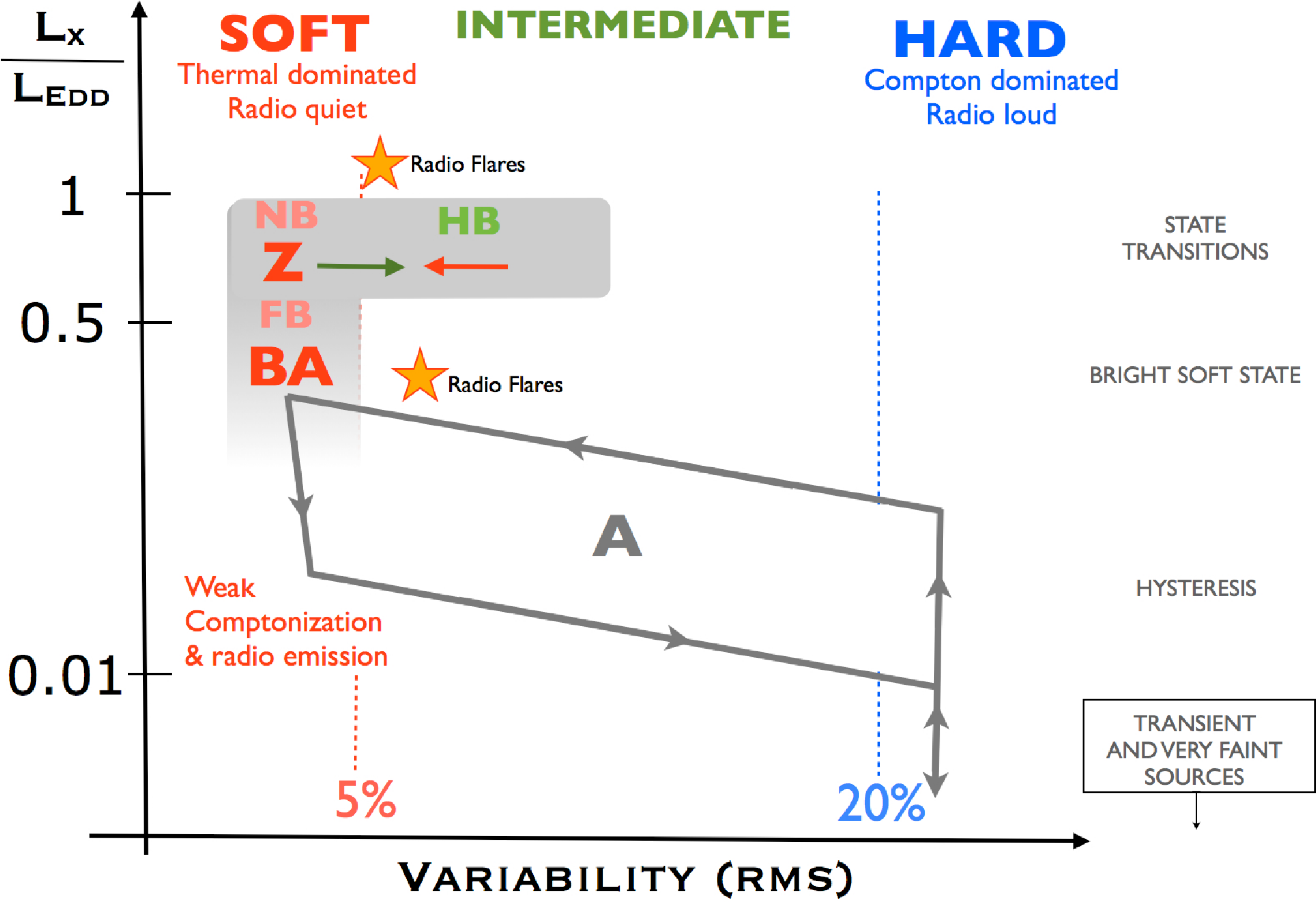}
\caption{Schematic RMS-intensity diagram  describing the position and behaviour of atoll (A),
bright atoll (A) and Z (Z) sources. A HID has qualitatively the same shape. \citep[From][]{Teo2014}. }.
\label{fig:teo}
\end{figure}

\subsection{Energy spectra}

The energy spectra of NS LMXBs are complex. The presence of the hard surface of
the NS introduces an extra component that is in the same energy range
than that (or those) from the accretion flow. As a result, two different
spectral models have been proposed at the end of the 1980s, which have hardly
been addressed until recent times, when a much larger number of spectra over
different states have been accumulated, mostly by the RossiXTE satellite. As one
can see from Figure\,\ref{fig:teo}, the observations indicate the presence of two
main states, which need to be interpreted within a common framework. These
states are connected by intermediate states, which add complications, but also
allow a connection between hard and soft states.  The two `old time' models are
traditionally referred to as the {\it Western} and the {\it Eastern} model. For
the hard state, both models are compatibile and consist of a hard Comptonised
component with an electron temperature of dozens of keV, plus a softer thermal
component with temperature $<$1\,keV that can be modelled as a black-body (BB) or
a disk-blackbody (DBB) \citep[see e.g.][]{GD2002}. These spectra are analogous
to those of BHBs in their hard state (see below) and as we will see their fast
timing properties are also very similar to those of BHBs. For the soft state,
the Eastern model \citep[][]{Mitsuda1989} includes a DBB (kT$\,\sim\,$0.5--2\,keV)
and a weakly Comptonised BB, while the Western model \citep[][]{White1988}
decomposes the spectra as a blackbody from the boundary layer and a Comptonised
disk component.

\cite{Lin2007} analyzed a large number of RossiXTE observations, with which they
could follow the spectral changes of two atoll systems across states and
concluded that not only all combinations of a thermal and a non-thermal
component are degenerate, but that it is impossible to reconcile observations in
different states with the two models outlined above. They proposed a third,
hybrid model: in the hard state the spectrum is deconvolved into a blackbody and
a Comptonised component, like in the older models. In the soft state, a
three-component model is needed, consisting of a DBB, a BB and a weakly
Comptonised component. This model has the advantage to maintain the
L$\propto$T$^4$ relation expected for a DBB, to have a constant area for the BB
component (corresponding to that of the boundary layer, around 15\,\% that of the
surface of the NS), and to have a Comptonised fraction compatible with that of
BHBs. This work was extended to the transitional system XTE\,J1701-462
\citep[][]{Lin2009} and to the persistent Z source GX\,17+2 \citep[][]{Lin2012},
i.e. to systems at higher accretion rate. The conclusion was that accretion rate
drives the secular changes in shape of the Z diagram, while the actual motion
along the branches of the diagram takes place at constant accretion rate and are
driven by other mechanisms such as change in the Comptonised fraction,
transition to a slim disk solution and changes in the inner disk radius.  All
these results can be linked very effectively with the characteristics observed
in BHBs, as shown below.

The availability of X-ray instruments with high energy resolution has 
allowed to detect and study additional components in the energy spectra of 
NS LMXBs, such as Compton reflection features (broad iron lines and 
continuum). Several systems have been studied \citep[see e.g.][and 
references therein]{Disalvo2015,Degenaar2015}, which led to constraints to 
the inner disk structure and inclination. 
I will discuss an application 
of these models in combination to timing models for mass measurement 
below. Additional non-thermal components have been detected in Z sources 
in the form of hard spectral tails \citep[see][and references 
therein]{migliari2007}. They can contribute up to 10\,\% of the total flux 
and appear to be in general stronger in the HB (see Figure\,\ref{fig:teo}). 
The possible interpretation proposed range from non-thermal electrons in a 
corona or at the base of a jet \citep[see][]{markoff2005} to bulk motion 
Comptonisation in a converging flow \citep[][]{farinelli2008}.

\subsection{Timing properties}\label{NStiming}

The phenomenology of the observed features in the power density spectra 
(PDS) of NS LMXBs is very extensive \citep[see][for a review]{Klis:2006}. 
Here I present the main points, in order to limit myself to the most basic 
picture, which can then be compared to BHBs.

\begin{figure}
\begin{tabular}{cc}
   \includegraphics[scale=0.25]{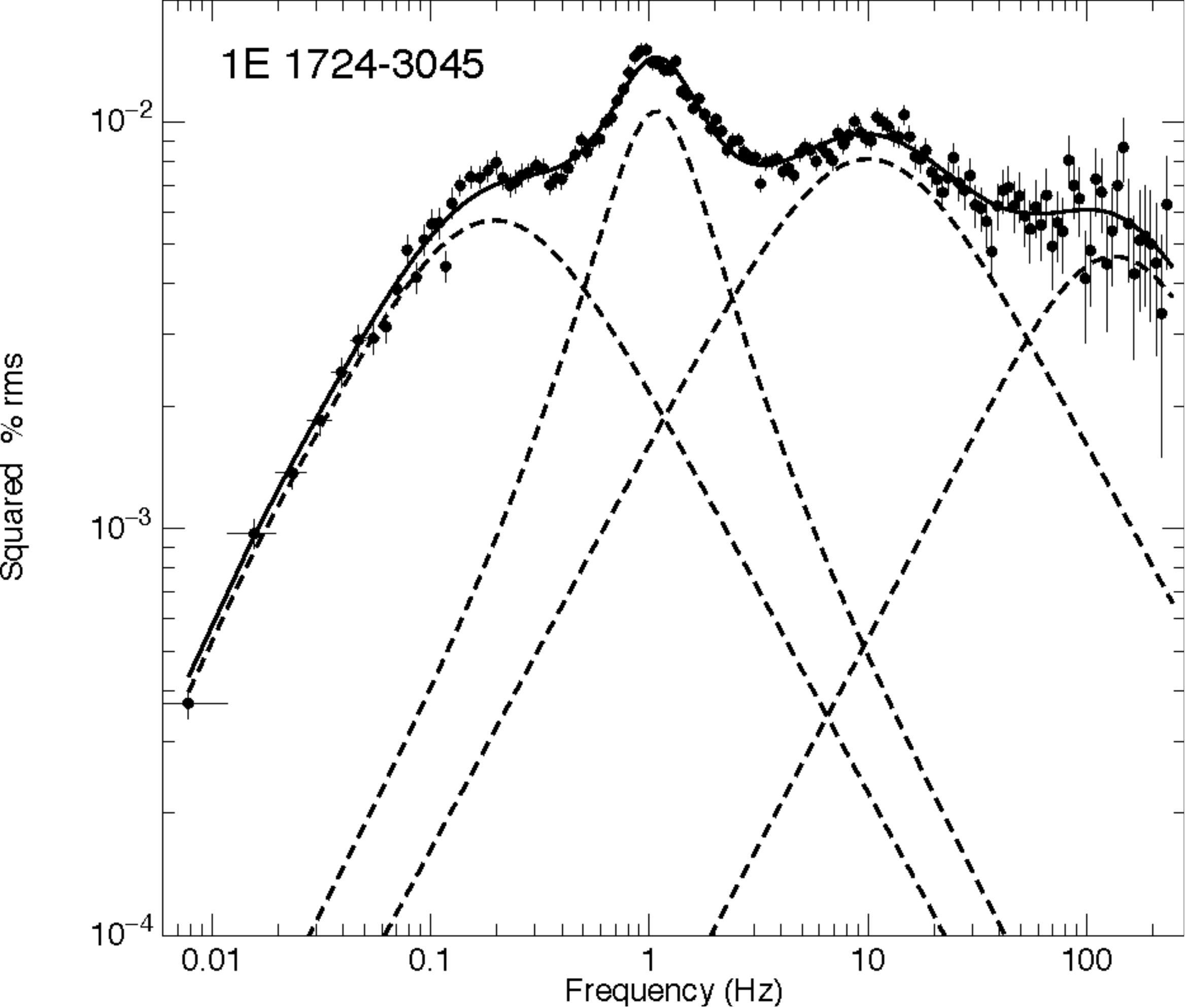} & 
   \includegraphics[scale=0.25]{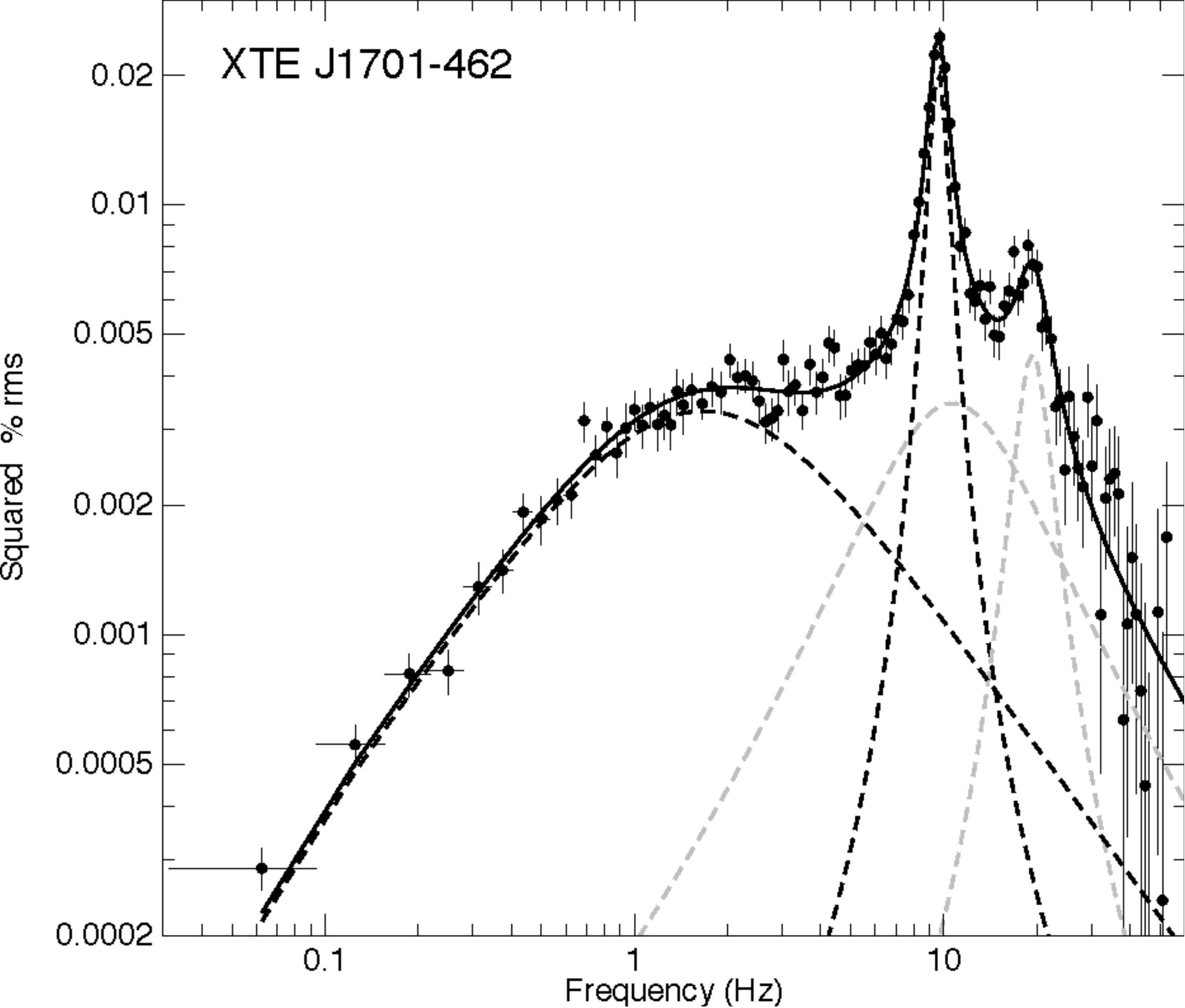} \\ 
\end{tabular}
\caption{Examples of PDS (in $\nu P_\nu$ form) from NS LMXBs. Left panel: typical PDS from a hard
state, from 1E\,1724-3045. Right panel: PDS of XTE\,J1701-462 featuring a strong HBO. The grey lines
indicate additional components (second harmonic and peaked noise, see text).}
\label{fig:ns_lfpds}
\end{figure}

{\bf Low frequencies.} At low frequencies ($\le$100\,Hz) the PDS of NS LMXBs are
complex. In the hard state (see Figure\,\ref{fig:teo}), at least four zero-centered
Lorentzian components are needed to model the variability (see the left panel in
Figure\,\ref{fig:ns_lfpds}) and the integrated fractional RMS is high, larger than
20\,\%. As one can see, with Lorentzian components no additional power-law
component is needed. One of the components (the one peaking around 1\,Hz in the
left panel of Figure\,\ref{fig:ns_lfpds}) is narrower and can be identified as a
quasi-periodic oscillation (QPO). The four Lorentzian components, from lower to higher characteristic
frequencies, are called $L_b$, $L_{LF}$, L$_\ell$ and $L_u$,with a corresponding
naming scheme for their characteristic frequency \cite[][]{bpk}. The full PDS is
flat-topped at frequencies below $\nu_b$, declines as $\nu^{-1}$ (with bumps),
then ends as $\nu^{-2}$ above $\nu_u$.

In the softer states, the characteristic frequencies increase and more 
obvious low-frequency QPOs appear. Three types of QPOs have been 
identified, each corresponding to one of the three branches in the `Z' 
path in the HID (see Figure\,\ref{fig:ns_lfpds}). Correspondingly, they have 
been dubbed horizontal branch oscillations (HBO), normal branch 
oscillations (NBO) and flaring branch oscillations (FBO). NBOs are 
observed always around 6\,Hz, FBOs are in the range $\sim\,4-20$\,Hz, while 
HBOs span a larger range, from $\sim\,$1\,Hz to dozens of hertz. 
\citep[see][]{Klis:2006}. HBOs are the most important for comparison with 
BH systems (although NBO/FBO also appear to have counterparts in BHBs): 
they are associated to a flat-top component which can be identified as a 
high-frequency extension of $L_b$.  An example from XTE\,J1701-462 is shown 
in the right panel of Figure\,\ref{fig:ns_lfpds}. The QPO often shows a 
second harmonic component and a peaked-noise component at around 
$\nu_{LF}$ (in grey in Figure\,\ref{fig:ns_lfpds}).

{\bf High frequencies.} At high frequencies ($\ge$100\,Hz) the picture is rather
different. While for hard states only the high-frequency end of $L_u$ can be
seen, the softer states are dominated by the so-called killohertz  QPOs. Their main
properties can be summarized in a few points: \begin{itemize}

\item they are QPO peaks observed in the frequency range 200--1200\,Hz, with a $Q$
factor (defined as the centroid frequency divided by the FWHM, an indication of
how coherent a peak is)  that can be as low as a few, but in some case as high
as 200 \citep[see][]{Barret2005a,Barret2005b}. In 4U\,1636-53, the $Q$ factor of
the lower kHz QPO has a maximum of 180\,Hz at $\sim\,$850\,Hz, while the upper kHz
QPO has always a $Q$ factor of a few \citep[][]{Barret2005b}.

\item there are two distinct peaks, often appearing together, referred to as the
{\it lower} and the {\it higher} kHz QPO (see left panel in Figure\,\ref{fig:khzQPO}). Their centroid frequencies vary on time-scales as short as
minutes, moving roughly in parallel, i.e. maintaining (roughly) the same
separation. The separation  is for all systems around $\sim\,$300\,Hz \citep[with
values between 200\,Hz and 400\,Hz, see][]{menbel2007}. There is a good
correlation also with spectral properties. In the right panel of Figure\,
\ref{fig:khzQPO} all kHz QPO detections in the {\it RXTE}/PCA data of 4U\,1636-53 are
shown in a plot of their centroid frequency versus hard colour. The two peaks are
clearly separated into two different tracks, on which they move keeping roughly
the same separation.

\item their centroid frequency is generally correlated with source luminosity on
short time-scales (hours), but the correlation disappears on long time-scales
and across different sources. In other words, while  luminosity can span orders
of magnitude, kHz QPOs remain in the 200--1200\,Hz range. A possible
interpretation of this phenomenon was presented by \citet{Klis:2001}

\item there are no special frequencies associated to single sources
\citep[][]{Belloni2007}. Their fractional RMS peaks at intermediate values and
decreases at low and high centroid frequencies \citep[][]{Barret2005b}. In the
well-studied case of 4U\,1636-53, the lower kHz QPO has a fractional RMS of
$\sim\,$9\,\% between 700 and 830\,Hz, to decrease above and below, while the upper
kHz QPO has a fractional RMS peaking at 15\,\% around 700\,Hz. The drop in
fractional RMS and coherence has been interpreted as caused by the accretion
flow reaching the innermost stable circular orbit
\citep[][]{Barret2005b,Barret2006,Barret2007} but an alternative scenario was
also suggested \citep[][]{mendez2006}.

While below 200\,Hz there are low-frequency components that complicate the
detection of a low-frequency kHz QPO, the sensitivity of power-spectral
techniques to broad components does not depend on the centroid frequency.
Therefore, there is no drop in sensitivity above 1200\,Hz, the maximum observed
frequency (although the sensitivity decreases for broader peaks and at equal $Q$
factor a higher centroid corresponds to a broader peak). 

\item both lower and upper kHz show phase lags, but of different sign. A study
of the properties of kHz QPOs in 4U\,1608-52 and 4U\,1636-53 showed that the lower
kHz peak shows soft lags of the order of a few tens of microseconds, while the
upper kHz peak shows hard lags of smaller amplitude \citep[][]{avellar2013}.

\item since their discovery, QPOs in NS LMXBs have been associated to the spin
of the compact object. Now we know the spin period of the NS in a good
number of systems and this can be tested. The results are puzzling. The first
accreting millisecond pulsar that was discovered, SAX\,J1808.4-3658 with a spin
frequency of 401\,Hz, showed two kHz QPO peaks at a separation of $\sim\,$195\,Hz,
consistent with half the spin frequency \citep[][]{Wijnands2003}. As new systems
became available with two kHz QPO peaks and a spin period, their difference
appeared to be close to $\nu_{spin}$ if $\nu_{spin} < 400$\,Hz, but close to
$\nu_{spin}/2$ if $\nu_{spin} > 400$\,Hz \cite[see][]{menbel2007}. However,
$\Delta\nu$ is not constant for any single source, so the association cannot be
exact. It was then shown that the existing data are also consistent with
$\Delta\nu$ being independent of spin and distributed around 300\,Hz in all
sources \citep[][see also Figure\,\ref{fig:deltanu}]{menbel2007}. However,
examining sources separately one can see that the values of $\Delta\nu$ as a
function of one of the two frequencies seem to `know' the value of
$\nu_{spin}$ or $\nu_{spin}/2$ \citep[see e.g.][]{Barret2005b,migliari2003}.

\begin{figure}
\begin{tabular}{cc}
   \includegraphics[scale=0.35]{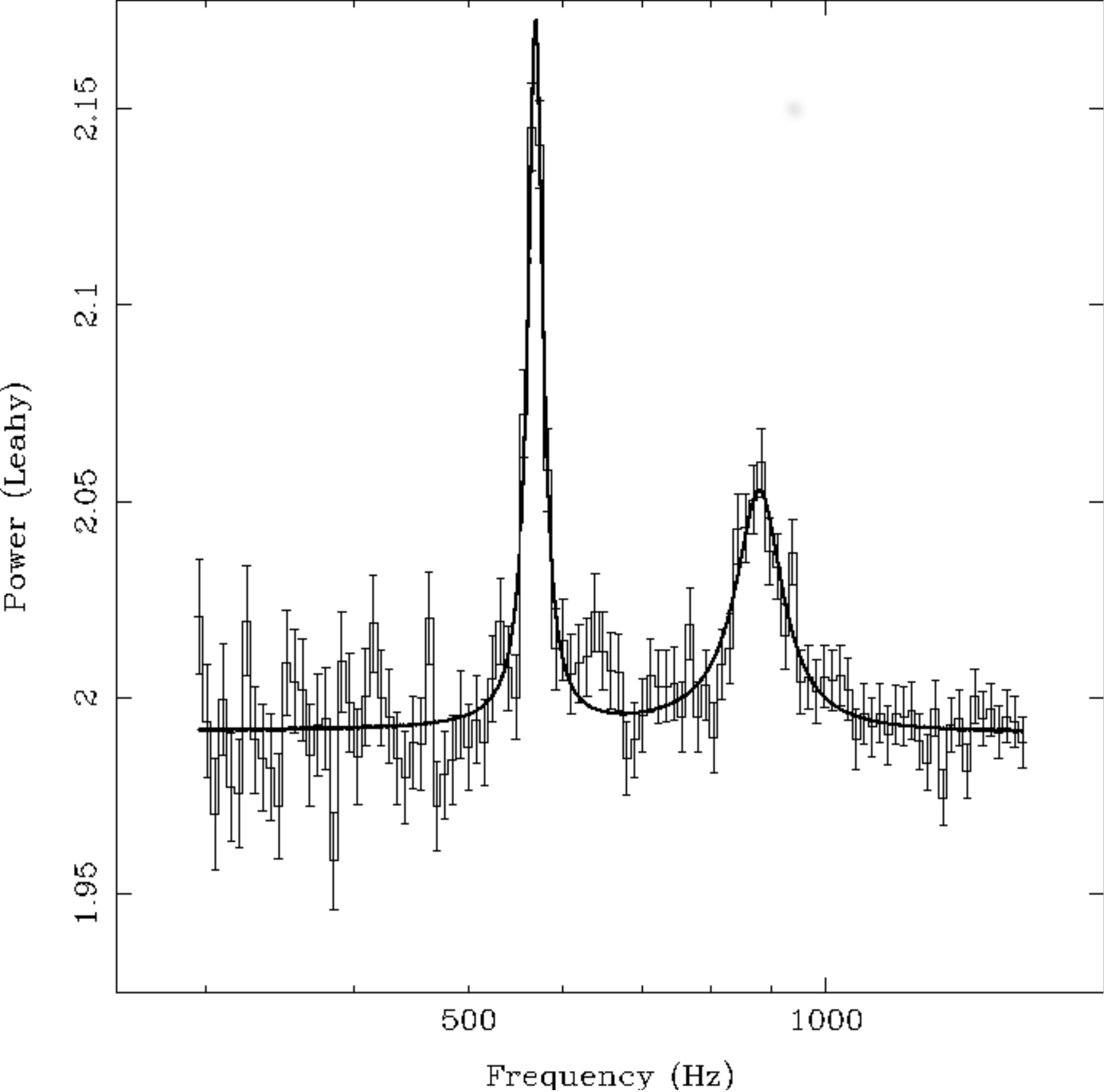} & 
   \includegraphics[scale=0.25]{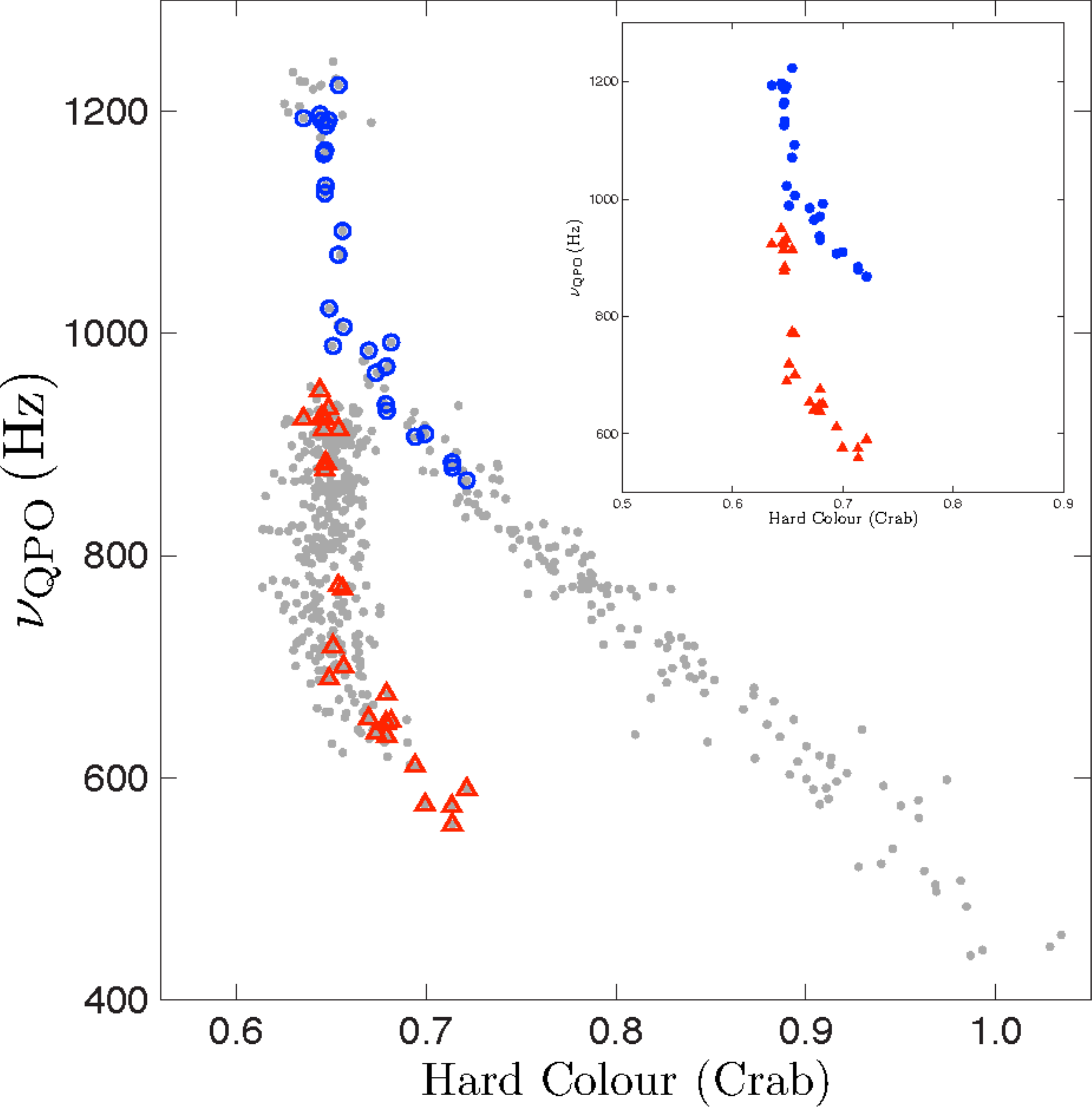} \\ 
\end{tabular}
\caption{ Left panel: PDS of the NS LMXB 4U\,1608-52 where the `twin peaks' of the kHz QPOs are
evident \citep[from][]{mendez1998}.  Right panel: kHz QPO centroid frequency as a function of X-ray
colour for all {\it RXTE} detections in 4U\,1636-53. Coloured dots correspond to pairs of simultaneously
detected lower (red) and upper (blue) kHz QPOs. Black points indicate single detections.
\citep[From][]{sanna2012}. 
\copyright AAS. Reproduced with permission.
}
\label{fig:khzQPO}
\end{figure}

{\bf Frequency correlations}

Excluding NBOs and FBOs, the characteristic frequencies of all components 
in the PDS of NS LMXBs, whether in the form of QPO or broad-band noise, 
correlate positively and different sources follow the same correlations 
(usually expressed as a function of the frequency of the upper kHz QPO, 
$\nu_u$ \citep[see][]{vanstraaten2005,linares2005}. There are two 
exceptions. The first is that the frequency of the so-called hecto-Hertz 
QPO component remains constant around $\sim\,$100\,Hz as $\nu_u$ varies as 
much as one order of magnitude. The second is that for accreting 
millisecond X-ray pulsars, some systems show displaced correlations that 
can be reconciled with the others only by multiplying the frequencies of 
both kHz QPO by a factor $\sim\,$1.5. Incidentally, with this factor the 
$\Delta\nu$ value for these sources would also become $\sim\,$300\,Hz 
\citep[][]{menbel2007}. Among these correlations, one is particularly 
important for theoretical models (see Section\,\ref{sec:5}): that between 
$\nu_{HBO}$ and $\nu_u$. It can be fitted rather well with a quadratic 
dependence \citep[][]{psaltisHBO}, although in the case of GX\,17+2 it was 
found to deviate for $\nu_u>$1000\,Hz \citep[][]{homan2002}.

While when examining kHz QPO the data are usually plotted in terms of 
$\Delta\nu$ vs. $\nu_u$, a more direct plot is that of $\nu_l$ vs. 
$\nu_u$, i.e. the frequency of the lower QPO vs. that of the upper QPO 
\citep[][]{Belloni2007b}. This plot for a set of sources can be seen in 
Figure\,\ref{fig:nu1nu2}. One can see that multiplying both frequencies by a 
factor 1.5 (see above) would bring the two accreting millisecond pulsars 
SAX\,J1808.4-3658 and XTE\,J1807-294 on the general correlation. The lines 
in the figure will be discussed below.

\begin{figure} 
   \includegraphics[scale=0.70]{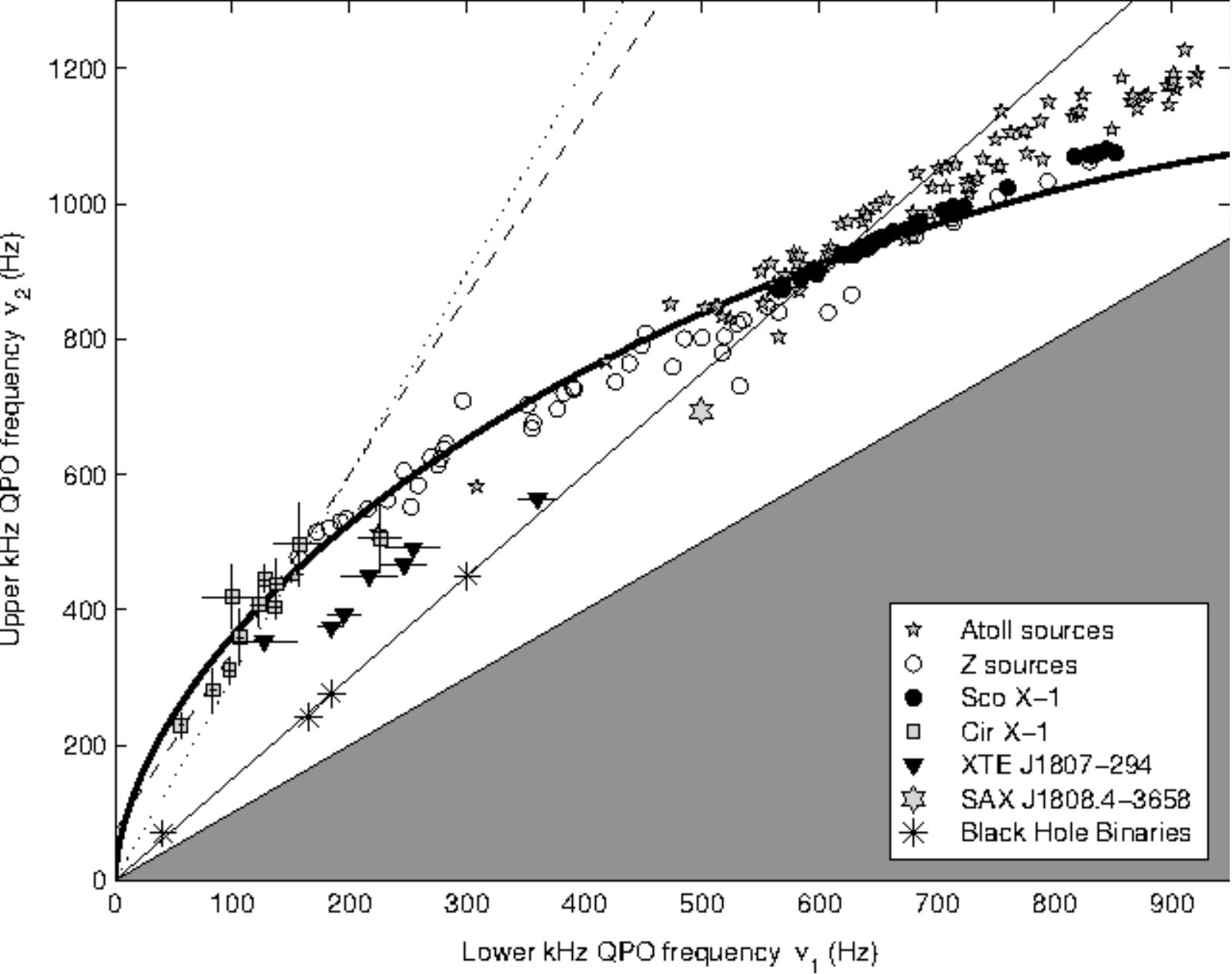} 
\caption{Correlation between lower and upper kHz QPO in NS LMXB. The points labeled 'Black Hole
Binaries' are included for comparison, although some of them have since then been reconsidered (see
text). The dark area correspond to lower frequency larger than upper frequency. The lines are: y=1.5x
(solid thin), y=3x (dashed) and the relation between the periastron-precession and the Keplerian
frequencies for a NS of 2\,M$_\odot$ (solid thick). \citep[From][]{Belloni2007b}. }
\label{fig:nu1nu2}
\end{figure}

\end{itemize}


\section{Black-Hole binaries}\label{sec:3}

\subsection{States and state-transitions}

Most of the known BHBs are LMXBs and most of them are transient. As in the 
case of NS systems, the large accretion rate swing in transients leads to 
a complex time evolution that can be subdivided into a number of separate 
states. These states are separated by marked transitions in either 
spectral or timing domain. Two main states were originally identified in 
the persistent system Cyg\,X-1 \citep[][]{tananbaum1972} and more were 
added when the first transient BHBs were discovered and observed 
\citep[][]{miyamoto1992,miyamoto1993,Belloni1996,Belloni1997,mendez1997}. 
As shown below, a more solid state classification and outburst evolution 
has emerged, with obvious links to that of NS LMXB presented above.

Most known BHBs are transient, spending most of the time in their 
`quiescent' state at a luminosity below $10^{33}$\,erg/s. At those low 
levels of accretion, X-ray observations yield few photons and detailed 
studies are not possible \citep[see][and references therein]{plotkin133}. 
Recent observations confirmed that the energy spectrum in quiescence is 
softer than that in the hard state (see below). Transients leave the 
quiescent state with a recurrence time that ranges from months to decades, 
as the accretion rate onto the BH increases by several orders of 
magnitude due to the onset of thermal-viscous instabilities 
\citep[see][for a review]{lasota}. During the outburst, the X-ray 
luminosity can approach the Eddington limit; typical durations are from a 
few days to several months, with a noticeable exception being the bright 
system GRS\,1915+105, whose outburst started in 1992 and at the time of 
writing has not ended.

\begin{figure} 
   \includegraphics[scale=0.50]{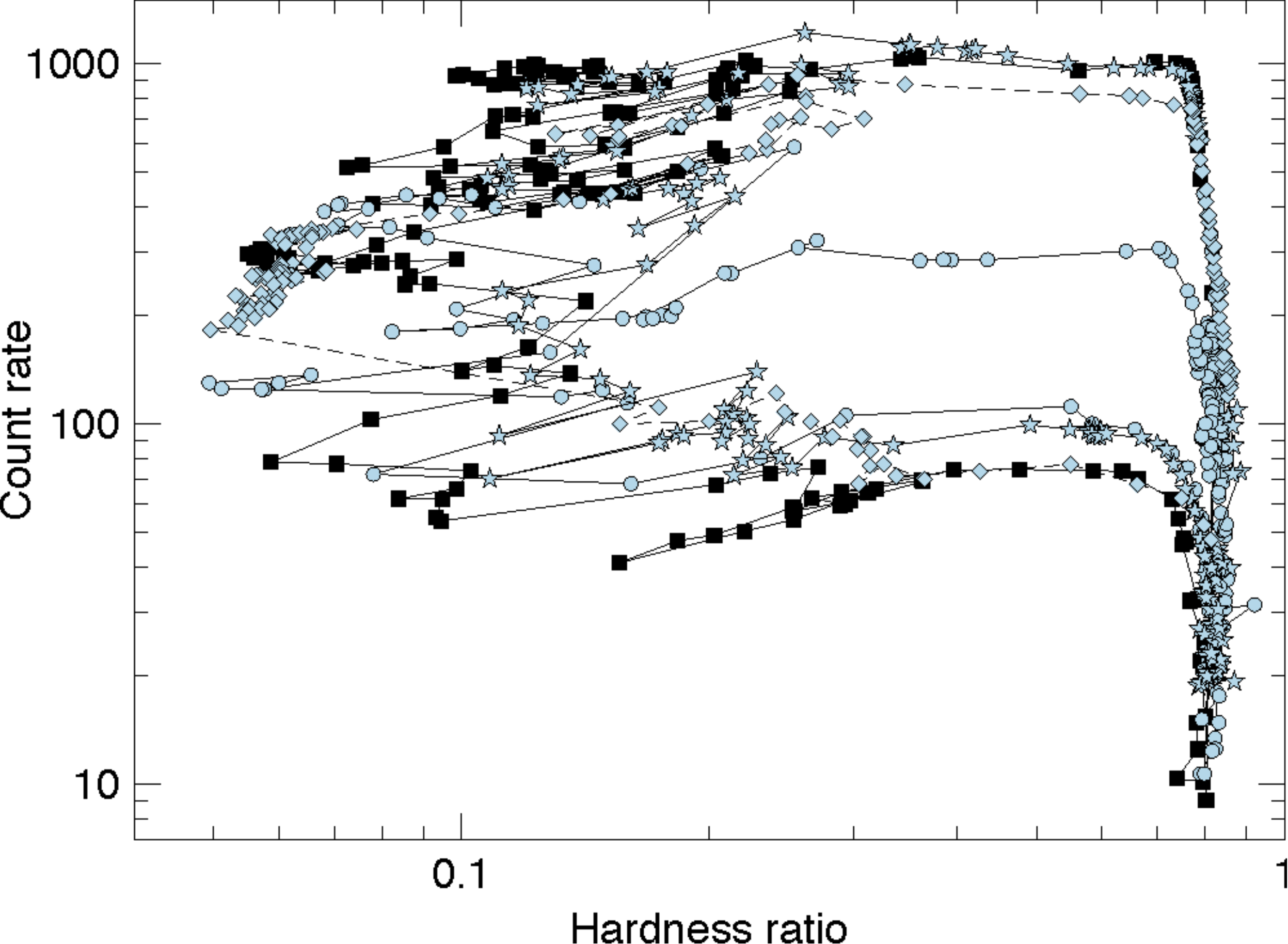} 
\caption{Hardness-Intensity diagram for four outbursts of GX\,339-4 observed
with {\it RXTE}. Each point corresponds to an observation and the lines connect them
in time sequence. The dashed line represents a rather large time gap.}
\label{fig:BHBHID}
\end{figure}

During an outburst, the spectral properties at X-ray energies change 
drastically, as does the fast variability: in addition to slow variations, 
rapid transitions are observed that led to the identification of source 
states, which are clearly related to the two original Cyg\,X-1 states. The 
sixteen years of dense observations with the RossiXTE satellite have led 
to a coherent picture to describe these variations in terms of few states, 
which have a clear correspondence to those of NS LMXBs shown above (whose 
outbursts are associated with the same instabilities). The X-ray light 
curves of different outbursts, even from the same recurrent source, can be 
very different, as they are influenced also by differences in the time 
evolution of the accretion rate. However, a clear pattern appears in most 
systems when a HID is produced. As in the case of NS LMXBs, different 
instruments will yield different HIDs and even different sources will be 
influenced by the interstellar absorption along their line of sight, but 
the pattern is strong and repeats in case of outbursts of the same source. 
The classical case is that of GX\,339-4: Figure\,\ref{fig:BHBHID} shows the 
HID for four consecutive outbursts of this prototypical system. It is 
clear that the evolution of the four outbursts is the same, only the 
levels at which the main transitions take place and the time-scales for 
the overall evolution change. The first outburst, from 2002, is marked in 
black to make it more visible. Following it, as time progresses the source 
follows the q-shaped path in counterclockwise direction, starting from the 
bottom right. The full outbursts starts in quiescence, well below the 
diagram in the figure. As mentioned above, the quiescent energy spectrum 
is softer in quiescence, which means that below the figure frame the stem 
of the `q' must bend to the left. The two original states correspond to 
the two vertical branches in the HID. The one at high hardness obviously 
corresponds to the low-hard state (LHS), the one on the left to the 
high-soft state (HSS). There are persistent systems that spend most of 
their time in the LHS, like Cyg\,X-1, and others that are found mostly in 
the HSS, like LMC\,X-3. Others, like LMC\,X-1, appear to be always in the 
same state, HSS in this case. The LHS branch is not really vertical, but 
shows a significant softening as the flux increases, while the large 
scatter observed in the HSS is partly caused by the magnification caused 
by the log scale. The similarity with the diagrams shown in 
Figures\,\ref{fig:ns_hid} and \ref{fig:teo} is evident. The same hysteretic 
behaviour is visible, leading to a loop-like path traveled 
counterclockwise. This means that consistently the hard to soft transition 
takes place at a higher flux (and possibly accretion rate) than the 
reverse soft to hard transition. Cyg X-1 does show transitions between 
hard and soft, but within the uncertainties the two transitions appear not 
to show hysteresis \citep[][]{belloni2010}.

Particularly interesting are the points in between the LHS and HSS, which 
have been classified into two intermediate states. They are observed as 
transitions from hard to soft and from soft to hard, but also shorter 
excursions from the HSS into these states can be observed. The exact 
positions of the transitions cannot be seen in Figure\,\ref{fig:BHBHID}, but 
can be measured precisely by monitoring other parameters such as the 
timing properties and infrared flux \citep[see][]{Belloni2005,homan2005}. 
Two intermediate states have been identified, called hard-intermediate 
state (HIMS) and soft-intermediate state (SIMS) 
\citep[see][]{belloni2010,Belloni2011,bellonimotta}. The boundary between 
them is very well defined in terms of timing properties and transitions 
have been observed as fast as a few seconds. Although not too evident in 
Figure\,\ref{fig:BHBHID} because of the large scatter in the HS points 
(partly due to the log scale), there are several small transitions between 
intermediate states, which can be seen clearly in the fast timing 
properties \citep[see below and][]{Belloni2005,Casella2004}. Not all 
sources display a perfectly shaped and repeatable `q' diagram, but the 
general evolution is similar. Two systems, GRO\,J1655-40 and XTE\,J1550-564 
have in addition a diagonal track in the upper part of the diagram, which 
has been called `anomalous' or 'hyperluminous' state. Here the energy 
spectrum and timing properties are complex and it seems to correspond to 
cases when accretion rate continues to increase after the bright HSS is 
reached.

There is substantial agreement on the fact that it is mass accretion rate 
driving the outburst and therefore the evolution along the HID. On a 
`clean' diagram like that in Figure\,\ref{fig:BHBHID}, the Y coordinate can 
be taken as a proxy for accretion rate: it increases from quiescence all 
the way to the upper left corner of the diagram, then decreases again down 
to quiescence. The vertical evolution is caused by accretion rate, but 
there is non agreement on what causes the horizontal evolution, in 
particular the two main transitions from hard to soft and from soft to 
hard. One possibility is that of a second parameter being involved, 
although a model based on the system memory has been proposed 
\citep[][]{kylafis2015}. A systematic difference between the shape of HIDs 
of low- and high-inclination has also been found \citep[][]{Teo2013}. Of 
course the interpretation of the evolution must rely ultimately upon 
detailed spectral analysis and the multiple components present in the 
energy spectra all contribute to the positioning in the HID; however, the 
HID is a clean way to present the evolution during an outburst and the 
inclination dependence would have been very difficult to discover from 
energy spectra alone \citep[see][]{kylafis2015,petrucci2008b}.

\begin{figure} 
   \includegraphics[scale=0.50]{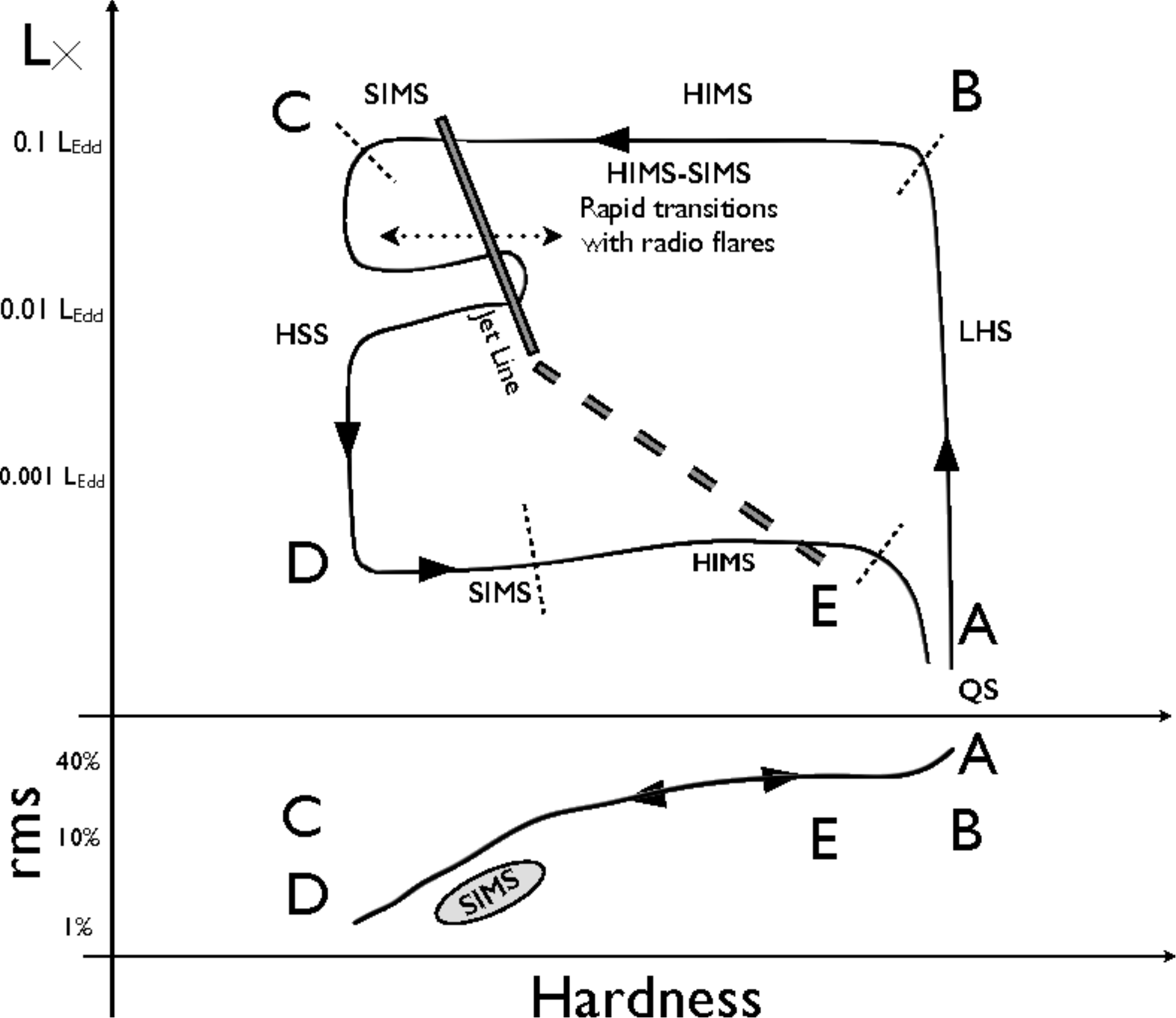} 
\caption{Schematic evolution of a BH transient in the HID (above) and HRD (below). The four states are
marked. Large letters indicate corresponding points in the two diagrams. The `jet line' marks the
ejection of superluminal jets at relativistic speeds 
\citep[From][]{bellonimotta}. }
\label{fig:evolution}
\end{figure}

In summary, a schematic HID for a BH transient is the one in Figure\,
\ref{fig:evolution}, where a second diagram is also included, the HRD, 
where the Y axis represents the integrated fractional variability in the 
Power Density Spectrum. It is evident that in this diagram there is no 
hysteresis and all points align along a general line from high variability 
in the LHS to very low levels of variability in the HSS. Only the points 
corresponding to the SIMS deviate from the general correlation, being 
characterised by lower variability (and very different timing features, 
see below)

\subsection{Energy spectra}

The Energy spectra of BHBs are also complex, but the absence of a solid 
surface simplifies their disentanglement.  The differences between spectra 
in different states are major and the transitions difficult to follow.

\begin{itemize}
\item{\it Low-Hard State} The energy spectrum is, obviously, hard and its main component 
can be roughly approximated with a power law with a photon index 1.5--1.8 and a 
high-energy cutoff  that can extend up to $\sim\,$100 keV (see Figure\,\ref{fig:bhspectra}). 
Even before the discovery of the cutoff, the presence of such a hard component was suggested 
as a direct evidence of a BH nature of the compact object, but a hard tail was then 
also observed in low-luminosity neutron-star binaries (see above). Most models associate 
this component to Comptonisation of softer photons from an optically thick accretion disk 
by electrons in a corona of either thermal or non-thermal nature \citep[see][]{gilfanov2010}, 
or possibly hybrid \citep[see e.g.][]{delsanto339}. Alternative models for the hard emission 
have been presented: bulk-motion Comptonisation, where the disk photons are energised by 
inverse Compton interaction with a fast converging flow 
\citep[see][and references therein]{turolla2002}. Alternatively, since we know that 
relativistic collimated jets are present in hard states, the corona can be replaced 
with the base of the jet and the direct contribution of the jet to the observed emission 
has been considered \citep[see][]{markoff2010}.
A common problem of Comptonisation models is that they depend on the geometry of the 
emitting region in a way that cannot be disentangled from intrinsic physical properties 
of the plasma, which means that additional information (such as fast timing variability 
or multi-wavelength observations) is needed in order to remove the degeneracy 
\citep[see e.g.][]{petrucci2008}.
In addition, a strong components from reflection of hard photons off a thermal disk is 
observed, in the form of a broad emission line and a continuum component at higher 
energies \citep[see][]{miller}, as in the case of NS systems (see above).

\begin{figure} 
   \includegraphics[scale=0.6]{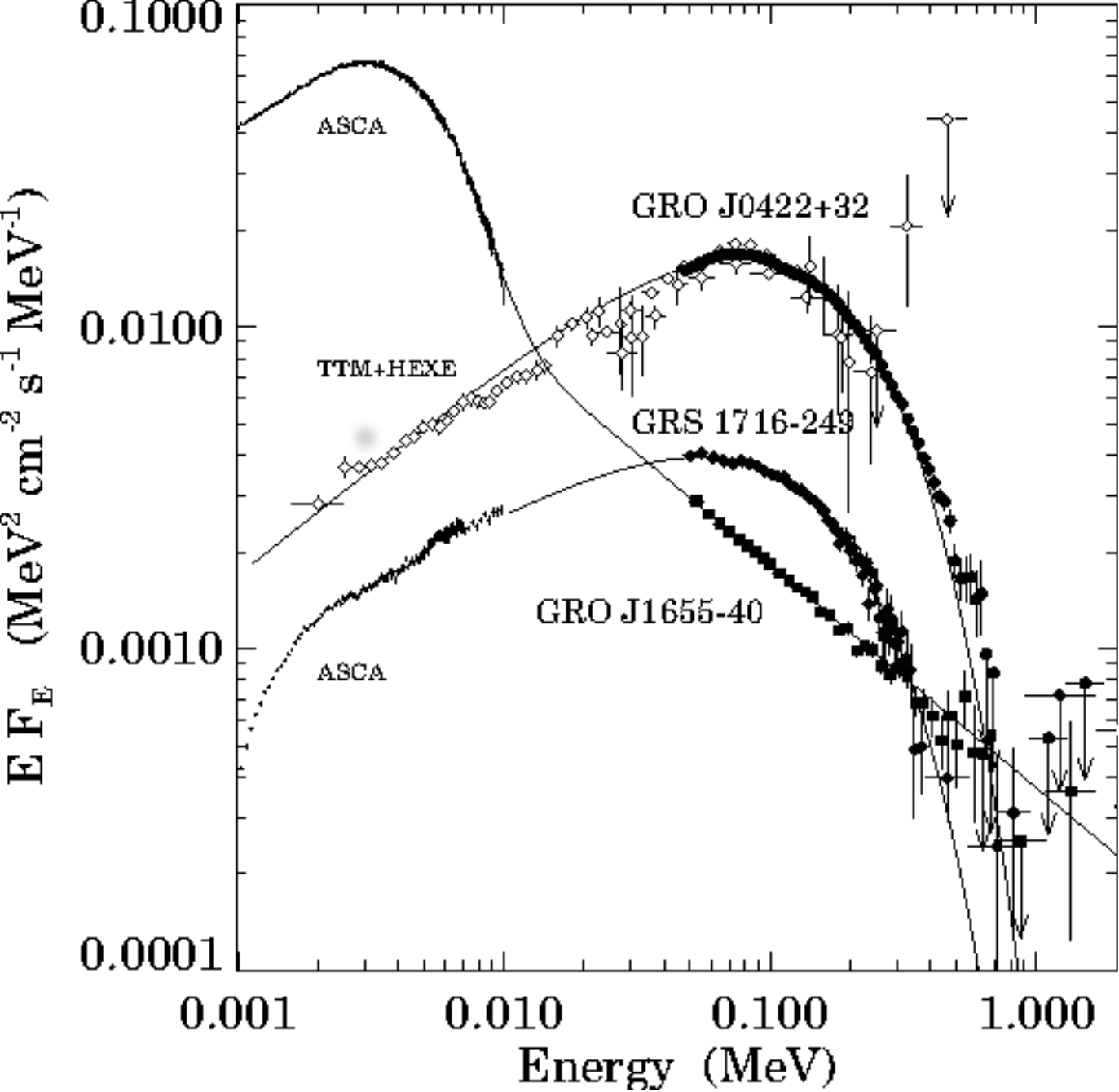} 
\caption{Examples of LHS (GRO\,J0422+32 and GRS\,1716-249) and HSS (GRO\,J1655-40) broad-band energy
spectra. \citep[From][]{grove}. \copyright AAS. Reproduced with permission.
}
\label{fig:bhspectra}
\end{figure}

For systems with low interstellar absorption, a thermal disk component is 
observed at low energies, modelled as a disk blackbody with a low 
temperature at the inner radius \citep[see e.g.][]{mcclintock1118}. 
However, there is disagreement as to what the inner radius of the disk is. 
Timing measurements suggest that the accretion disk is truncated at rather 
large radii, but the presence of strong reflection components leads to the 
opposite conclusion, namely that the disk extends to radii close to the 
BH \citep[see][and references therein]{bellonimotta}. Notice that 
the fits to the disk component are complicated by the presence of 
absorption and by the fact that at a fixed temperature the inner radius of 
the disk is determined by the normalization of the fit component 
\citep[see][]{kolehmainen}.

The LHS has been studied in detail in the persistent system Cyg\,X-1, which 
spends most of its time in that state \citep[][]{wilms2006} and in 
transients. As shown above, this state occurs at the beginning of the 
outburst, but also at the end, when the flux is lower, extending into the 
quiescent state. The ouburst decay properties also have been studied 
extensively \citep[see][]{kalemci2013,dincer2014}. The number of available 
observations at lower luminosity levels is much lower, due to the need of 
more sensitive instruments like those on board {\it Chandra} and {\it XMM-Newton}. In 
the bright LHS, the spectrum hardens as the source dims, but as luminosity 
goes down below $\sim\,$10$^{-2}$L$_{\rm Edd}$ it starts softening until the 
spectral photon index reaches a `plateau' value of $\sim\,$2 
\citep[see][]{sobolewska,plotkin}.

\item{\it High-Soft State-} In the HSS, the energy spectrum is dominated by a 
thermal disk component, with an inner temperature of the order of 1\,keV and 
extending to radii close to the BH. The original model of an optically 
thick and geometrically thin disc represents well the observations \citep[][]{ss}. 
However, the high statistics makes it possible to fit the spectrum with a more 
complex accretion disc model in order to take into account of all emission 
properties and estimate a precise value for the inner radius \citep[see][]{davis}. 
Assuming the radius is at the innermost stable circular orbit, this leads to the 
measurement of BH spins \citep[see][]{middleton}. As the HSS is observed 
over a wide range of luminosities and the inner radius is not observed to vary, 
the decay in mass accretion rate leads to a progressive softening of the spectrum. 
At high energies, an additional weak hard component is observed, often variable 
between observations (see Figure\,\ref{fig:bhspectra}). The characterisation of 
this component is difficult due to its faintness and the dilution by the disk 
component at low energies \citep[][]{grove,zdz2012}. Reflection features have been 
observed also in the HSS, where again measurements are complicated by the direct 
disk emission \citep[][]{steiner2016}.

\begin{figure} 
   \includegraphics[scale=0.30,angle=90]{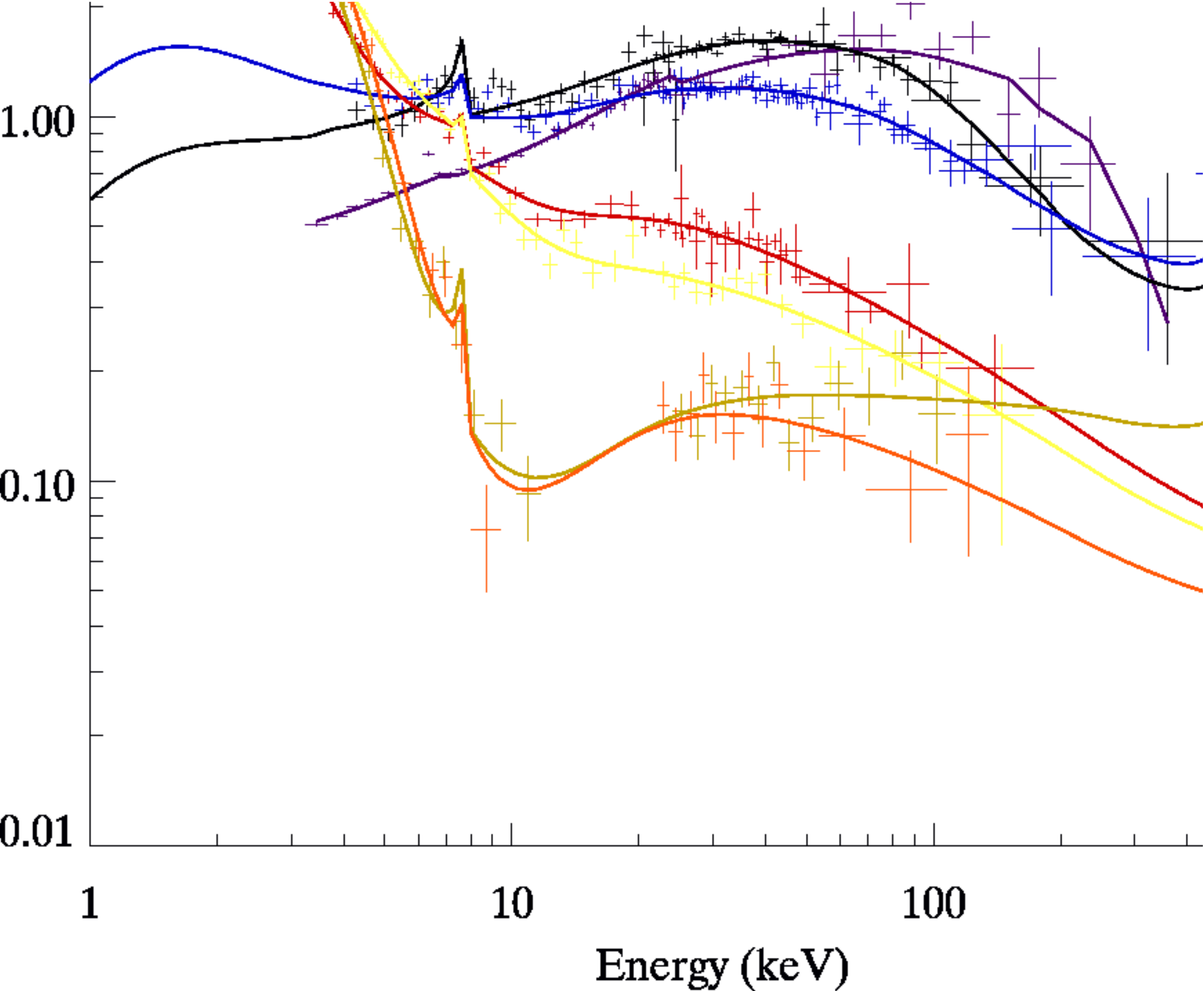} 
\caption{Examples of intermediate-state energy spectra from GX\,339-4. The violet 
spectrum is an example of LHS for comparison. The spectral fits are with a hybrid 
Comptonisation model, a thermal disk and a reflection complex. 
\citep[From][]{delsanto339}}
\label{fig:hims}
\end{figure}

\item{\it Intermediate States} The energy spectrum during intermediate states is
indeed intermediate. The transition from a hard spectrum dominated by a
Comptonisation component featuring a weak disk component to a soft spectrum
dominated by a disk component is gradual (see Figure\,\ref{fig:hims}). Fits with
hybrid Comptonisation models indicate that the spectral evolution is drive by a
decrease in the ratio between electron heating and soft cooling flux from the
disk \citep[][]{delsanto339}. In GX\,339-4, the high-energy cutoff in the
spectrum, which decreased from 120\,keV to 60\,keV as the flux increased in the
LHS, showed a marked increase in the HIMS up to above 100\,keV, after which it
became too high to be detected by {\it RXTE} \citep[][]{mottacutoff}. Whether it was
still detected in the only SIMS observation is unclear, as only one observation
was made in that state. The same behaviour was observed in GRO\,J1655-40
\citep[][]{mottacutoff,joinet}. On the return track from soft to hard state,
sources go through the same intermediate states (see Figures\,\ref{fig:BHBHID} and
\ref{fig:evolution}). The main spectral differences compared to the high-flux
branch are a softer disk component and a flatter hard component, whose photon
index reaches LHS values at the start of the HIMS \citep[][]{stiele}.

\end{itemize}

\subsection{Timing properties}

As mentioned above, the movement on the HID is determined by spectral variations, but 
the discrete states are defined in terms of timing properties, which show marked 
variations corresponding to specific hardness levels. In analogy of what done above for 
NS systems, I will present the main characteristics of timing features distinguishing 
low and high frequencies \citep[see][for a recent review]{mottaAN}. 

\vspace{5mm}

{\bf Low frequencies}

In addition to noise components that can be very 
strong and completely dominate the variability, 
three LFQPOs have been identified, corresponding to 
different source states \citep[see][]{Casella2005}. 
Their properties are different and they have been 
detected simultaneously in a few cases, confirming 
that they are different signals, although of course 
this does not rule out the same physical origin.

\begin{itemize}

\item {Type-C QPO}. These are the most common low-frequency oscillations,
observed in the HIMS and in the bright LHS. They have been detected also in the
HSS \citep[see e.g.][]{motta2012}. Its centroid frequency is anti-correlated
with hardness: in the LHS it can be below 0.1\,Hz, in the HIMS it is typically
1--10\,Hz and in the HSS it can reach 30\,Hz
\citep[see][]{belloni2010,bellonimotta}. Type-C QPOs are rather strong, reaching
up to 20\,\% fractional RMS, and narrow, with a Q factor $\sim\,$10. Their centroid
frequency varies over a large range (0.1--15\,Hz) and they appear associated to
band-limited noise components (see Figure\,\ref{fig:typeBC}). Often a second
harmonic peak is detected, rarely a third harmonic. In many cases, a
sub-harmonic feature at half the QPO frequency is also seen \cite[see
e.g.][]{takizawa,Casella2004,rao2010,mottaAN}. The frequency of type-C QPO
correlates positively with the characteristic frequencies of noise components
(see below).  A strong anti-correlation is observed between the QPO centroid
frequency and the fractional RMS measured over a broad frequency range
\citep[not the RMS of the QPO itself,][]{Casella2005,motta2011}.

\begin{figure}
\begin{tabular}{cc}
   \includegraphics[scale=0.2]{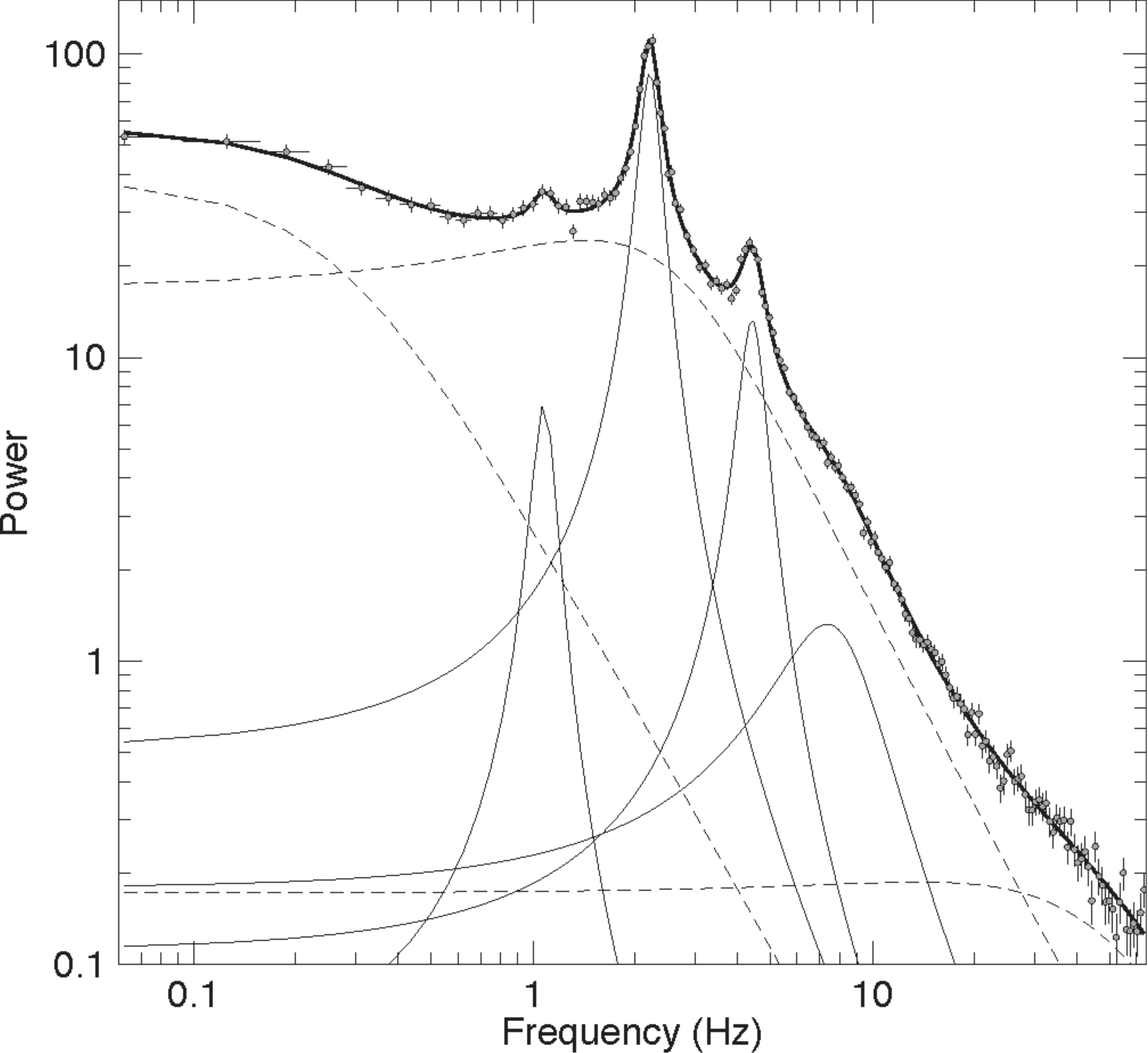} & 
   \includegraphics[scale=0.2]{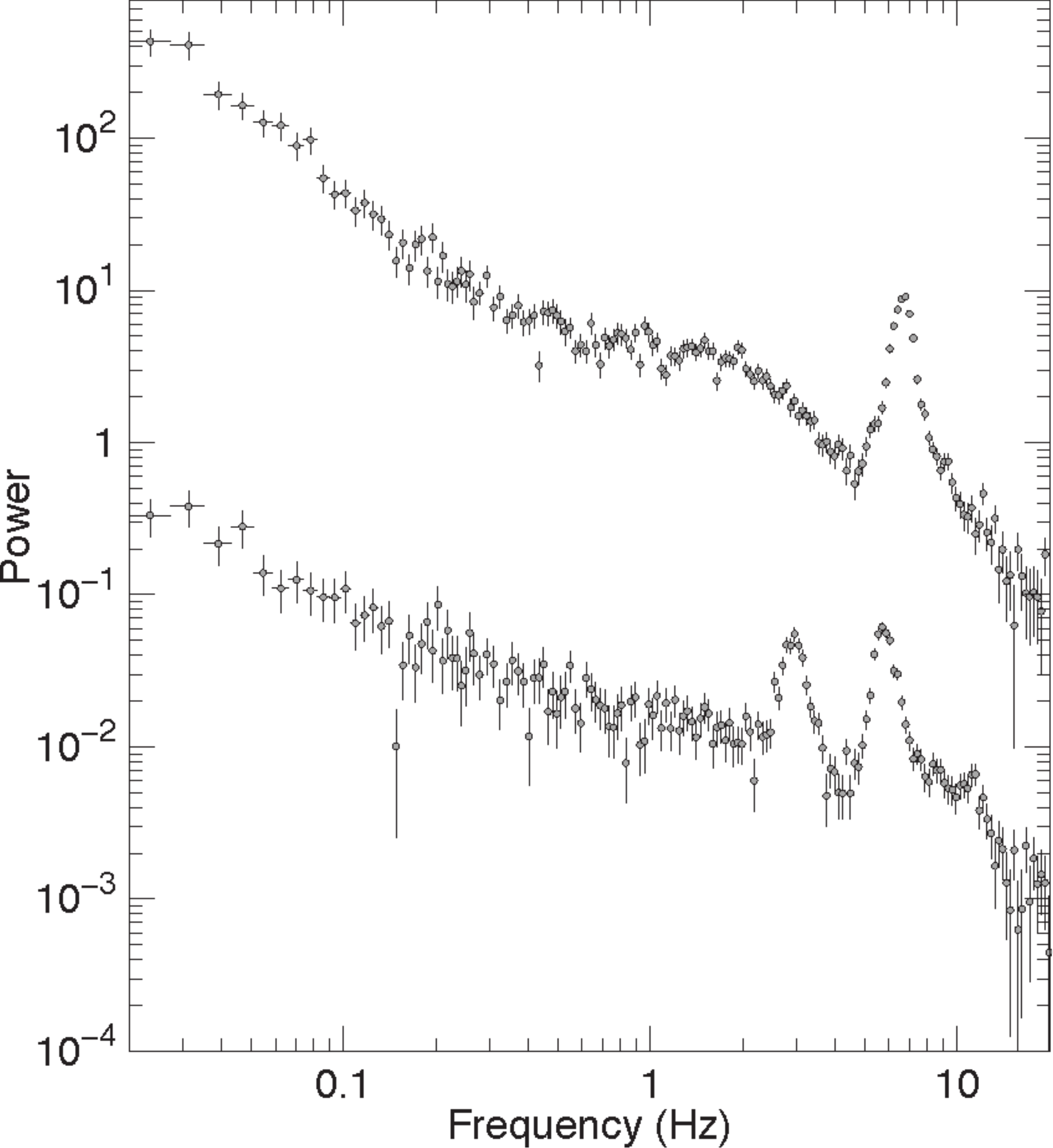} \\ 
\end{tabular}
\caption{ Left panel: PDS from GRS\,1915+105 in its C state, corresponding to the HIMS
\citep[from][]{ratti}. Right panel: two examples of type-B QPO from GX\,339-4 in the SIMS, shifted in
power for clarity. In the top one the QPO does not have harmonics, in the second one the subharmonic
peak is as strong as the fundamental (the so-called `cathedral' QPO). }
\label{fig:typeBC}
\end{figure}

\item {Type-B QPO}. These oscillations are the defining trait of the SIMS. The
peaks are relatively strong around $\sim\,4-5$\,\% RMS and with $Q$ larger than 5.
In contrast with type C, they are observed over a limited range of frequencies
($\sim\,1-7$\,Hz). They are sometimes observed with a second harmonic and a
sub-harmonic, which at times can be as strong as the fundamental \citep[the
so-called `cathedral' QPO, see][and Figure\,\ref{fig:typeBC}, left
panel]{Casella2004}. They are associated with a power-law noise with a few \,\%
fractional RMS, which means the QPO dominates the observed variability 
(see Figure\,\ref{fig:typeBC}). The centroid frequency is observed to vary rapidly with a
typical time-scale of $\sim\,10$\,s.
\citep[][]{nespoli2003,Casella2004,motta2011}.  Rapid transitions have been
observed between this QPO and the other types, with transition times as short as
a few seconds \citep[][]{nespoli2003,Casella2004,motta2011}. In GX\,339-4, a
strong correlation was found between centroid frequency and source flux above 6
keV, where the disk contribution is negligible. For type-C QPOs there is a
general anti-correlation, although not so tight \citep[][]{motta2011}. A few
observations at very high flux of GX\,339-4 have led to the simultaneous
detection of a type-B and a type-C QPO, confirming that the two oscillations are
different \citep[][]{motta2012}.

\item {Type-A QPO}. This QPO is also associated with weak power-law noise, but the
peak is much more blunt, with Q$\leq 3$. Its centroid frequency is always around
6--8\,Hz. Given these characteristics that make it more difficult to detect and
characterise, not many details are know. No harmonic content appears to be
present, although the low coherence of the peak make it difficult to study
additional peaks. These QPOs are observed in the HSS, close to transitions from
the SIMS \citep[see][]{Casella2004,motta2012}.

\item {Noise components}

\begin{figure}
   \includegraphics[scale=0.30]{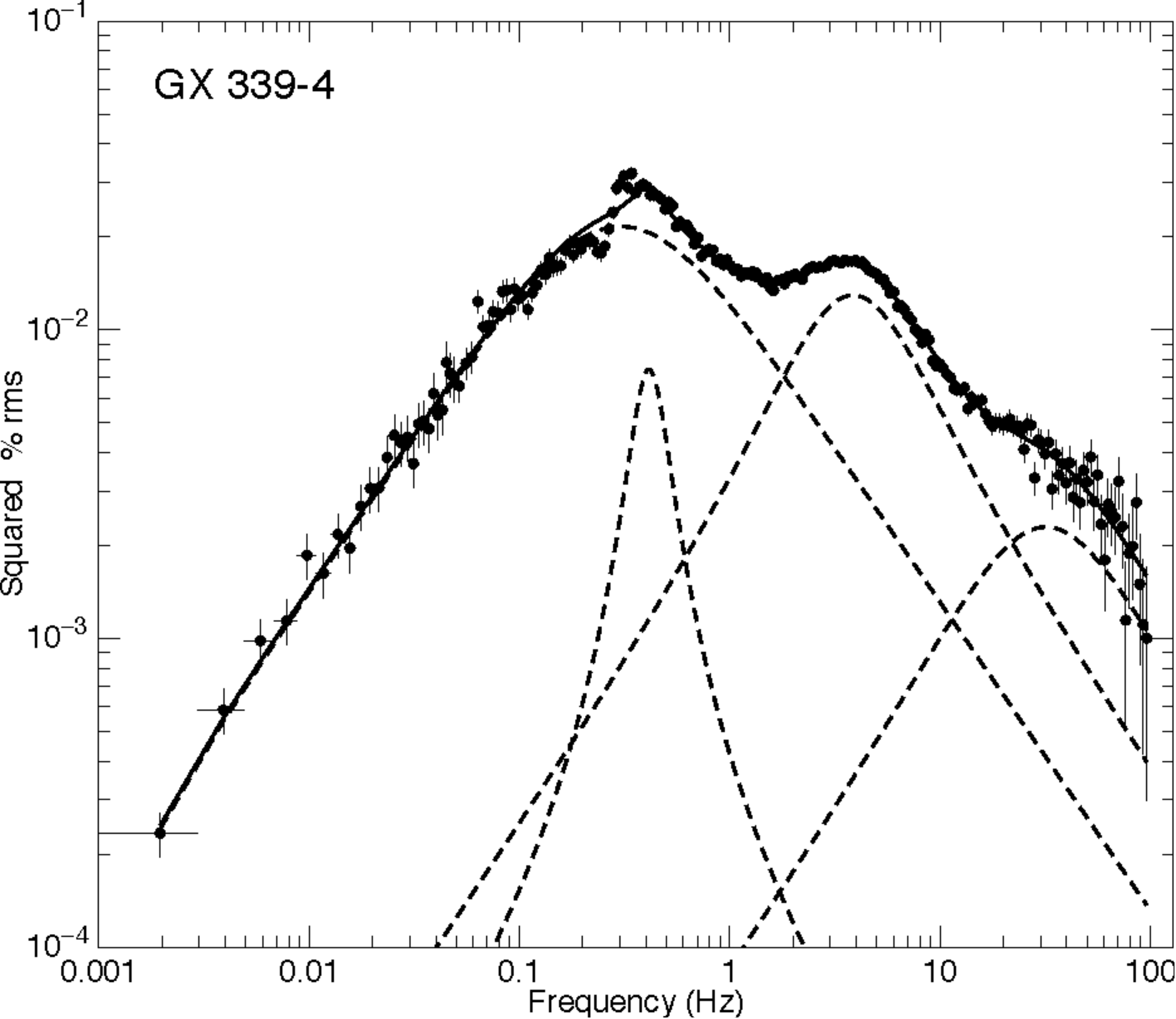}
\caption{Power density spectrum of a set of {\it RXTE} observations of GX\,339-4 in hard state. The best fit
with four Lorentzian components is shown. The quality of the fit in the 0.1--1\,Hz range is poor because
of variability between the observations \citep[see][]{nowak2000}. }
\label{fig:lhs}
\end{figure}

In the LHS, the variability is very strong and dominated by band-limited 
noise components. As shown in the HRD in Figure\,\ref{fig:evolution}, the 
total fractional RMS can be as high as 40\,\%. As in the case of NS LMXBs 
presented above, the PDS can be decomposed into the sum of at least four 
separate components, called called $L_b$, $L_{LF}$, L$_\ell$ and $L_u$ 
\cite[see][and Figure\,\ref{fig:lhs}, left panel]{bpk}. The first fits with 
flat-top Lorentzian components were made by \citet{olive1998} on a NS and 
a BH system, but the first systematic attempt of replacing power laws with 
broad Lorentzians was presented by \citet{nowak2000} and further 
generalized by \citet{bpk}. The $L_b$, L$_\ell$ and $L_u$ are broad, 
usually represented by zero-centered Lorentzians, while $L_{LF}$ is the 
type-C LFQPO more evident in the HIMS. A study of a large number of {\it RXTE} 
observations of Cyg\,X-1, most of which taken in the LHS, revealed a 
complex relationship between the parameters of the four Lorentzians in 
this source \citep[][]{pottschmidt2003}. In X-ray transients, as flux 
increases and the spectrum softens in the LHS, the characteristic 
frequencies of the four components increase and their fractional RMS 
decreases, until at the onset of the HIMS L$_\ell$ and $L_u$ are normally 
not detected anymore due to their frequency being too high and their RMS 
too low. Since these results come from the analysis of {\it RXTE} data, the 
{\it RXTE}/PCA instrument has a lower energy threshold of $\sim\,$3\,keV and most 
X-ray transients suffer of high levels of interstellar absorption, in most 
cases the PCA only detects photons from the high-energy component and 
cannot see the thermal disk component (see above). Therefore, the observed 
variability can only originate from the Comptonised component. However, 
observations of GX\,339-4 with {\it XMM-Newton} have led to the discovery that in 
the LHS on time-scales longer than a second the variability of the thermal 
disk, detected below 3 keV, {\it leads} that of the hard component, while 
the reverse happens on faster time-scales. The conclusion is that in the 
LHS the thermal disk variability is responsible for the hard-component 
one, while on short time-scales disk heating effects reverse the casual 
connection \citep[][]{uttley2011}. This result is supported by the 
observation that on long time-scales the thermal-disk variability is 
stronger than the one from the Comptonised component 
\cite[][]{Wilkinson09}. Another important observational results, which 
appears to hold in other states other than the LHS, is the so-called 
`RMS-flux' relation: the total (not fractional) RMS is linearly 
correlated with the observed count rate.  This property cannot be 
explained in terms of older models involving the superposition of randomly 
occurring `shots' and is naturally interpreted in terms of 
propagating-fluctuation models \cite[see][]{lyubarskii}, in which the 
slower variations from the outer parts of the accretion disk propagate 
inwards and are then further modulated at higher frequencies by more 
internal portions of the disk.

In the HIMS the noise is dominated by the high-frequency extension of 
$L_b$ and the characteristic frequencies continue to increase as the 
energy spectrum softens. A thermal disk component starts contributing to 
the {\it RXTE} energy band, but the fractional RMS at those energies decreases 
in a way compatible with all variability being associated to the hard 
component, while the detected thermal-disk flux is not variable. This is 
in marked contrast with what discussed above for the LHS, meaning that the 
disk varies more when not observable, but becomes `quiet' when 
dominating the flux. Broader studies have shown that the RMS-flux relation 
originally observed for the LHS is present across states, with rather 
complex properties \citep[see][]{teoRID,heil2012}.

After the transition to the SIMS the band-limited noise components 
disappear and are replaced by a steep power-law component. The variability 
can be as low as $\sim\,$1 fractional RMS.
\end{itemize}

{\bf High Frequencies}

The first High-frequency QPO (HFQPO) from a BHB was detected during the first year of 
{\it RXTE} operation in the bright and peculiar transient GRS\,1915+105 \citep[][]{morgan}. 
Its frequency was around 65--67\,Hz, with a $Q$ value of $\sim\,$15 and a fractional RMS 
amplitude of 1--1.6\,\%. However, after sixteen years of observations of BHBs, only few 
detections in other systems have been found \citep[see][]{belloniHFQPO,mottaAN}. 
These are 
XTE\,J1550-564 \citep[][]{remillard1999a,homan2001,miller2001,remillard2002}, 
GRO\,J1655-40 \citep[][]{remillard1999b,strohmayer2001a}, 
XTE\,J1650-500 \citep[][]{homan2003}, 
H1743-322 \citep[][]{homan2005b,Remillard2006b},
and IGR\,J17091-3624 \citep[][]{altamiranobelloni2012}.
Detections were reported also for 
XTE\,J1859+226 \citep[][]{klein}
and 4U\,1630-47 \citep[][]{klein}
although their significance appears to be low. 

From this limited number of detections we have now the following 
observational picture, where GRS\,1915+105 is dealt with as a separate 
case.

\begin{figure}
\begin{tabular}{cc}
   \includegraphics[scale=0.22]{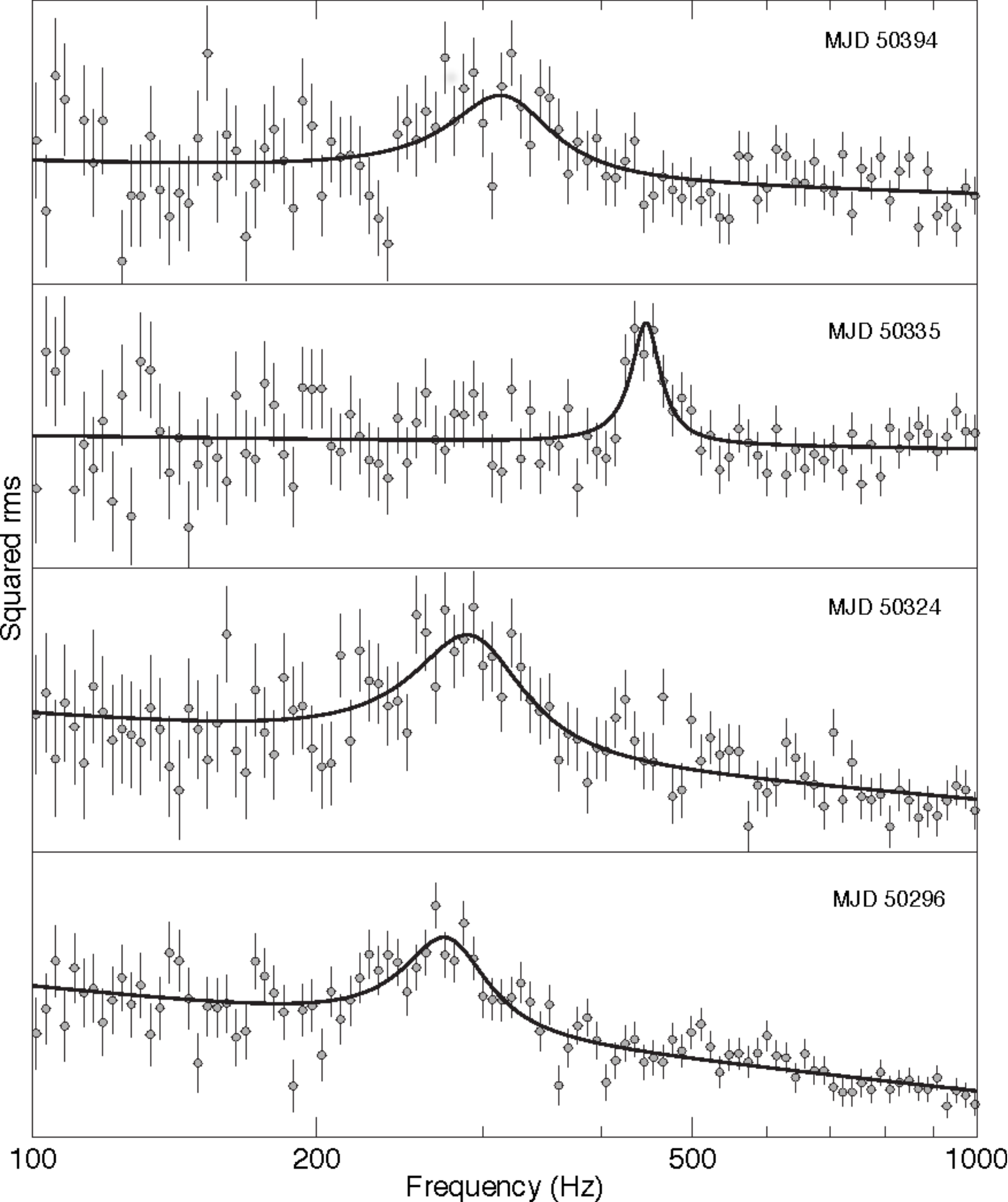} & 
   \includegraphics[scale=0.20]{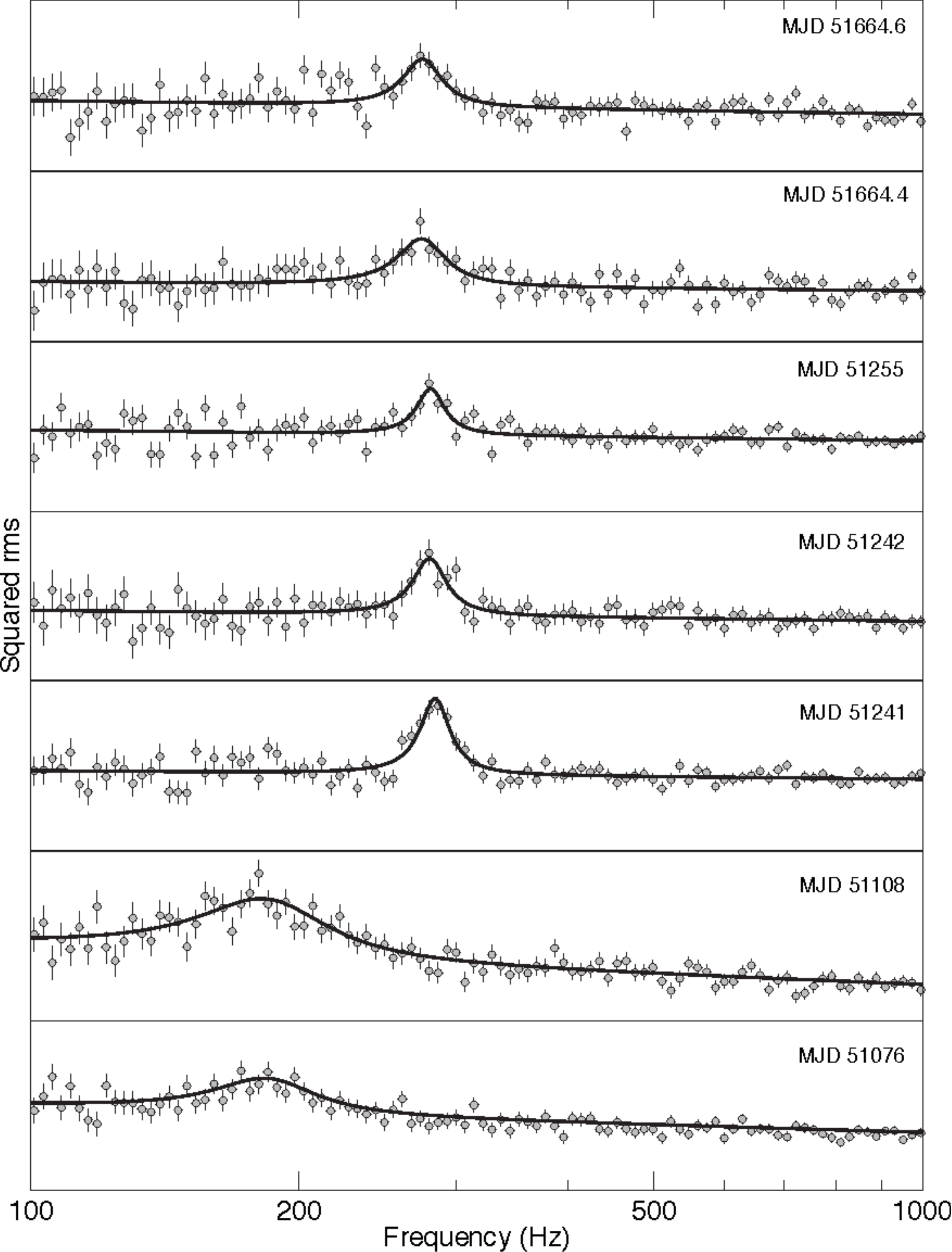} \\ 
\end{tabular}
\caption{ Left panel: detections of single HFQPO peaks in GRO\,J1655-40, peaking at $\sim\,$300\,Hz and
$\sim\,$450\,Hz. Right panel: detections of single HFQPO peaks in XTE\,J1550-564, peaking at $\sim\,$180\,Hz
and $\sim\,$280\,Hz \citep[from][]{belloniHFQPO}. }
\label{fig:hfqpo1a}
\end{figure}

\begin{itemize} 

\item In one source, GRO\,J1655-40,  there is a significant detection of two simultaneous peaks, 
at $\sim\,300$\,Hz and $\sim\,440$\,Hz, with other detections of single peaks around the same value 
(see Fig.\ref{fig:hfqpo1a}, left panel), although statistically different 
\citep[][]{strohmayer2001a,belloniHFQPO,motta2014}. In XTE\,J1550-564 two separate peaks 
have been detected around 180\,Hz and 280\,Hz (see \ref{fig:hfqpo1a}, right panel), but not 
at the same time, although in one case there is a hint of a second peak \citep[][]{miller2001}. 
It is tempting to consider them two separate peaks, although comparisons based on phase lags 
suggest that they might be the same component changing frequency \citep[][]{mendez2013}. 
Besides this consideration, the variations in the centroid frequency of all HFQPOs 
are minor, suggesting they are associated to fixed frequencies. 

\begin{figure}
\begin{tabular}{cc}
   \includegraphics[scale=0.22,angle=270]{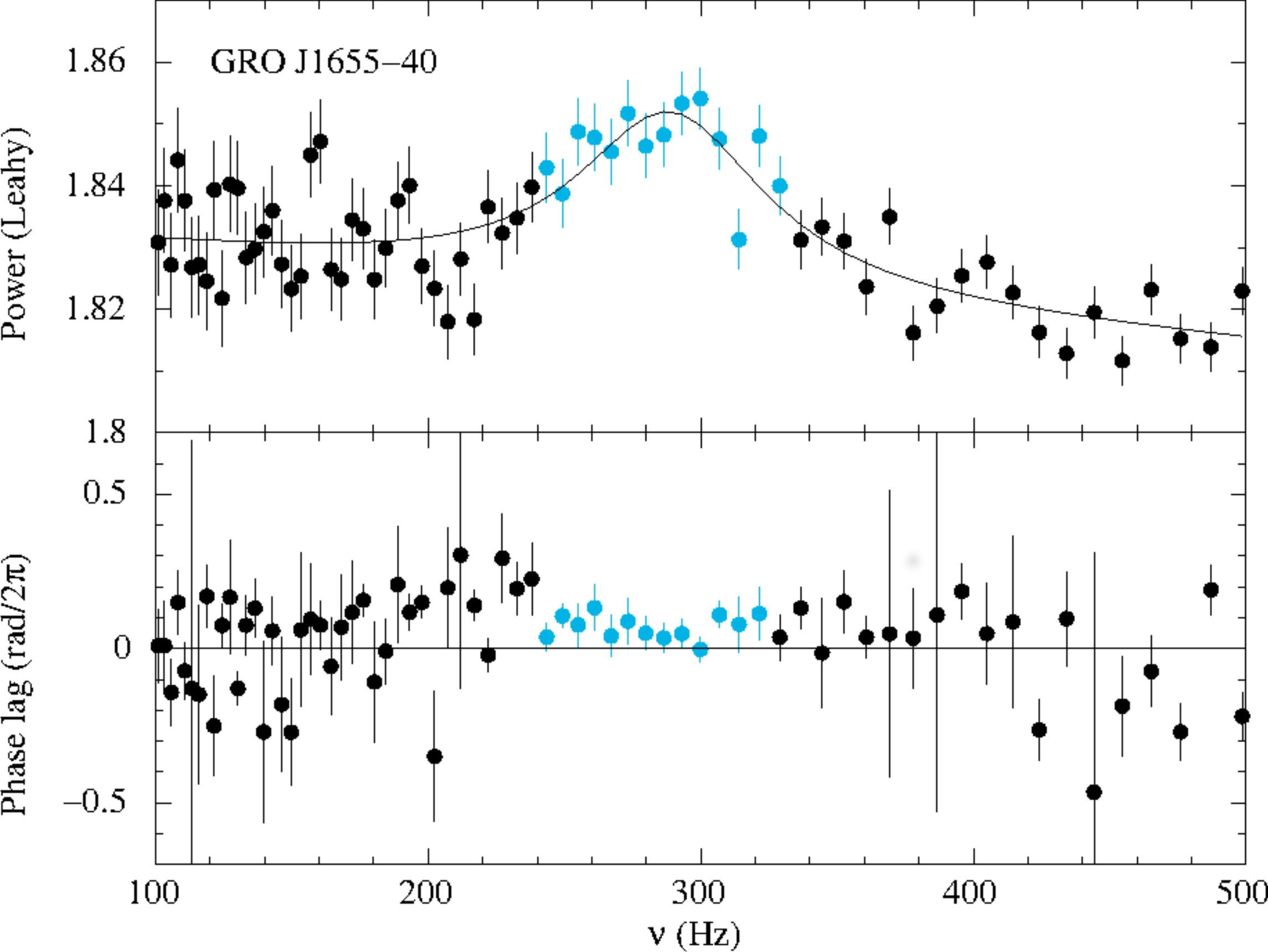} & 
   \includegraphics[scale=0.22,angle=270]{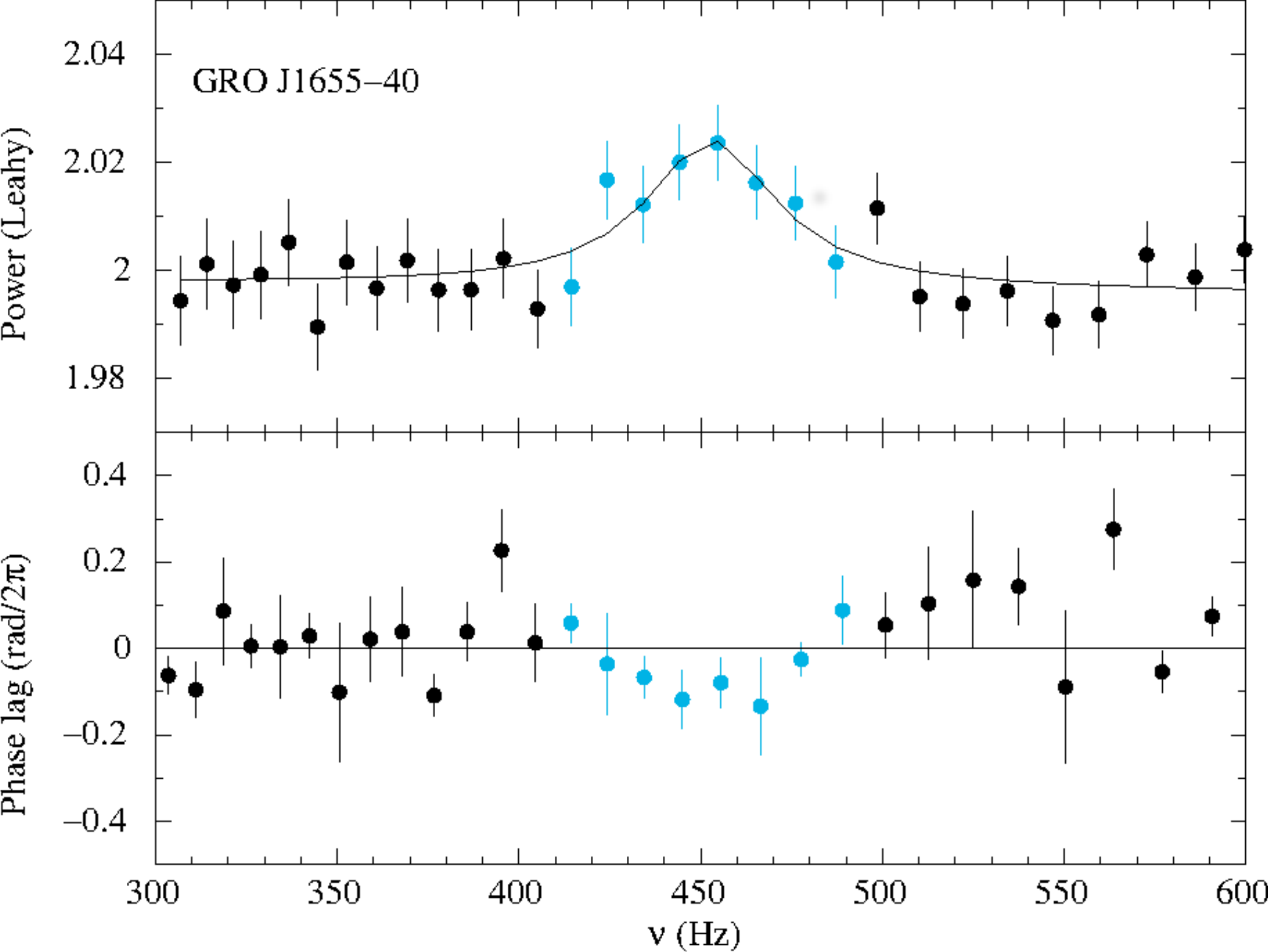} \\ 
\end{tabular}
\caption{ Top panels: detections of single HFQPO peaks in GRO\,J1655-40, peaking
at $\sim\,$300\,Hz and $\sim\,$450\,Hz. Bottom panels: corresponding phase-lag
spectra. \citep[from][]{mendez2013}. }
\label{fig:hfqpo1b}
\end{figure}

\item The fractional RMS of HFQPOs is between 1 and 6, but it is
strongly energy dependent, being much stronger at high energies. Quality factors
are in the range 1--5.

\item All detections correspond to observations at high flux. While this could
be due to a sensitivity bias for weak signals, in GRO\,J1655-40 and XTE\,J1550-564
it is clear that all detections correspond to the `anomalous' or 
`hyperluminous'
state, which can be observed at higher fluxes than those shown in the diagrams
above \citep[see][]{belloni2010,belloniHFQPO}. No detection is reported in the
LHS and in the HSS, indicating that they are associated to intermediate states,
while connected to the hard component as their RMS is highest at energies $>$10
keV.

\item When two peaks are observed simultaneously, in GRO\,J1655-40 and GRS\,1915+105, the lower 
peaks has soft lags and the upper peak has hard lags
\citep[see Figure\,\ref{fig:hfqpo1b};][]{mendez2013}. For XTE\,J1550-564, as mentioned,
the 180\,Hz and 280\,Hz peaks have similar lags, which suggests they might be the
same peak at different frequency. 

\item In the case of GRO\,J1655-40 and XTE\,J1550-564, the frequencies of the two
peaks are around a 3:2 ratio \cite[see][]{abrklu2001}.

\begin{figure}
\begin{tabular}{cc}
   \includegraphics[scale=0.20]{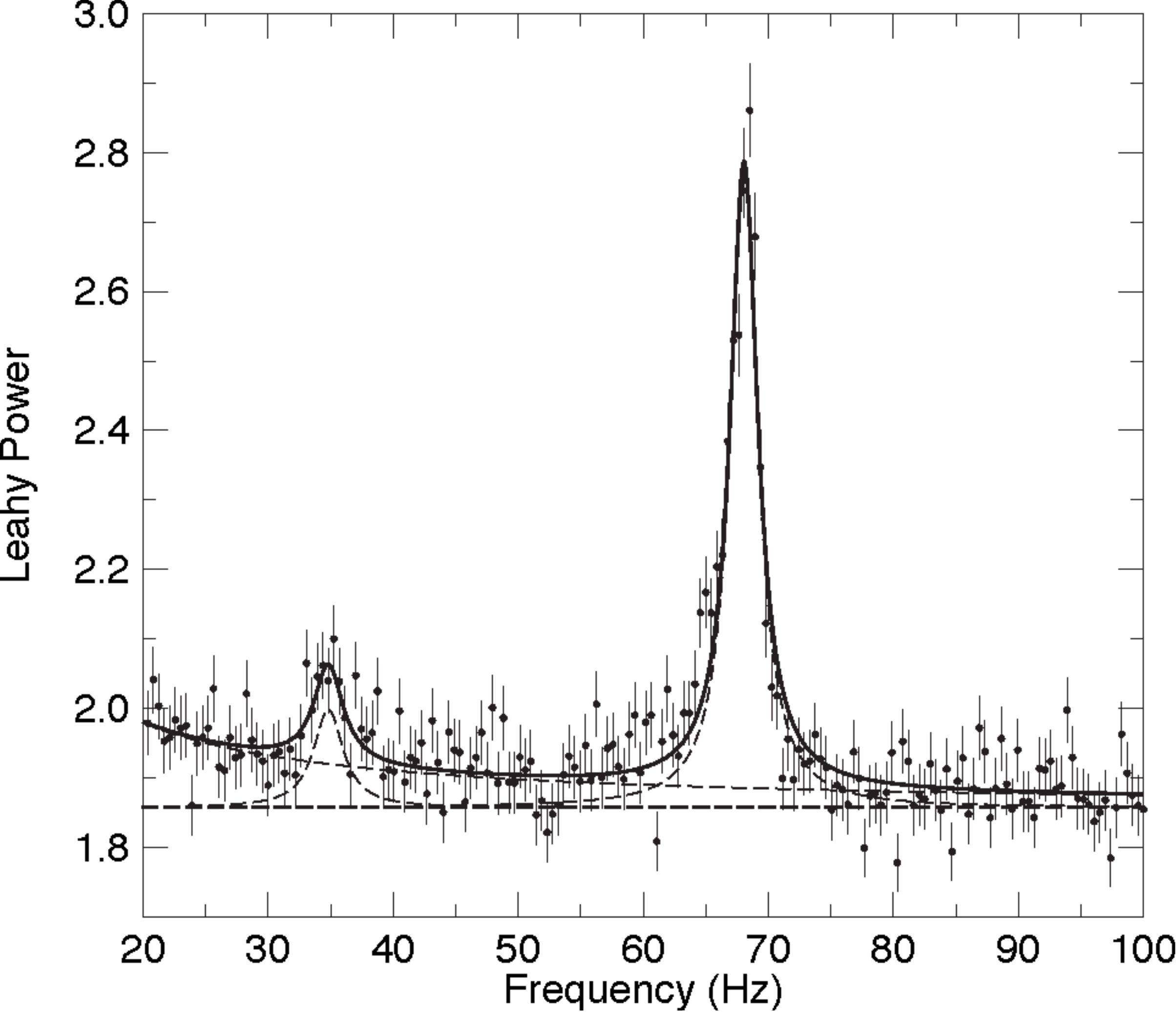} & 
   \includegraphics[scale=0.25]{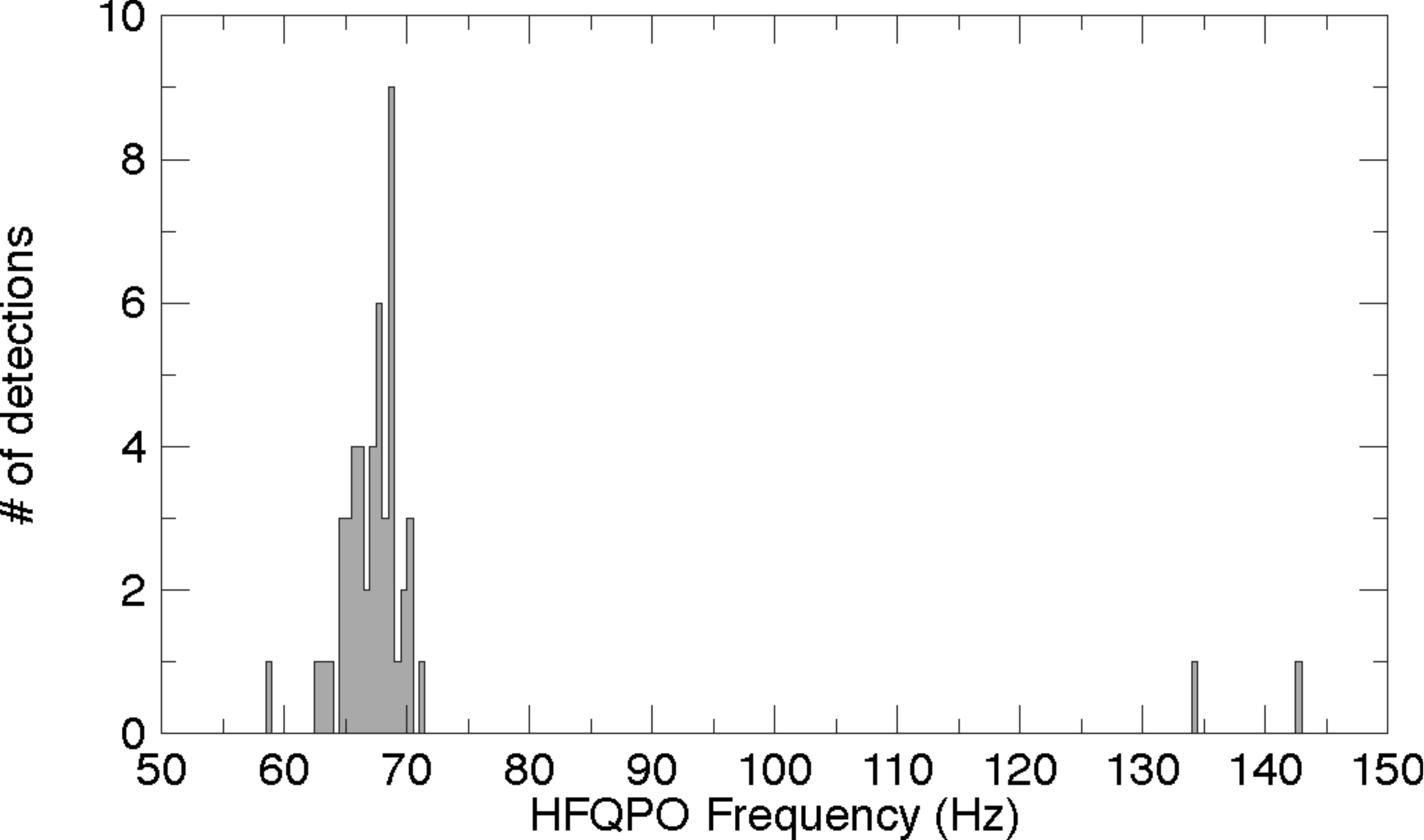} \\ 
\end{tabular}
\caption{ Left panel: double HFQPO in GRS\,1915+105
\citep[from][]{altamiranobelloni2013b}. Right panel: distribution of the
detection of single HFQPOs in GRS\,1915+105 \citep[from][]{bellonialtamirano}. }
\label{fig:hfqpo2}
\end{figure}

\item The case of GRS\,1915+105, as in many other respects, needs to be treated
separately. The original detection was of a peak at 65--67\,Hz \citep[][]{morgan}.
Since the observations where the oscillation was found showed very strong
variability on time-scales of 10--20\,s, a flux-selected analysis led to the
discovery of another peak at 27\,Hz, not simultaneous to the 67\,Hz one
\citep[][]{bellonicelia}. Shortly afterwards, a 41\,Hz peak simultaneous to the
67\,Hz one was discovered from a subset of {\it RXTE} observations
\citep[][]{strohmayer2001b}. A semi-automatic study of all {\it RXTE} observations for
a total exposure time of 5\,Ms led to the detection of 51 QPO peaks in as many
observations. Most of the centroid frequencies were in the 63--71\,Hz range (see
Figure\,\ref{fig:hfqpo2}, left panel), indicating that $\sim\, $67\,Hz (the average
value) must be a fundamental frequency in the system
\citep[][]{bellonialtamirano}. A time-resolved analysis indicates that the
detections are confined to a restricted region of the HID of GRS\,1915+105 (which
has a peculiar and complex shape), again corresponding to the highest count
rates \citep[][]{bellonialtamirano}. An additional peak at 34\,Hz (see
\ref{fig:hfqpo2}, right panel), simultaneous with the 67\,Hz one and consistent
with half of its frequency, was found in data corresponding to a subregion in
the HID, just as the 41\,Hz one corresponded to a separate subregion
\citep[][]{altamiranobelloni2013b}. This system showed therefore multiple
centroid frequencies: 67\,Hz, 27\,Hz, a pair 41--67\,Hz and a pair 34--67\,Hz. The
sequence 27:41:67 corresponds to 2:3:5, while 34:67 is 1:2. Independent of the
model, 67\,Hz is too low to be the Keplerian frequency at the innermost stable
circular orbit for GRS\,1915+105, being too slow even for the highest mass
allowed for the BH in the system \citep[][]{reid2014} and zero spin.

\end{itemize}


\section{Comparison between NS and BH systems}{Comparison}\label{sec:4}

When a new galactic transient X-ray source is observed in the sky, even in
absence of detected pulsations or X-ray bursts, unambiguous telltales of the
presence of a NS, in most cases the X-ray properties are sufficient
for the identification of the nature of the compact object, although not a
conclusive one as no definitive criterion for the presence of a BH in
the system has yet been found. Below, I will outline the main differences and
similarities in X-ray properties between classes of sources.

\subsection{Quiescent emission}

\begin{figure}
   \includegraphics[scale=0.40]{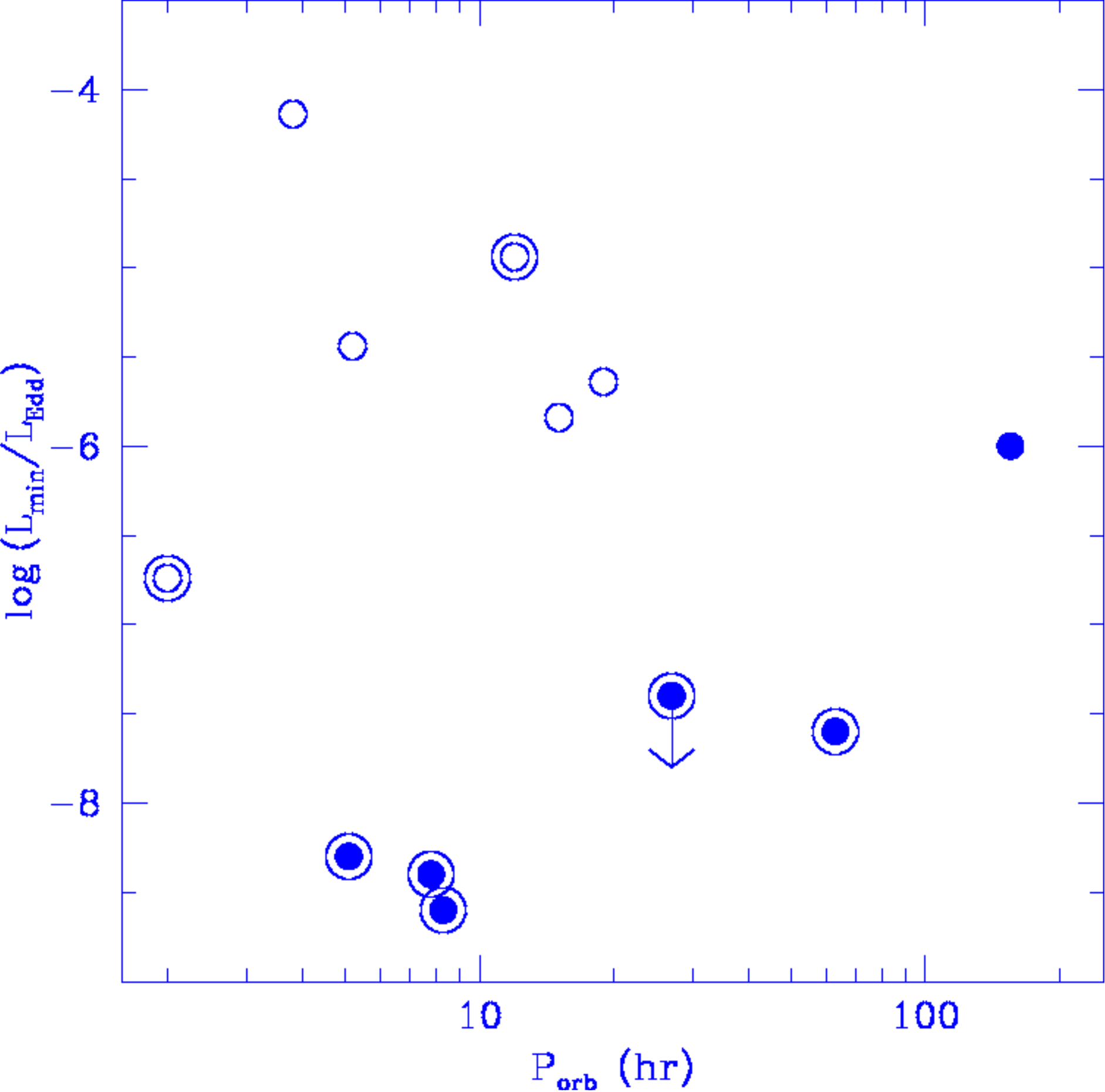}
\caption{ Quiescent luminosity for X-ray transients as a function of orbital period. Only the lowest
observed value for each system is shown. Filled symbols represent BH binaries, empty symbols
neutron-star binaries. \citep[From][]{garcia}.
\copyright AAS. Reproduced with permission.
}
\label{fig:quiescence}
\end{figure}

BH transients in quiescence have a very low X-ray emission that 
contrasts with the relative optical brightness of the accretion disk, a 
discrepancy that can be solved with the presence of an advection-dominated 
accretion flow (ADAF), where the flow is radiatively inefficient 
\citep[see][]{adaf}. However, in the case of a NS, the advection 
 energy stored in heat in the ADAF is not lost through the event 
horizon and must be released on the star's surface. Therefore, the 
quiescent luminosity of a NS transient is expected to be lower than in the 
BH case, given the same accretion rate.  Accretion rate depends on the 
binary orbit, with tighter systems being fainter \citep[see][]{menou1999}, 
but is not expected to be different between different classes of systems. 
Deep observations with current instruments confirm both statements (see 
Figure\,\ref{fig:quiescence}). The theoretical situation is however more 
complicated and the claim that this results is an indication of the 
existence of event horizons has been critically discussed by 
\citet{marek2002}.

\subsection{X-ray bursts}

Thermonuclear X-ray bursts are X-emission originating on the surface of a 
NS and as such a direct evidence of the presence of said 
surface. It follows that, by definition, none of the currently known BHBs 
has shown an X-ray burst. However, it has been suggested that the absence 
of X-ray bursts can be taken as an indication of the absence of a surface 
and therefore as an indication of the BH nature of a source 
\citep[][]{narayan2002}. A statistical analysis of {\it RXTE} data in relation 
to burst models confirmed numerically this statement 
\citep[][]{Remillard2006c}. Also in this case, \citet{marek2002} have 
criticised this approach as a detection of the presence of an event 
horizon.

\subsection{Energy spectra}

As mentioned in the introduction, several spectral `signatures' 
have been proposed in the past to identify
the  presence of a BH in an X-ray binary  The
very  soft thermal component observed from the accretion disk in soft states
cannot be  considered as a signature, as low-B neutron-star binaries can have
accretion disks  extending to roughly the same inner radius as BHs. Measurements
of inner disk  radii through high-resolution spectroscopy are also unlikely to
yield results that would  allow a strong identification. A good discussion of
this comparison throughout the full  spectrum can be found in \citet{Lin2007},
but no clear dividing line between classes of  systems is found.

\begin{figure}
   \includegraphics[scale=0.60]{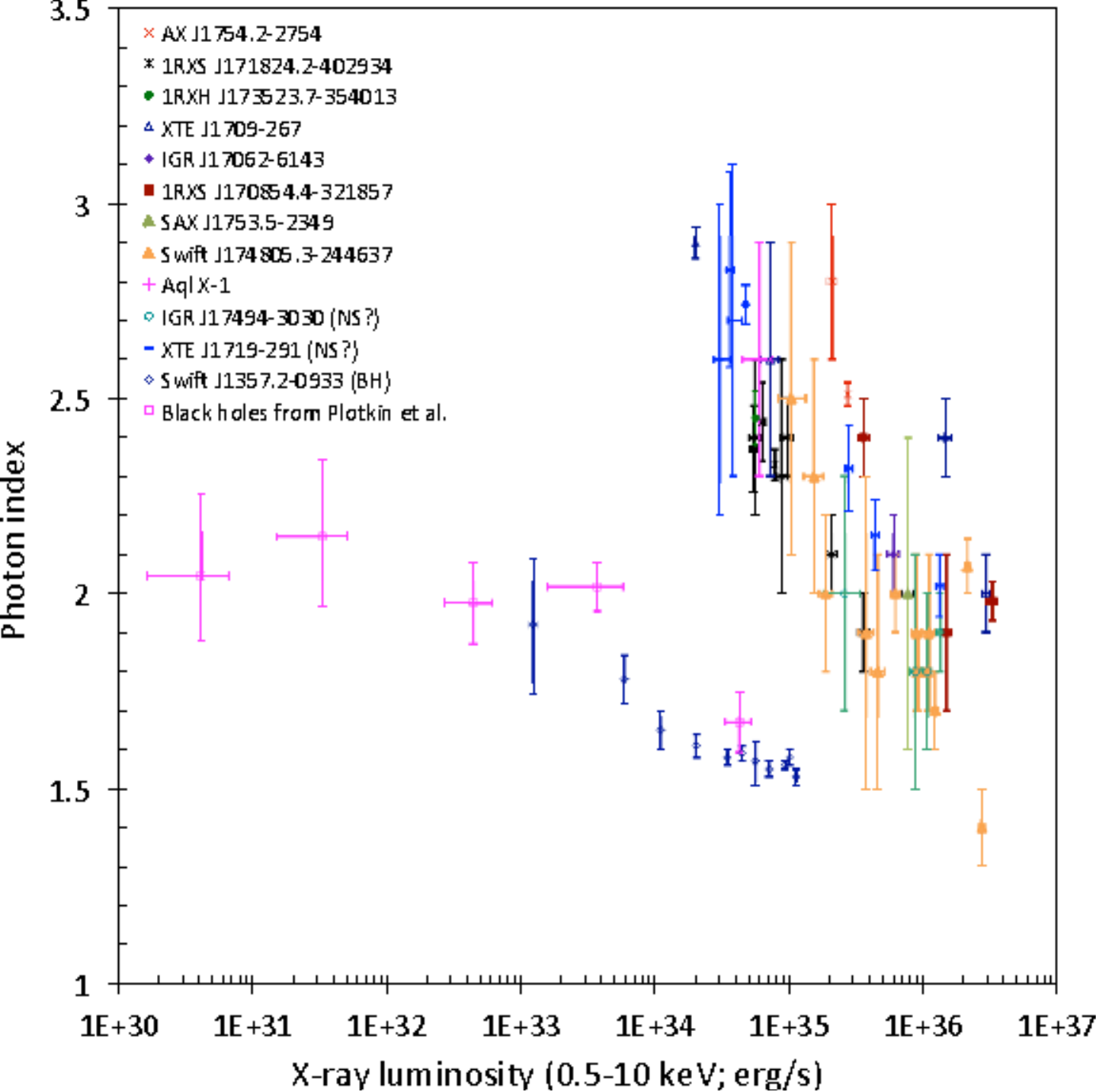}
\caption{ Plot of photon index - luminosity for a sample of NS and BH X-ray binaries
\citep[from][]{Wijnands2015}. 
\copyright AAS. Reproduced with permission.
}
\label{fig:wijnands}
\end{figure}

The hard component (often referred to as `hard tail'), which in the past 
was compared simply in terms of flux or power-law index, can now be also 
analysed in terms of more complex physical models. Recently, 
\citet{Wijnands2015} reported the results of the spectral analysis of a 
sample of NS LMXBs showing that at lower luminosities, in the range 
$10^{34} -- 10^{35}$\,erg/s, BH systems have systematically harder (lower 
power-law photon index) spectra than NS (see \ref{fig:wijnands}). The 
separation between classes of sources appears to be marked and additional 
data will be able to test this method. Another recent proposal for a clear 
distinction between classes of sources was presented by \citet{burke2016}. 
Fits with Comptonisation models to {\it RXTE} data of systems in the hard state show 
that BH and NS systems display different values of parameters. BHs 
correspond to higher values of the Comptonisation parameter $y$ and higher 
amplification ratio $A$ (hard flux divided by seed flux). While the $y-A $ 
relation appears to be common, there is a clear segregation 
\citep[][]{burke2016}.

Comparing the evolution of both transient and persistent systems in the 
HID, one can see that there is no major difference, indicating that the 
underlying physics must be, as expected, the same \citep[see][]{Teo2014}. 
At the same time, results like those of \citet{burke2016} and 
\citet{Lin2007} indicate that the same models can be applied, with 
significant differences in physical parameters.

\subsection{Time variability}

The comparison of fast-timing properties in BH and NS sources is complex, 
but very promising. The deconvolution of PDS into a combination of 
Lorentzian components \citep[][]{olive1998,bpk} has allowed a homogeneous 
treatment of all systems, allowing to link different classes of sources 
across different states (see Figures\,\ref{fig:ns_lfpds} and \ref{fig:lhs}). 
Comparing the two figures one can see that in the hard states the power 
distribution appears similar: a more peaked (QPO) component, a broad 
component at lower frequencies ($L_b$) and two broad components at higher 
frequencies ($L_\ell$ and $L_u$). The main difference between the NS and 
BH case is that the relative power of $L_\ell$ and in particular $L_u$ are 
higher in the NS case. Indeed, the difference in power above 10--50\,Hz has 
been proposed as an empirical method to distinguish BH from NS binaries 
\citep[][]{sunyaev}. In the hard states, no counter-example has been 
found. \citet{sunyaev} interpret it as an effect of the presence of a 
boundary layer in NS binaries, although the presence of the same number of 
components in the PDS seems not to be consistent with this idea.

\begin{figure}
   \includegraphics[scale=0.40,angle=270]{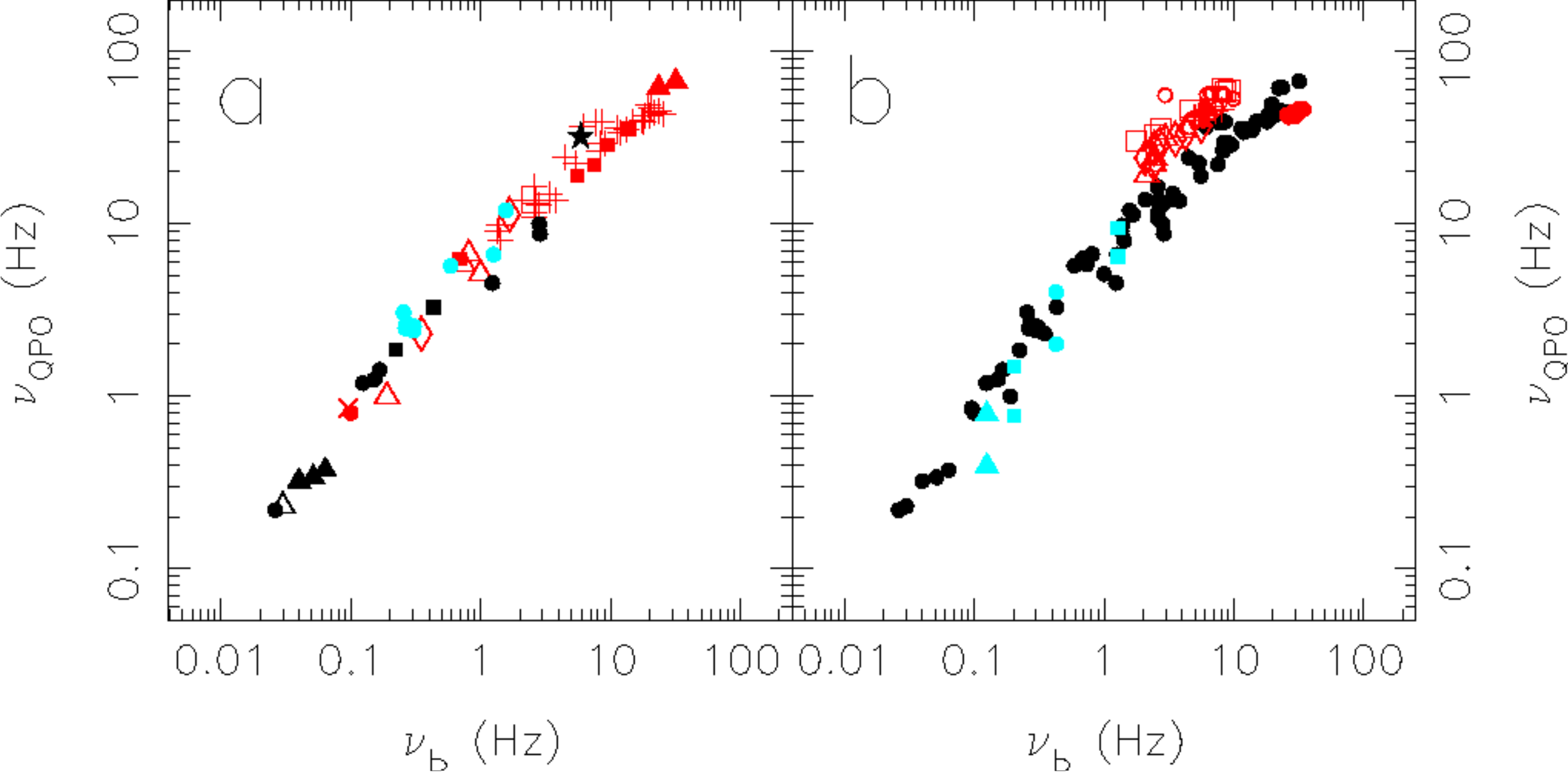}
\caption{Frequency of type-C QPO and HBO vs. break frequency for a sample of NS and BH  systems. Left
panel: BHB in black, atoll sources in red and accreting millisecond pulsars in cyan. Panel: all points
from the left panel in black, Z sources in red and sources with more than one QPO is observed in cyan.
\citep[From][]{WK}. }
\label{fig:wk}
\end{figure}

The identification of similar components in both classes of systems 
\citep[see][]{bpk} has allowed a detailed comparison. Identifying the most 
common QPO in BHBs (type-C) with the HBO in NS systems (see below) and 
comparing its frequency with that of the $L_b$ component has led to a 
rather tight correlation \citep[called WK correlation, see Figure\,
\ref{fig:wk} and][]{WK}. Z sources appear to deviate from the correlation, 
which otherwise extends over three orders of magnitude \citep[see][for an 
updated version]{bu2015}. \citet{bpk} added data to the correlation and 
noticed that there is a parallel correlation following a 1:1 relation, 
when characteristic frequencies are considered. However, the noise 
component related to these points was found to be an additional one 
($L_h$). The WK correlation connects two homogeneous observables through 
different classes of systems and indicates that the underlying phenomenon 
does not depend on the nature of the compact object. There is the 
indication of a segregation of BHB points to lower frequencies, but this 
is expected on the basis of mass scaling.

\begin{figure} 
\includegraphics[scale=0.06]{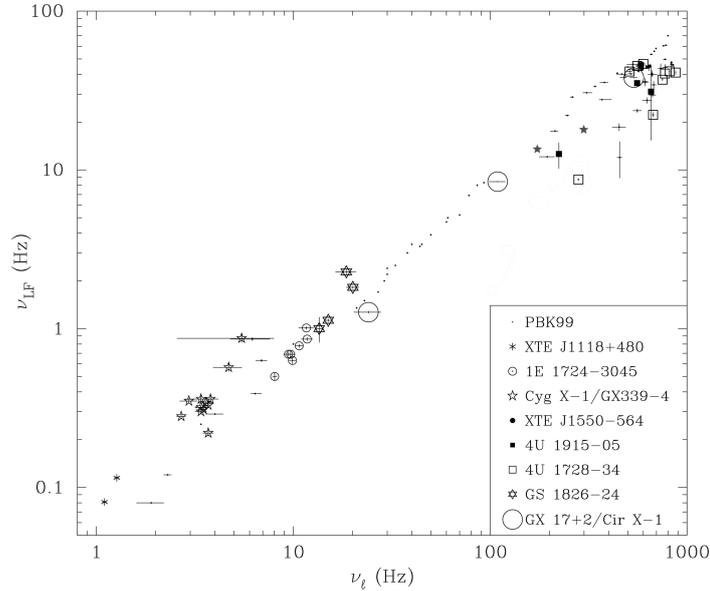} 
\caption{PB correlation as published by \citet{bpk}, with HFQPO points removed
as many of them were later found not to be significant. Two points were added
(five-point stars in the upper right), corresponding to the two simultaneous
detections of type-C and HFQPO analysed in \citet{motta2014,motta2014b}.  } 
\label{fig:pbk} 
\end{figure}

There is another correlation that links different classes of systems over 
three orders of magnitude, the so-called PBK correlation 
\citep[][]{pbk,bpk}. Unlike the WK correlation, here non-homogeneous 
quantities are included. The version of the correlation by \citet{bpk} is 
shown in Figure\,\ref{fig:pbk}, modified to remove incorrect points and to 
add the two points from \citet{motta2014,motta2014b} (see below). For NS 
binaries with kHz QPO, what is plotted in Figure\,\ref{fig:pbk} is the 
frequency of the HBO versus that of the lower kHz QPO. For hard state 
sources, both containing BH and NS, the plotted quantities are the type-C 
QPO for BH or the low-frequency QPO for NS versus
 the characteristic frequency of the lower of the high-frequency broad 
component $L_\ell$. Added to the plot are the two only {\it RXTE} detections of 
a HFQPO and a simultaneous type-C QPO for BHBs from 
\citet{motta2014,motta2014b}. Therefore at x$<$20\,Hz the y axis includes a 
broad component, and at x$>$100\,Hz it includes the lower kHz QPO. The 
20--100\,Hz interval is covered by points from the anomalous Z-source 
Cir\,X-1, which bridges the gap in the plot. Although there seems to be a 
parallel correlation at very high frequencies, the correlation is very 
well defined and links not only BH and NS sources, but also broad and 
narrow components, suggesting that the broad features observed in the hard 
state mark the same physical frequencies as kHz QPOs (and HFQPOs) in 
softer states. Interestingly, an extension of this correlation by three 
more orders of magnitude at low frequencies has been presented when adding 
the frequencies of dwarf-nova oscillations and quasi-periodic 
oscillations observed in the optical band from cataclysmic variables 
\citep[][]{Warner03}. The extension of the correlation looks impressive, 
although it is not clear how to relate the signals observed in a different 
band from white dwarf systems, where the inner region of the accretion flow 
cannot exist due to the physical size of the compact star (more below).

As presented above, at low frequencies NS binaries feature three 
`flavors' of QPOs: HBOs, NBOs, and FBOs and three are the `flavors' of 
BH QPOs: type A,B, and C. We have seen compelling evidence that links HBOs 
and type-Cs. \citet{Casella2005} compared in detail the QPO types and 
suggested a one-to-one correspondence between them, with NBOs 
corresponding to type-B and HBOs to type-C. More work is needed to build 
on these similarities. \citet{motta2011} on the basis of the analysis of a 
large sample of observations of GX\,339-4 suggested the possibility that 
type-A QPOs are high-frequency extensions of type-C QPOs.

Of course it is natural to explore the possibility that the two kHz QPOs 
and the two HFQPOs are the same signal observed in the vicinity of neutron 
stars and BHs respectively. The comparison is not simple, as very 
few HFQPO have been detected until now. There are clear phenomenological 
differences: kHz QPOs are observed to span a rather broad range of 
centroid frequencies, while the frequencies HFQPOs appear to vary only 
slightly if they vary at all. Moreover, the fact that kHz QPOs have been 
observed frequently and HFQPOs have not, implies that the typical 
fractional RMS of the latter must be lower. Until new observations with 
more sensitive instruments are available, it is difficult to make a full 
comparison.

\subsection{Mass measurements from X-rays}

Spectral analysis, whether from the continuum emission or from 
fluorescence lines, does not yield a mass for the compact object, which 
must be assumed on the basis of optical observations. Adding timing 
information can deliver this information, but of course a model must be 
assumed. The problem with past and current observatories is that they were 
optimised for either timing or high-resolution spectral analysis, which 
means that coordinated observations are needed, introducing further 
complication.  A comparison was attempted on the NS binary 4U\,1636-53 with 
simultaneous observations with {\it RXTE} and {\it XMM-Newton} \citep[][]{sanna2014}. 
The inner disk radius obtained by the fitting of a relativistically 
distorted profile to the iron emission line was combined with the 
information from the upper kHz QPO under the assumption (see below) that 
it represents the Keplerian frequency at the same radius. Six separate 
couples of simultaneous observations yielded inconsistent results on the 
mass, indicating that either one of the model is not correct or the two 
signatures do not originate from the same radius.

Fast time variability offers a potentially unique possibility for the 
study of effects of General Relativity in the strong-field regime and to 
discriminate the nature of the compact object. The information from timing 
signals like QPO centroid frequencies is essentially model-independent and 
must originate in the innermost regions of the accretion flow. In the 
recent years, several models have been proposed for the interpretation of 
these signals and almost all of them include the presence of frequencies 
from General Relativity. In particular, in almost all cases the highest 
detected feature corresponds to the Keplerian frequency at a certain 
radius, which can be the ISCO or larger (see above). Ideally, a successful 
model must be able to apply to both the NS and BH cases and explain the 
main observable facts outlined in the previous sections.  One particular 
model (called relativistic precession model, hereafter RPM) has received 
particular attention. The model identifies the three observed frequencies 
(type-C and HFQPOs for BH binaries, HBO and kHz QPOs for neutron 
star binaries) to a combination of fundamental frequencies set by General 
Relativity. The low-frequency peaks are identified with the nodal 
precession frequency (Lense-Thirring), while the high frequency peaks 
would be the periastron precession frequency and the Keplerian frequency, 
all corresponding to the same orbit, which identifies a `special' radius 
in the accretion flow 
\cite[see][]{StellaVietri1998,StellaVietri1999,StellaVietriMorsink1999}. 
This model has the advantage of being extremely simple, relying on basic 
relativistic frequencies, and at same time with the disadvantage of being 
extremely simple, as it associates the frequencies to a special radius, 
but does not address the issues of how the oscillation is produced and why 
that (variable) radius is special (it cannot be the ISCO, as the two 
highest frequencies are identical at ISCO.

\begin{figure}
\begin{tabular}{cc}
   \includegraphics[scale=0.20]{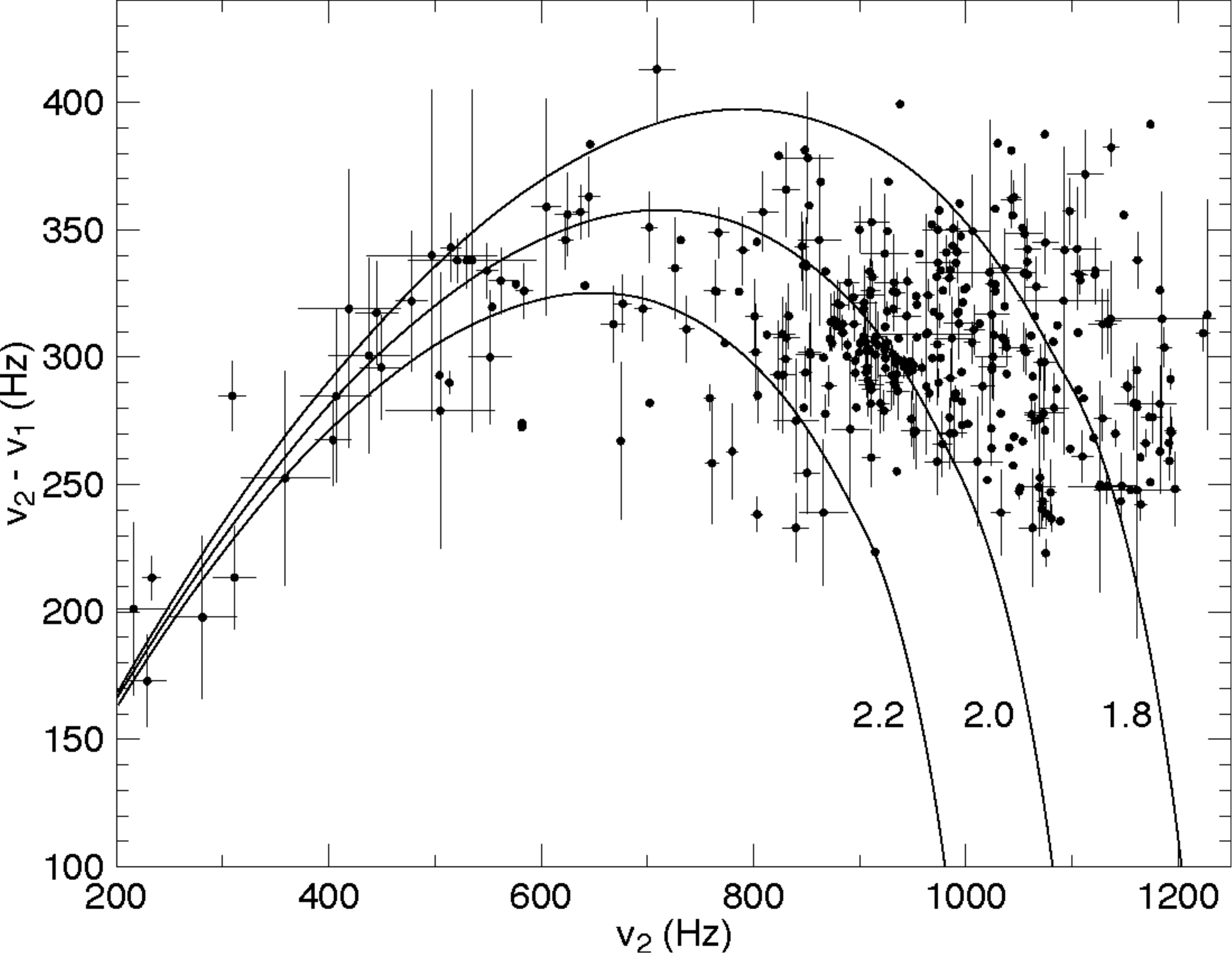} & 
   \includegraphics[scale=0.20]{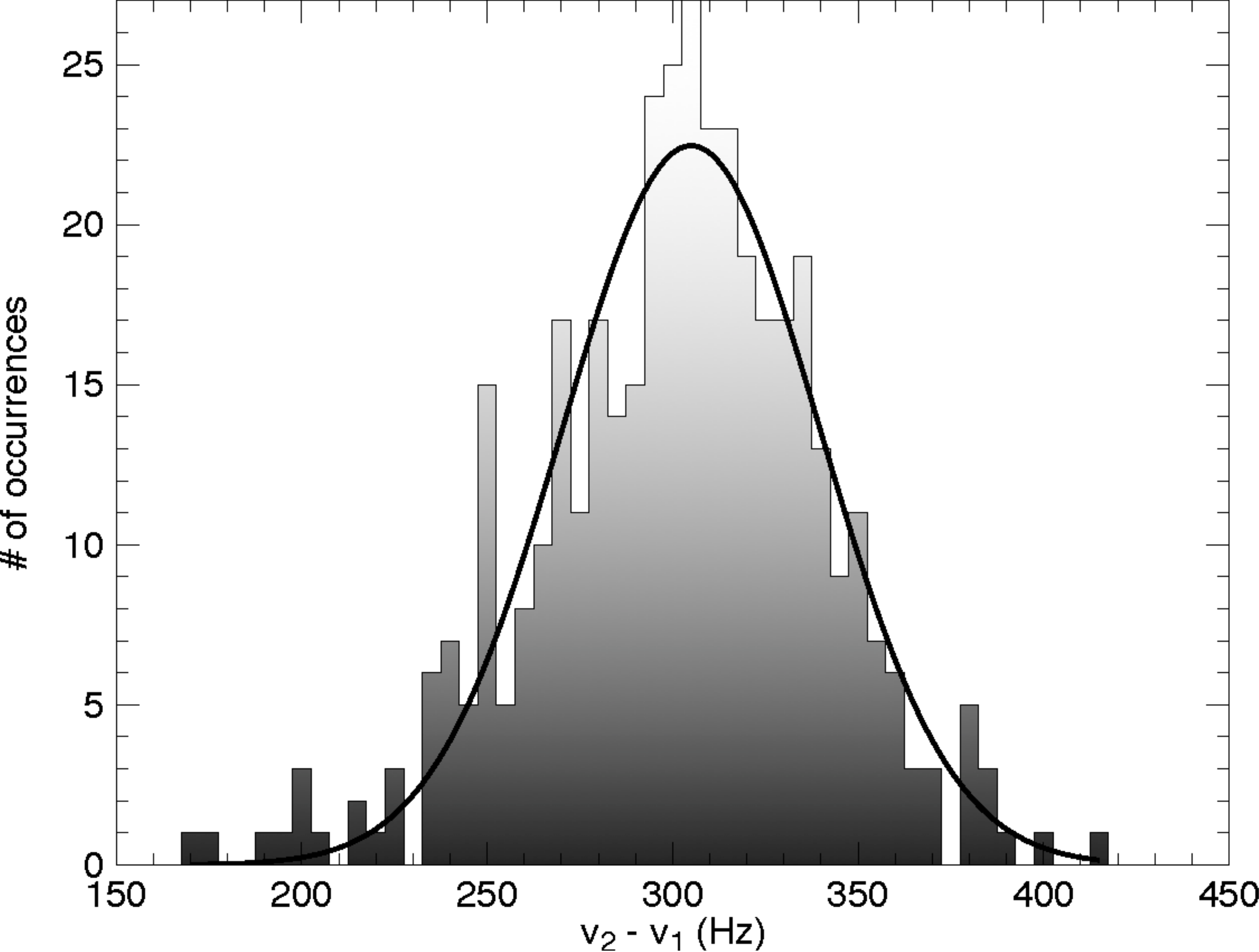} \\ 
\end{tabular}
\caption{ Left panel: plot of the difference $\Delta\nu$ between the centroid frequencies of kHz QPO
pairs as a function to the centroid of the upper peak \citep[after][]{StellaVietri1999,Boutloukos2006}
for all {\it RXTE} detections in the literature from NS LMXBs (excluding accreting millisecond pulsars). The
lines are predictions from the RPM for three NS masses (indicated in solar masses). Right
panel: distribution of all $\Delta\nu$ values on the Y axis of the plot in the left panel
\citep[after][]{menbel2007}. The line is a Gaussian fit with centroid $\sim\,$305\,Hz. }
\label{fig:deltanu}
\end{figure}

Originally applied to a sample of pairs of kHz QPOs, this model did not 
provide a precise fit, but provided frequencies in the correct range and 
with roughly the correct dependence \citep[see the left panel in Figure\,
\ref{fig:deltanu} and][]{StellaVietri1999,Boutloukos2006}. Applied to Figure\,
\ref{fig:deltanu} (where accreting millisecond pulsars are not included), 
the model makes three predictions, all of which observed: that the 
difference $\Delta\nu$ between the centroids of kHz QPOs must increase at 
low frequencies, decrease at high frequencies and have a maximum around 
350--400\,Hz. The increase part is covered only by data from Cir\,X-1, a 
rather anomalous Z source \citep[][]{Boutloukos2006}. The decrease part is 
more scattered, as expected if not all NS have the same mass, although the 
source with the highest statistics, Sco\,X-1, shows a decrease, steeper 
than expected \citep[see][]{StellaVietri1999}.  Notice that a more direct 
way to examine the relation between the frequencies of the  two kHz QPO peaks 
is plotting one versus the other (see Figure\,\ref{fig:nu1nu2}, where 
the RPM prediction for 2\,M$_\odot$ is shown). In this plot, Cir\,X-1 and Sco 
X-1 are marked separately, as are accreting millisecond X-ray pulsars, 
which are systematically below the correlation and can be brought back to 
it only by multiplying both frequencies by a factor 1.5 (see Section\,
\ref{NStiming}). The RPM was found also to interpret naturally the PBK 
correlation, which also contains very broad features 
\citep[][]{StellaVietriMorsink1999}, although it would not be applicable 
to the white-dwarf case because of the size of the compact star being too 
large.

Within the RPM there is no direct dependence of kHz QPOs from the spin of 
the NS (although the model lines would be slightly modified by 
rotation) and indeed there is evidence that the spin does not have 
influence (see Section\,\ref{NStiming}): Figure\,\ref{fig:deltanu} 
\citep[extended from that in][]{menbel2007} shows the distribution of all 
published $\Delta\nu$ values (again without millisecond accreting pulsars) 
with a simple Gaussian fit, indicating that the typical $\Delta\nu$ value 
is around 300\,Hz. Why accreting millisecond pulsars, the only systems 
where we directly observe the pulsation, are shifted by a factor 1.5 is 
unclear, although also considering correlations involving the HBO suggest 
that both kHz peaks are shifted, but not the HBO 
\citep[see][]{vanstraaten2005,linares2005}.

To test the validity of the model and the relation to the spin frequency 
one needs an extreme source, either very slow or very fast. In 2010, an 
accreting accretion-powered pulsar with a rotation frequency of 11\,Hz was 
discovered in outburst \citep[][]{bordas}. Unfortunately, only one kHz QPO 
was observed and it was not possible to discriminate between the 
(in-)dependence on spin ($\Delta\nu$ was expected to be $\sim\,$11\,Hz or 
$\sim\,$300\,Hz in the two scenarios). However a low-frequency QPO was 
observed in six observations, moving in the range 35--50\,Hz, which was 
identified as a HBO because of its intrinsic properties and positioning in 
the correlation diagrams \citep[][]{altamiranoHBO}. Together with the 
observed kHz QPO, this results to be fast to be associated to with 
Lense-Thirring frequency at the same radius. More observations will be 
needed to firmly confirm the lack of association.

The curves in the left panel of Figure\,\ref{fig:deltanu} can be calculated 
also for BHs, i.e. higher masses, and obviously would lie to lower 
frequencies, merging with the others at their low end. For a 10\,M$_\odot$ 
BH, the curve would peak around 120\,Hz. Of course for low-frequency 
broad features the model does not allow to discriminate the compact 
object, but notice that Figure\,\ref{fig:deltanu} indicates that the RPM 
identifies all sources excluding Cir\,X-1 as NSs.

\begin{figure}
   \includegraphics[scale=0.20]{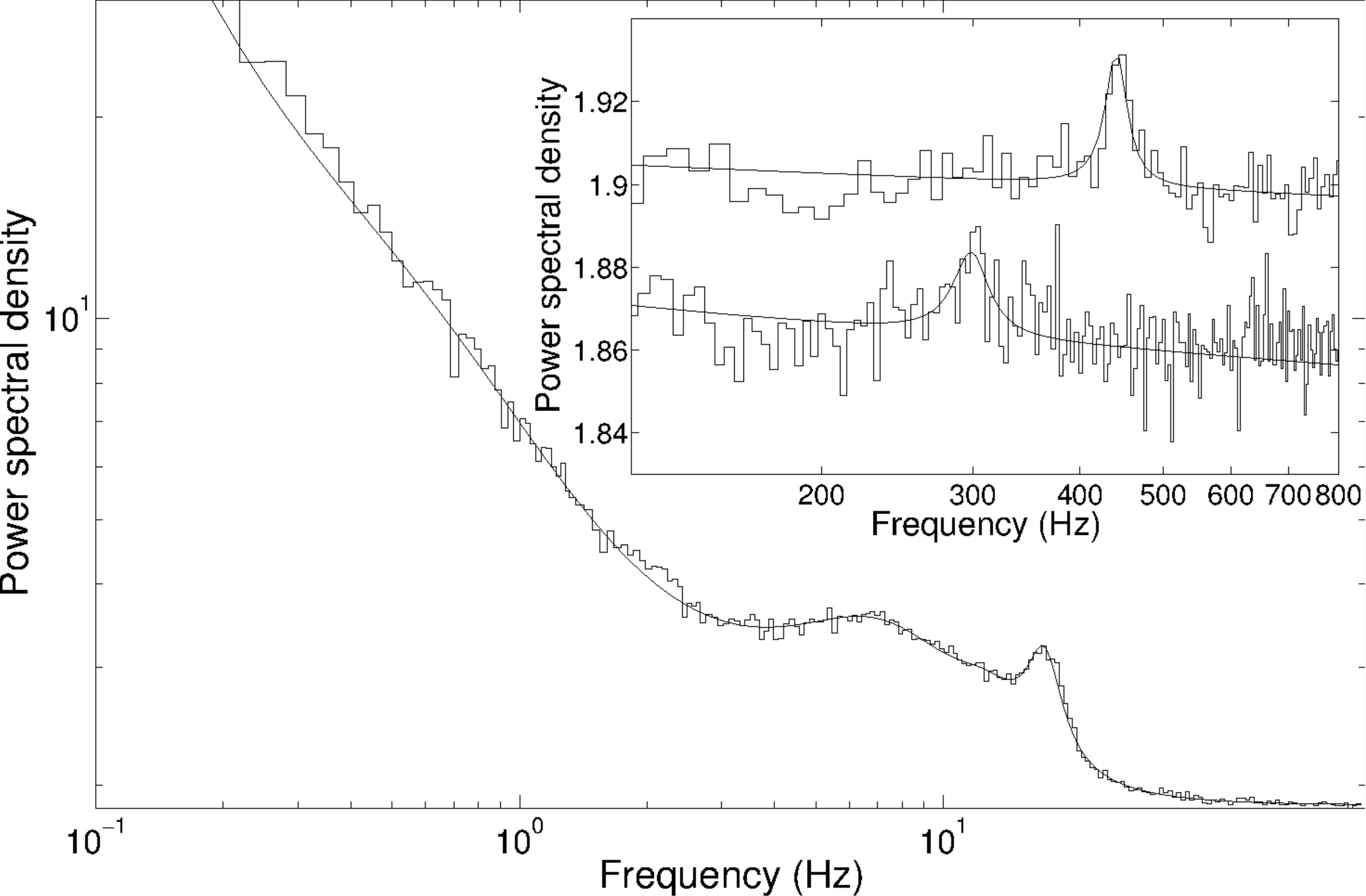}
\caption{ Power density spectrum of the only {\it RXTE} observation of a BHB with a triplet of QPOs. The main
panel shows the 18\,Hz type-C QPO, the inset shows the high-frequency part in two separate energy bands
to highlight the two HFQPOs. \citep[From][]{motta2014}. }
\label{fig:sara1}
\end{figure}

\begin{figure}
   \includegraphics[scale=0.2]{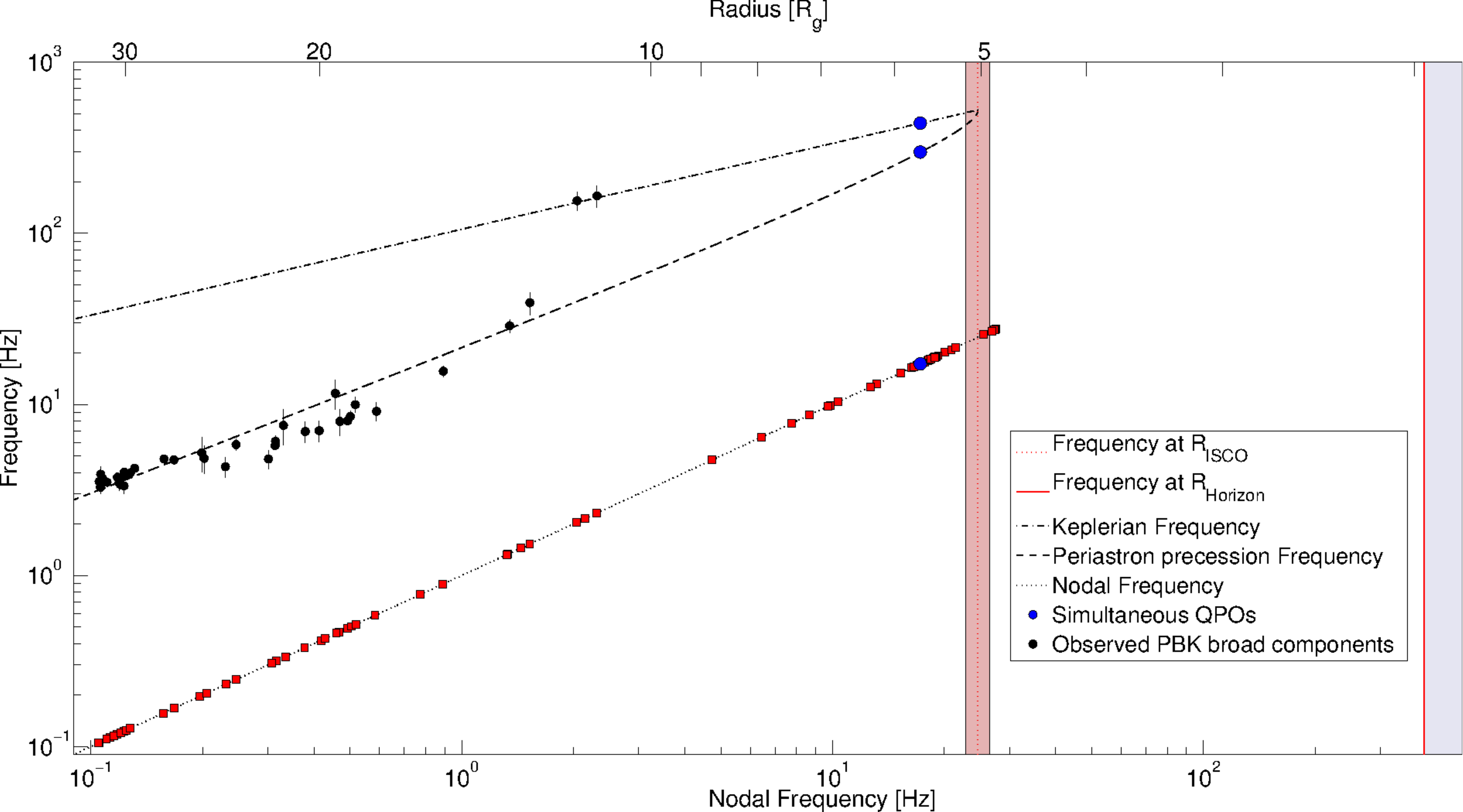}
\caption{Frequency correlations in the full sample of {\it RXTE} observations of GRO\,J1655-40. The red points
represent type-C vs. type-C, therefore lie on a 1:1 line. The three blue points are the frequencies of
the triplet of QPOs used to determine mass and spin of the BH. The black points are the
$\nu_\ell$ and $\nu_u$ frequencies as a function of type-C QPO ($\nu_\ell$ vs. type-C is the PBK
correlation).The lines are the dependencies predicted by the RPM for Keplerian, periastron precession
and nodal precession (from top to bottom) for the best fit parameters from the blue points. The red
band indicates the maximum nodal frequency set by the presence of an ISCO in the system.
\citep[From][]{motta2014}. }
\label{fig:sara2}
\end{figure}

As mentioned above, there are only a 
few detections of HFQPOs and fewer detections of pairs of 
them. Moreover, the presence of HFQPOs seems to be mutually exclusive with that of 
type-C QPOs. However, there is a single observation of GRO\,J1655-40, the only {\it RXTE} 
observation in twelve years of operations, in which all three frequencies (one type-C 
and two HFQPOs) have been observed \citep[see][and 
Figure\,\ref{fig:sara1}]{strohmayer2001a}. From these three values, measured with 
considerable precision, it has been possible to apply the RPM relations and obtain three 
physical parameters: the radius $R$ of the orbit to which they are related, the mass $M$ 
and the spin $a$ of the compact object \citep[][]{motta2014}. The derived parameters are 
$R$=5.677$\pm$0.035 r$_g$, $M$=5.307$\pm$0.066M$_\odot$, and $a$=0.286$\pm$0.003. 
Although with a small error bar, the derived mass is compatible with the precise 
dynamical measurement of $M$=5.4$\pm$0.4M$_\odot$ \citep[][]{beer}. Being based only on 
one detection, this result will have to be corroborated by more observations, but in 
principle this would constitute the first direct evidence of the presence of a black 
hole in an X-ray binary, as well as a precise measurement of its spin. In addition, 
since $a$ and $M$ define fully the system, it is possible to measure the radius of the 
ISCO and therefore the maximum values that the frequencies of the three peaks can reach. 
While the few detections of HFQPOs are all around the same values, a large number of 
type-C frequencies were measured throughout the two outbursts of this system, and their 
frequencies are distributed between 0.1\,Hz and 28\,Hz, the latter only marginally above 
the maximum value allowed by the model \citep[red points in Figure\,\ref{fig:sara2}, 
see][]{motta2014}. If confirmed by more data, this constitutes a direct measurement of 
the presence of the ISCO. Finally, adding all detections of broad components in the PDS, 
identified as $\nu_\ell$ and $\nu_u$ (black points in Figure\,\ref{fig:sara2}, the PBK 
correlation), they line along the prediction of the model.

The same analysis can be done on the only observation of XTE\,J1550-564 that shows a type-C and a single
HFQPO \citep[][]{motta2014b}. Here the third parameter needed to solve the system of GR equations
cannot be the second HFQPO, so the mass measured from optical observations was used. The same analysis
led to the measurement of a spin of $a$=0.34$\pm$0.01, with all type-C frequencies being below the
maximum ISCO values and the broad components fitting the model. 

The identification of the physical origin of the observed frequencies is not sufficient for a solid
model and a modulation process is necessary. A promising model has been proposed for the 
BH case, based on a thermal disk with an inner truncation radius, below which the accretion flow is
geometrically thick model consistent with spectral models \citep[see e.g.][]{poutanen}. This model
interprets type-C QPO as arising from the frame-dragging precession of the inner flow, where the flux
is modulated by effects like self-occultation, relativistic effects and changes in projected area.
Although each radius has a different precession frequency, the overall result is an oscillation at a
frequency intermediate between those at the outer and inner boundary of the geometrically thick
precessing region \citep[][]{ingram2009}, with the broad-band noise arising from propagation effects
\citep[see above and][]{ingram2010,ingram2011,ingramvdk}. This model does not address (yet) the
high-frequency oscillations nor the case of NSs, but it is being investigated further through
the search and detection of modulation of the effects of the inner-flow precession on the properties of
the radiation reflected by the geometrically thin accretion disk. A modulation at the QPO period of the
iron line due to reflection has been detected, strengthening the applicability of the model
\citep[][]{ingram2016}.

Finally, an alternative model for low-frequency QPOs, the transition layer model, interprets type-C
QPOs as viscous magneto-acoustic oscillations of a spherical transition layer between the Keplerian
flow and a sub-Keplerian region near the BH \citep[][]{titarchukfiorito}. This model predicts a
relationship between the photon index of the hard spectral component and the frequency of the type-C
QPO, which is used to measure the mass of the BH \citep[see][]{gliozzi}.

\section{Conclusions}\label{sec:5}

In the sections above,  I have presented the current standpoint in a selected
number of topics regarding X-ray emission from BH and NS binaries.  The past two
decades have seen a dramatic increase in our knowledge of the process of
accretion onto compact objects. We now know that the presence of collimated
relativistic jets and powerful wind outflows from X-ray binaries, which I did
not touch here, is impossible to ignore if we want to understand the physics of
accretion. At the same time, high-resolution spectral information, long-term
coverage of transient systems and high-sensitivity timing analysis have proven
to be essential, in particular when combined. With current instruments, we now
have the tools to start disentangling the effects of accretion that are in
common between NS and BH sources and those connected to the nature of the
compact object. This will allow us to use these powerful sources to study both
astrophysics and basic physics, making the best use of these cosmic
laboratories.


\section{Acknowledgements} 

I thank S. Motta, M. M\'endez, D. Altamirano and L. Stella or useful discussions
and the University of Southampton for providing more than once the ideal
environment to write this chapter.  I also thank the editors for their work on
the Winter School and on the book, and last but not least for bearing with me.
Finally, I would like to thank Dr. M. Mantica for getting rid of the only
transient QPO signal that I was definitely not interested in.






\begin{thebibliography}{149}
\expandafter\ifx\csname natexlab\endcsname\relax\def\natexlab#1{#1}\fi
\expandafter\ifx\csname selectlanguage\endcsname\relax
  \def\selectlanguage#1{\relax}\fi

\bibitem[\protect\citename{{Abbott} {et~al.}, }2016]{abbott}
{Abbott}, B.~P., {Abbott}, R., {Abbott}, T.~D., {Abernathy}, M.~R., {Acernese},
  F., {Ackley}, K., {Adams}, C., {Adams}, T., {Addesso}, P., {Adhikari}, R.~X.,
  and et~al. 2016.
\newblock {Observation of Gravitational Waves from a Binary Black Hole Merger}.
\newblock {\em Physical Review Letters}, {\bf 116}(6), 061102.

\bibitem[\protect\citename{{Abramowicz} and {Klu{\'z}niak}, }2001]{abrklu2001}
{Abramowicz}, M.~A., and {Klu{\'z}niak}, W. 2001.
\newblock {A precise determination of black hole spin in GRO J1655-40}.
\newblock {\em \aap}, {\bf 374}(Aug.), L19--L20.

\bibitem[\protect\citename{{Abramowicz} {et~al.}, }2002]{marek2002}
{Abramowicz}, M.~A., {Klu{\'z}niak}, W., and {Lasota}, J.-P. 2002.
\newblock {No observational proof of the black-hole event-horizon}.
\newblock {\em \aap}, {\bf 396}(Dec.), L31--L34.

\bibitem[\protect\citename{{Altamirano} and {Belloni},
  }2012]{altamiranobelloni2012}
{Altamirano}, D., and {Belloni}, T. 2012.
\newblock {Discovery of High-frequency Quasi-periodic Oscillations in the Black
  Hole Candidate IGR J17091-3624}.
\newblock {\em \apjl}, {\bf 747}(Mar.), L4.

\bibitem[\protect\citename{{Altamirano} {et~al.}, }2012]{altamiranoHBO}
{Altamirano}, D., {Ingram}, A., {van der Klis}, M., {Wijnands}, R., {Linares},
  M., and {Homan}, J. 2012.
\newblock {Low-frequency Quasi-periodic Oscillation from the 11 Hz Accreting
  Pulsar in Terzan 5: Not Frame Dragging}.
\newblock {\em \apjl}, {\bf 759}(Nov.), L20.

\bibitem[\protect\citename{{Barret} {et~al.}, }2005a]{Barret2005b}
{Barret}, D., {Olive}, J.-F., and {Miller}, M.~C. 2005a.
\newblock {An abrupt drop in the coherence of the lower kHz quasi-periodic
  oscillations in 4U 1636-536}.
\newblock {\em \mnras}, {\bf 361}(Aug.), 855--860.

\bibitem[\protect\citename{{Barret} {et~al.}, }2005b]{Barret2005a}
{Barret}, D., {Klu{\'z}niak}, W., {Olive}, J.~F., {Paltani}, S., and {Skinner},
  G.~K. 2005b.
\newblock {On the high coherence of kHz quasi-periodic oscillations}.
\newblock {\em \mnras}, {\bf 357}(Mar.), 1288--1294.

\bibitem[\protect\citename{{Barret} {et~al.}, }2006]{Barret2006}
{Barret}, D., {Olive}, J.-F., and {Miller}, M.~C. 2006.
\newblock {The coherence of kilohertz quasi-periodic oscillations in the X-rays
  from accreting neutron stars}.
\newblock {\em \mnras}, {\bf 370}(Aug.), 1140--1146.

\bibitem[\protect\citename{{Barret} {et~al.}, }2007]{Barret2007}
{Barret}, D., {Olive}, J.-F., and {Miller}, M.~C. 2007.
\newblock {Supporting evidence for the signature of the innermost stable
  circular orbit in Rossi X-ray data from 4U 1636-536}.
\newblock {\em \mnras}, {\bf 376}(Apr.), 1139--1144.

\bibitem[\protect\citename{{Beer} and {Podsiadlowski}, }2002]{beer}
{Beer}, M.~E., and {Podsiadlowski}, P. 2002.
\newblock {The quiescent light curve and the evolutionary state of GRO
  J1655-40}.
\newblock {\em \mnras}, {\bf 331}(Mar.), 351--360.

\bibitem[\protect\citename{{Belloni} {et~al.}, }1996]{Belloni1996}
{Belloni}, T., {Mendez}, M., {van der Klis}, M., {Hasinger}, G., {Lewin},
  W.~H.~G., and {van Paradijs}, J. 1996.
\newblock {An Intermediate State of Cygnus X-1}.
\newblock {\em \apjl}, {\bf 472}(Dec.), L107.

\bibitem[\protect\citename{{Belloni} {et~al.}, }1997]{Belloni1997}
{Belloni}, T., {van der Klis}, M., {Lewin}, W.~H.~G., {van Paradijs}, J.,
  {Dotani}, T., {Mitsuda}, K., and {Miyamoto}, S. 1997.
\newblock {Energy dependence in the quasi-periodic oscillations and noise of
  black hole candidates in the very high state.}
\newblock {\em \aap}, {\bf 322}(June), 857--867.

\bibitem[\protect\citename{{Belloni} {et~al.}, }2001]{bellonicelia}
{Belloni}, T., {M{\'e}ndez}, M., and {S{\'a}nchez-Fern{\'a}ndez}, C. 2001.
\newblock {The high-frequency QPOs in GRS 1915+105}.
\newblock {\em \aap}, {\bf 372}(June), 551--556.

\bibitem[\protect\citename{{Belloni} {et~al.}, }2002]{bpk}
{Belloni}, T., {Psaltis}, D., and {van der Klis}, M. 2002.
\newblock {A Unified Description of the Timing Features of Accreting X-Ray
  Binaries}.
\newblock {\em \apj}, {\bf 572}(June), 392--406.

\bibitem[\protect\citename{{Belloni} {et~al.}, }2005]{Belloni2005}
{Belloni}, T., {Homan}, J., {Casella}, P., {van der Klis}, M., {Nespoli}, E.,
  {Lewin}, W.~H.~G., {Miller}, J.~M., and {M{\'e}ndez}, M. 2005.
\newblock {The evolution of the timing properties of the black-hole transient
  GX 339-4 during its 2002/2003 outburst}.
\newblock {\em \aap}, {\bf 440}(Sept.), 207--222.

\bibitem[\protect\citename{{Belloni} {et~al.}, }2007a]{Belloni2007b}
{Belloni}, T., {M{\'e}ndez}, M., and {Homan}, J. 2007a.
\newblock {On the kHz QPO frequency correlations in bright neutron star X-ray
  binaries}.
\newblock {\em \mnras}, {\bf 376}(Apr.), 1133--1138.

\bibitem[\protect\citename{{Belloni} {et~al.}, }2007b]{Belloni2007}
{Belloni}, T., {Homan}, J., {Motta}, S., {Ratti}, E., and {M{\'e}ndez}, M.
  2007b.
\newblock {Rossi XTE monitoring of 4U1636-53 - I. Long-term evolution and kHz
  quasi-periodic oscillations}.
\newblock {\em \mnras}, {\bf 379}(July), 247--252.

\bibitem[\protect\citename{{Belloni}, }2010]{belloni2010}
{Belloni}, T.~M. 2010 (Mar.).
\newblock {States and Transitions in Black Hole Binaries}.
\newblock {Page ~53 of:} {Belloni}, T. (ed), {\em Lecture Notes in Physics,
  Berlin Springer Verlag}.
\newblock Lecture Notes in Physics, Berlin Springer Verlag, vol. 794.

\bibitem[\protect\citename{{Belloni} and {Altamirano},
  }2013a]{altamiranobelloni2013b}
{Belloni}, T.~M., and {Altamirano}, D. 2013a.
\newblock {Discovery of a 34 Hz quasi-periodic oscillation in the X-ray
  emission of GRS 1915+105}.
\newblock {\em \mnras}, {\bf 432}(June), 19--22.

\bibitem[\protect\citename{{Belloni} and {Altamirano},
  }2013b]{bellonialtamirano}
{Belloni}, T.~M., and {Altamirano}, D. 2013b.
\newblock {High-frequency quasi-periodic oscillations from GRS 1915+105}.
\newblock {\em \mnras}, {\bf 432}(June), 10--18.

\bibitem[\protect\citename{{Belloni} and {Motta}, }2016]{bellonimotta}
{Belloni}, T.~M., and {Motta}, S.~E. 2016.
\newblock {Transient Black Hole Binaries}.
\newblock {Page ~61 of:} {Bambi}, C. (ed), {\em Astrophysics of Black Holes:
  From Fundamental Aspects to Latest Developments}.
\newblock Astrophysics and Space Science Library, vol. 440.

\bibitem[\protect\citename{{Belloni} {et~al.}, }2011]{Belloni2011}
{Belloni}, T.~M., {Motta}, S.~E., and {Mu{\~n}oz-Darias}, T. 2011.
\newblock {Black hole transients}.
\newblock {\em Bulletin of the Astronomical Society of India}, {\bf 39}(Sept.),
  409--428.

\bibitem[\protect\citename{{Belloni} {et~al.}, }2012]{belloniHFQPO}
{Belloni}, T.~M., {Sanna}, A., and {M{\'e}ndez}, M. 2012.
\newblock {High-frequency quasi-periodic oscillations in black hole binaries}.
\newblock {\em \mnras}, {\bf 426}(Nov.), 1701--1709.

\bibitem[\protect\citename{{Bordas} {et~al.}, }2010]{bordas}
{Bordas}, P., {Kuulkers}, E., {Alfonso-Garz{\'o}n}, J., {Beckmann}, V., {Bird},
  T., {Chenevez}, S.~B.~J., {Courvoisier}, T., {Del Santo}, M., {Domingo}, A.,
  {Ebisawa}, K., {Ferrigno}, C., {Jonker}, P., {Kretschmar}, P., {Markwardt},
  C., {Oosterbroek}, T., {Paizis}, A., {Pottschmidt}, K.,
  {S{\'a}nchez-Fern{\'a}ndez}, C., and {Wijnands}, R. 2010.
\newblock {A hard X-ray transient in the direction of Terzan 5 detected by
  INTEGRAL}.
\newblock {\em The Astronomer's Telegram}, {\bf 2919}(Oct.).

\bibitem[\protect\citename{{Boutloukos} {et~al.}, }2006]{Boutloukos2006}
{Boutloukos}, S., {van der Klis}, M., {Altamirano}, D., {Klein-Wolt}, M.,
  {Wijnands}, R., {Jonker}, P.~G., and {Fender}, R.~P. 2006.
\newblock {Discovery of Twin kHz QPOs in the Peculiar X-Ray Binary Circinus
  X-1}.
\newblock {\em \apj}, {\bf 653}(Dec.), 1435--1444.

\bibitem[\protect\citename{{Brown} and {Cumming}, }2009]{BrownCumming2009}
{Brown}, E.~F., and {Cumming}, A. 2009.
\newblock {Mapping Crustal Heating with the Cooling Light Curves of
  Quasi-Persistent Transients}.
\newblock {\em \apj}, {\bf 698}(June), 1020--1032.

\bibitem[\protect\citename{{Bu} {et~al.}, }2015]{bu2015}
{Bu}, Q.-c., {Chen}, L., {Li}, Z.-s., {Qu}, J.-l., {Belloni}, T.~M., and
  {Zhang}, L. 2015.
\newblock {Correlations in Horizontal Branch Oscillations and Break Components
  in XTE J1701-462 and GX 17+2}.
\newblock {\em \apj}, {\bf 799}(Jan.), 2.

\bibitem[\protect\citename{{Burgay} {et~al.}, }2014]{burgay}
{Burgay}, M., {Kramer}, M., and {McLaughlin}, M.~A. 2014.
\newblock {The Double Pulsar J0737-3039A/B: a decade of surprises}.
\newblock {\em Bulletin of the Astronomical Society of India}, {\bf 42}(Sept.),
  101--119.

\bibitem[\protect\citename{{Burke} {et~al.}, }2016]{burke2016}
{Burke}, M.~J., {Gilfanov}, M., and {Sunyaev}, R. 2016.
\newblock {A dichotomy between the hard state spectral properties of black hole
  and neutron star X-ray binaries}.
\newblock {\em \mnras}, Oct.

\bibitem[\protect\citename{{Casares} and {Jonker}, }2014]{casjon2014}
{Casares}, J., and {Jonker}, P.~G. 2014.
\newblock {Mass Measurements of Stellar and Intermediate-Mass Black Holes}.
\newblock {\em \ssr}, {\bf 183}(Sept.), 223--252.

\bibitem[\protect\citename{{Casella} {et~al.}, }2004]{Casella2004}
{Casella}, P., {Belloni}, T., {Homan}, J., and {Stella}, L. 2004.
\newblock {A study of the low-frequency quasi-periodic oscillations in the
  X-ray light curves of the black hole candidate; XTE J1859+226}.
\newblock {\em \aap}, {\bf 426}(Nov.), 587--600.

\bibitem[\protect\citename{{Casella} {et~al.}, }2005]{Casella2005}
{Casella}, P., {Belloni}, T., and {Stella}, L. 2005.
\newblock {The ABC of Low-Frequency Quasi-periodic Oscillations in Black Hole
  Candidates: Analogies with Z Sources}.
\newblock {\em \apj}, {\bf 629}(Aug.), 403--407.

\bibitem[\protect\citename{{Davis} {et~al.}, }2005]{davis}
{Davis}, S.~W., {Blaes}, O.~M., {Hubeny}, I., and {Turner}, N.~J. 2005.
\newblock {Relativistic Accretion Disk Models of High-State Black Hole X-Ray
  Binary Spectra}.
\newblock {\em \apj}, {\bf 621}(Mar.), 372--387.

\bibitem[\protect\citename{{de Avellar} {et~al.}, }2013]{avellar2013}
{de Avellar}, M.~G.~B., {M{\'e}ndez}, M., {Sanna}, A., and {Horvath}, J.~E.
  2013.
\newblock {Time lags of the kilohertz quasi-periodic oscillations in the
  low-mass X-ray binaries 4U 1608-52 and 4U 1636-53}.
\newblock {\em \mnras}, {\bf 433}(Aug.), 3453--3463.

\bibitem[\protect\citename{{Degenaar} {et~al.}, }2015]{Degenaar2015}
{Degenaar}, N., {Miller}, J.~M., {Chakrabarty}, D., {Harrison}, F.~A., {Kara},
  E., and {Fabian}, A.~C. 2015.
\newblock {A NuSTAR observation of disc reflection from close to the neutron
  star in 4U 1608-52}.
\newblock {\em \mnras}, {\bf 451}(July), L85--L89.

\bibitem[\protect\citename{{Del Santo} {et~al.}, }2008]{delsanto339}
{Del Santo}, M., {Malzac}, J., {Jourdain}, E., {Belloni}, T., and {Ubertini},
  P. 2008.
\newblock {Spectral variability of GX339-4 in a hard-to-soft state transition}.
\newblock {\em \mnras}, {\bf 390}(Oct.), 227--234.

\bibitem[\protect\citename{{Di Salvo} {et~al.}, }2015]{Disalvo2015}
{Di Salvo}, T., {Iaria}, R., {Matranga}, M., {Burderi}, L., {D'A{\'{\i}}}, A.,
  {Egron}, E., {Papitto}, A., {Riggio}, A., {Robba}, N.~R., and {Ueda}, Y.
  2015.
\newblock {Suzaku broad-band spectrum of 4U 1705-44: probing the reflection
  component in the hard state}.
\newblock {\em \mnras}, {\bf 449}(May), 2794--2802.

\bibitem[\protect\citename{{Din{\c c}er} {et~al.}, }2014]{dincer2014}
{Din{\c c}er}, T., {Kalemci}, E., {Tomsick}, J.~A., {Buxton}, M.~M., and
  {Bailyn}, C.~D. 2014.
\newblock {Complete Multiwavelength Evolution of Galactic Black Hole Transients
  during Outburst Decay. II. Compact Jets and X-Ray Variability Properties}.
\newblock {\em \apj}, {\bf 795}(Nov.), 74.

\bibitem[\protect\citename{{Farinelli} {et~al.}, }2008]{farinelli2008}
{Farinelli}, R., {Titarchuk}, L., {Paizis}, A., and {Frontera}, F. 2008.
\newblock {A New Comptonization Model for Weakly Magnetized, Accreting Neutron
  Stars in Low-Mass X-Ray Binaries}.
\newblock {\em \apj}, {\bf 680}(June), 602--614.

\bibitem[\protect\citename{{Garcia} {et~al.}, }2001]{garcia}
{Garcia}, M.~R., {McClintock}, J.~E., {Narayan}, R., {Callanan}, P., {Barret},
  D., and {Murray}, S.~S. 2001.
\newblock {New Evidence for Black Hole Event Horizons from Chandra}.
\newblock {\em \apjl}, {\bf 553}(May), L47--L50.

\bibitem[\protect\citename{{Gierli{\'n}ski} and {Done}, }2002]{GD2002}
{Gierli{\'n}ski}, M., and {Done}, C. 2002.
\newblock {The X-ray spectrum of the atoll source 4U 1608-52}.
\newblock {\em \mnras}, {\bf 337}(Dec.), 1373--1380.

\bibitem[\protect\citename{{Gilfanov}, }2010]{gilfanov2010}
{Gilfanov}, M. 2010 (Mar.).
\newblock {X-Ray Emission from Black-Hole Binaries}.
\newblock {Page ~17 of:} {Belloni}, T. (ed), {\em Lecture Notes in Physics,
  Berlin Springer Verlag}.
\newblock Lecture Notes in Physics, Berlin Springer Verlag, vol. 794.

\bibitem[\protect\citename{{Gliozzi} {et~al.}, }2011]{gliozzi}
{Gliozzi}, M., {Titarchuk}, L., {Satyapal}, S., {Price}, D., and {Jang}, I.
  2011.
\newblock {Testing a Scale-independent Method to Measure the Mass of Black
  Holes}.
\newblock {\em \apj}, {\bf 735}(July), 16.

\bibitem[\protect\citename{{Grove} {et~al.}, }1998]{grove}
{Grove}, J.~E., {Johnson}, W.~N., {Kroeger}, R.~A., {McNaron-Brown}, K.,
  {Skibo}, J.~G., and {Phlips}, B.~F. 1998.
\newblock {Gamma-Ray Spectral States of Galactic Black Hole Candidates}.
\newblock {\em \apj}, {\bf 500}(June), 899--908.

\bibitem[\protect\citename{{Hasinger} and {van der Klis}, }1989]{HK:1989}
{Hasinger}, G., and {van der Klis}, M. 1989.
\newblock {Two patterns of correlated X-ray timing and spectral behaviour in
  low-mass X-ray binaries}.
\newblock {\em \aap}, {\bf 225}(Nov.), 79--96.

\bibitem[\protect\citename{{Heil} {et~al.}, }2012]{heil2012}
{Heil}, L.~M., {Vaughan}, S., and {Uttley}, P. 2012.
\newblock {The ubiquity of the rms-flux relation in black hole X-ray binaries}.
\newblock {\em \mnras}, {\bf 422}(May), 2620--2631.

\bibitem[\protect\citename{{Homan} {et~al.}, }2001]{homan2001}
{Homan}, J., {Wijnands}, R., {van der Klis}, M., {Belloni}, T., {van Paradijs},
  J., {Klein-Wolt}, M., {Fender}, R., and {M{\'e}ndez}, M. 2001.
\newblock {Correlated X-Ray Spectral and Timing Behavior of the Black Hole
  Candidate XTE J1550-564: A New Interpretation of Black Hole States}.
\newblock {\em \apjs}, {\bf 132}(Feb.), 377--402.

\bibitem[\protect\citename{{Homan} {et~al.}, }2002]{homan2002}
{Homan}, J., {van der Klis}, M., {Jonker}, P.~G., {Wijnands}, R., {Kuulkers},
  E., {M{\'e}ndez}, M., and {Lewin}, W.~H.~G. 2002.
\newblock {RXTE Observations of the Neutron Star Low-Mass X-Ray Binary GX 17+2:
  Correlated X-Ray Spectral and Timing Behavior}.
\newblock {\em \apj}, {\bf 568}(Apr.), 878--900.

\bibitem[\protect\citename{{Homan} {et~al.}, }2003]{homan2003}
{Homan}, J., {Klein-Wolt}, M., {Rossi}, S., {Miller}, J.~M., {Wijnands}, R.,
  {Belloni}, T., {van der Klis}, M., and {Lewin}, W.~H.~G. 2003.
\newblock {High-Frequency Quasi-periodic Oscillations in the Black Hole X-Ray
  Transient XTE J1650-500}.
\newblock {\em \apj}, {\bf 586}(Apr.), 1262--1267.

\bibitem[\protect\citename{{Homan} {et~al.}, }2005a]{homan2005b}
{Homan}, J., {Miller}, J.~M., {Wijnands}, R., {van der Klis}, M., {Belloni},
  T., {Steeghs}, D., and {Lewin}, W.~H.~G. 2005a.
\newblock {High- and Low-Frequency Quasi-periodic Oscillations in the X-Ray
  Light Curves of the Black Hole Transient H1743-322}.
\newblock {\em \apj}, {\bf 623}(Apr.), 383--391.

\bibitem[\protect\citename{{Homan} {et~al.}, }2005b]{homan2005}
{Homan}, J., {Buxton}, M., {Markoff}, S., {Bailyn}, C.~D., {Nespoli}, E., and
  {Belloni}, T. 2005b.
\newblock {Multiwavelength Observations of the 2002 Outburst of GX 339-4: Two
  Patterns of X-Ray-Optical/Near-Infrared Behavior}.
\newblock {\em \apj}, {\bf 624}(May), 295--306.

\bibitem[\protect\citename{{Homan} {et~al.}, }2007]{Homan2007}
{Homan}, J., {van der Klis}, M., {Wijnands}, R., {Belloni}, T., {Fender}, R.,
  {Klein-Wolt}, M., {Casella}, P., {M{\'e}ndez}, M., {Gallo}, E., {Lewin},
  W.~H.~G., and {Gehrels}, N. 2007.
\newblock {Rossi X-Ray Timing Explorer Observations of the First Transient Z
  Source XTE J1701-462: Shedding New Light on Mass Accretion in Luminous
  Neutron Star X-Ray Binaries}.
\newblock {\em \apj}, {\bf 656}(Feb.), 420--430.

\bibitem[\protect\citename{{Homan} {et~al.}, }2010]{Homan2010}
{Homan}, J., {van der Klis}, M., {Fridriksson}, J.~K., {Remillard}, R.~A.,
  {Wijnands}, R., {M{\'e}ndez}, M., {Lin}, D., {Altamirano}, D., {Casella}, P.,
  {Belloni}, T.~M., and {Lewin}, W.~H.~G. 2010.
\newblock {XTE J1701-462 and Its Implications for the Nature of Subclasses in
  Low-magnetic-field Neutron Star Low-mass X-ray Binaries}.
\newblock {\em \apj}, {\bf 719}(Aug.), 201--212.

\bibitem[\protect\citename{{Ingram} and {Done}, }2010]{ingram2010}
{Ingram}, A., and {Done}, C. 2010.
\newblock {A physical interpretation of the variability power spectral
  components in accreting neutron stars}.
\newblock {\em \mnras}, {\bf 405}(July), 2447--2452.

\bibitem[\protect\citename{{Ingram} and {Done}, }2011]{ingram2011}
{Ingram}, A., and {Done}, C. 2011.
\newblock {A physical model for the continuum variability and quasi-periodic
  oscillation in accreting black holes}.
\newblock {\em \mnras}, {\bf 415}(Aug.), 2323--2335.

\bibitem[\protect\citename{{Ingram} and {van der Klis}, }2013]{ingramvdk}
{Ingram}, A., and {van der Klis}, M. 2013.
\newblock {An exact analytic treatment of propagating mass accretion rate
  fluctuations in X-ray binaries}.
\newblock {\em \mnras}, {\bf 434}(Sept.), 1476--1485.

\bibitem[\protect\citename{{Ingram} {et~al.}, }2009]{ingram2009}
{Ingram}, A., {Done}, C., and {Fragile}, P.~C. 2009.
\newblock {Low-frequency quasi-periodic oscillations spectra and Lense-Thirring
  precession}.
\newblock {\em \mnras}, {\bf 397}(July), L101--L105.

\bibitem[\protect\citename{{Ingram} {et~al.}, }2016]{ingram2016}
{Ingram}, A., {van der Klis}, M., {Middleton}, M., {Done}, C., {Altamirano},
  D., {Heil}, L., {Uttley}, P., and {Axelsson}, M. 2016.
\newblock {A quasi-periodic modulation of the iron line centroid energy in the
  black hole binary H1743-322}.
\newblock {\em \mnras}, {\bf 461}(Sept.), 1967--1980.

\bibitem[\protect\citename{{Joinet} {et~al.}, }2008]{joinet}
{Joinet}, A., {Kalemci}, E., and {Senziani}, F. 2008.
\newblock {Hard X-Ray Emission of the Microquasar GRO J1655-40 during the Rise
  of Its 2005 Outburst}.
\newblock {\em \apj}, {\bf 679}(May), 655--663.

\bibitem[\protect\citename{{Kalemci} {et~al.}, }2013]{kalemci2013}
{Kalemci}, E., {Din{\c c}er}, T., {Tomsick}, J.~A., {Buxton}, M.~M., {Bailyn},
  C.~D., and {Chun}, Y.~Y. 2013.
\newblock {Complete Multiwavelength Evolution of Galactic Black Hole Transients
  during Outburst Decay. I. Conditions for ''Compact'' Jet Formation}.
\newblock {\em \apj}, {\bf 779}(Dec.), 95.

\bibitem[\protect\citename{{Klein-Wolt} {et~al.}, }2004]{klein}
{Klein-Wolt}, M., {Homan}, J., and {van der Klis}, M. 2004.
\newblock {High frequency features in the 1998 outburst of 4U 1630-47}.
\newblock {\em Nuclear Physics B Proceedings Supplements}, {\bf 132}(June),
  381--386.

\bibitem[\protect\citename{{Kolehmainen} {et~al.}, }2014]{kolehmainen}
{Kolehmainen}, M., {Done}, C., and {D{\'{\i}}az Trigo}, M. 2014.
\newblock {The soft component and the iron line as signatures of the disc inner
  radius in Galactic black hole binaries}.
\newblock {\em \mnras}, {\bf 437}(Jan.), 316--326.

\bibitem[\protect\citename{{Kong} {et~al.}, }2002]{kong2002}
{Kong}, A.~K.~H., {McClintock}, J.~E., {Garcia}, M.~R., {Murray}, S.~S., and
  {Barret}, D. 2002.
\newblock {The X-Ray Spectra of Black Hole X-Ray Novae in Quiescence as
  Measured by Chandra}.
\newblock {\em \apj}, {\bf 570}(May), 277--286.

\bibitem[\protect\citename{{Kylafis} and {Belloni}, }2015]{kylafis2015}
{Kylafis}, N.~D., and {Belloni}, T.~M. 2015.
\newblock {Accretion and ejection in black-hole X-ray transients}.
\newblock {\em \aap}, {\bf 574}(Feb.), A133.

\bibitem[\protect\citename{{Lasota}, }2016]{lasota}
{Lasota}, J.-P. 2016.
\newblock {Black Hole Accretion Discs}.
\newblock {Page ~1 of:} {Bambi}, C. (ed), {\em Astrophysics of Black Holes:
  From Fundamental Aspects to Latest Developments}.
\newblock Astrophysics and Space Science Library, vol. 440.

\bibitem[\protect\citename{{Lin} {et~al.}, }2007]{Lin2007}
{Lin}, D., {Remillard}, R.~A., and {Homan}, J. 2007.
\newblock {Evaluating Spectral Models and the X-Ray States of Neutron Star
  X-Ray Transients}.
\newblock {\em \apj}, {\bf 667}(Oct.), 1073--1086.

\bibitem[\protect\citename{{Lin} {et~al.}, }2009]{Lin2009}
{Lin}, D., {Remillard}, R.~A., and {Homan}, J. 2009.
\newblock {Spectral States of XTE J1701 - 462: Link Between Z and Atoll
  Sources}.
\newblock {\em \apj}, {\bf 696}(May), 1257--1277.

\bibitem[\protect\citename{{Lin} {et~al.}, }2012]{Lin2012}
{Lin}, D., {Remillard}, R.~A., {Homan}, J., and {Barret}, D. 2012.
\newblock {The Spectral Evolution along the Z Track of the Bright Neutron Star
  X-Ray Binary GX 17+2}.
\newblock {\em \apj}, {\bf 756}(Sept.), 34.

\bibitem[\protect\citename{{Linares} {et~al.}, }2005]{linares2005}
{Linares}, M., {van der Klis}, M., {Altamirano}, D., and {Markwardt}, C.~B.
  2005.
\newblock {Discovery of Kilohertz Quasi-periodic Oscillations and Shifted
  Frequency Correlations in the Accreting Millisecond Pulsar XTE J1807-294}.
\newblock {\em \apj}, {\bf 634}(Dec.), 1250--1260.

\bibitem[\protect\citename{{Lyubarskii}, }1997]{lyubarskii}
{Lyubarskii}, Y.~E. 1997.
\newblock {Flicker noise in accretion discs}.
\newblock {\em \mnras}, {\bf 292}(Dec.), 679.

\bibitem[\protect\citename{{Markoff}, }2010]{markoff2010}
{Markoff}, S. 2010 (Mar.).
\newblock {From Multiwavelength to Mass Scaling: Accretion and Ejection in
  Microquasars and AGN}.
\newblock {Page  143 of:} {Belloni}, T. (ed), {\em Lecture Notes in Physics,
  Berlin Springer Verlag}.
\newblock Lecture Notes in Physics, Berlin Springer Verlag, vol. 794.

\bibitem[\protect\citename{{Markoff} {et~al.}, }2005]{markoff2005}
{Markoff}, S., {Nowak}, M.~A., and {Wilms}, J. 2005.
\newblock {Going with the Flow: Can the Base of Jets Subsume the Role of
  Compact Accretion Disk Coronae?}
\newblock {\em \apj}, {\bf 635}(Dec.), 1203--1216.

\bibitem[\protect\citename{{McClintock} {et~al.}, }2003]{mcclintock1118}
{McClintock}, J.~E., {Narayan}, R., {Garcia}, M.~R., {Orosz}, J.~A.,
  {Remillard}, R.~A., and {Murray}, S.~S. 2003.
\newblock {Multiwavelength Spectrum of the Black Hole XTE J1118+480 in
  Quiescence}.
\newblock {\em \apj}, {\bf 593}(Aug.), 435--451.

\bibitem[\protect\citename{{M{\'e}ndez}, }2006]{mendez2006}
{M{\'e}ndez}, M. 2006.
\newblock {On the maximum amplitude and coherence of the kilohertz
  quasi-periodic oscillations in low-mass X-ray binaries}.
\newblock {\em \mnras}, {\bf 371}(Oct.), 1925--1938.

\bibitem[\protect\citename{{M{\'e}ndez} and {Belloni}, }2007]{menbel2007}
{M{\'e}ndez}, M., and {Belloni}, T. 2007.
\newblock {Is there a link between the neutron-star spin and the frequency of
  the kilohertz quasi-periodic oscillations?}
\newblock {\em \mnras}, {\bf 381}(Oct.), 790--796.

\bibitem[\protect\citename{{M{\'e}ndez} and {van der Klis}, }1997]{mendez1997}
{M{\'e}ndez}, M., and {van der Klis}, M. 1997.
\newblock {The EXOSAT Data on GX 339-4: Further Evidence for an
  ``Intermediate'' State}.
\newblock {\em \apj}, {\bf 479}(Apr.), 926--932.

\bibitem[\protect\citename{{M{\'e}ndez} {et~al.}, }1998]{mendez1998}
{M{\'e}ndez}, M., {van der Klis}, M., {Wijnands}, R., {Ford}, E.~C., {van
  Paradijs}, J., and {Vaughan}, B.~A. 1998.
\newblock {Kilohertz Quasi-periodic Oscillation Peak Separation Is Not Constant
  in the Atoll Source 4U 1608-52}.
\newblock {\em \apjl}, {\bf 505}(Sept.), L23--L26.

\bibitem[\protect\citename{{M{\'e}ndez} {et~al.}, }2013]{mendez2013}
{M{\'e}ndez}, M., {Altamirano}, D., {Belloni}, T., and {Sanna}, A. 2013.
\newblock {The phase lags of high-frequency quasi-periodic oscillations in four
  black hole candidates}.
\newblock {\em \mnras}, {\bf 435}(Nov.), 2132--2140.

\bibitem[\protect\citename{{Menou} {et~al.}, }1999]{menou1999}
{Menou}, K., {Esin}, A.~A., {Narayan}, R., {Garcia}, M.~R., {Lasota}, J.-P.,
  and {McClintock}, J.~E. 1999.
\newblock {Black Hole and Neutron Star Transients in Quiescence}.
\newblock {\em \apj}, {\bf 520}(July), 276--291.

\bibitem[\protect\citename{{Middleton}, }2016]{middleton}
{Middleton}, M. 2016.
\newblock {Black Hole Spin: Theory and Observation}.
\newblock {Page ~99 of:} {Bambi}, C. (ed), {\em Astrophysics of Black Holes:
  From Fundamental Aspects to Latest Developments}.
\newblock Astrophysics and Space Science Library, vol. 440.

\bibitem[\protect\citename{{Migliari} {et~al.}, }2003]{migliari2003}
{Migliari}, S., {van der Klis}, M., and {Fender}, R.~P. 2003.
\newblock {Evidence of a decrease of kHz quasi-periodic oscillation peak
  separation towards low frequencies in 4U 1728-34 (GX 354-0)}.
\newblock {\em \mnras}, {\bf 345}(Nov.), L35--L39.

\bibitem[\protect\citename{{Migliari} {et~al.}, }2007]{migliari2007}
{Migliari}, S., {Miller-Jones}, J.~C.~A., {Fender}, R.~P., {Homan}, J., {Di
  Salvo}, T., {Rothschild}, R.~E., {Rupen}, M.~P., {Tomsick}, J.~A.,
  {Wijnands}, R., and {van der Klis}, M. 2007.
\newblock {Linking Jet Emission, X-Ray States, and Hard X-Ray Tails in the
  Neutron Star X-Ray Binary GX 17+2}.
\newblock {\em \apj}, {\bf 671}(Dec.), 706--712.

\bibitem[\protect\citename{{Miller}, }2007]{miller}
{Miller}, J.~M. 2007.
\newblock {Relativistic X-Ray Lines from the Inner Accretion Disks Around Black
  Holes}.
\newblock {\em \araa}, {\bf 45}(Sept.), 441--479.

\bibitem[\protect\citename{{Miller} {et~al.}, }2001]{miller2001}
{Miller}, J.~M., {Wijnands}, R., {Homan}, J., {Belloni}, T., {Pooley}, D.,
  {Corbel}, S., {Kouveliotou}, C., {van der Klis}, M., and {Lewin}, W.~H.~G.
  2001.
\newblock {High-Frequency Quasi-Periodic Oscillations in the 2000 Outburst of
  the Galactic Microquasar XTE J1550-564}.
\newblock {\em \apj}, {\bf 563}(Dec.), 928--933.

\bibitem[\protect\citename{{Mitsuda} {et~al.}, }1989]{Mitsuda1989}
{Mitsuda}, K., {Inoue}, H., {Nakamura}, N., and {Tanaka}, Y. 1989.
\newblock {Luminosity-related changes of the energy spectrum of X1608-522}.
\newblock {\em \pasj}, {\bf 41}, 97--111.

\bibitem[\protect\citename{{Miyamoto} {et~al.}, }1992]{miyamoto1992}
{Miyamoto}, S., {Kitamoto}, S., {Iga}, S., {Negoro}, H., and {Terada}, K. 1992.
\newblock {Canonical time variations of X-rays from black hole candidates in
  the low-intensity state}.
\newblock {\em \apjl}, {\bf 391}(May), L21--L24.

\bibitem[\protect\citename{{Miyamoto} {et~al.}, }1993]{miyamoto1993}
{Miyamoto}, S., {Iga}, S., {Kitamoto}, S., and {Kamado}, Y. 1993.
\newblock {Another canonical time variation of X-rays from black hole
  candidates in the very high flare state?}
\newblock {\em \apjl}, {\bf 403}(Jan.), L39--L42.

\bibitem[\protect\citename{{Morgan} {et~al.}, }1997]{morgan}
{Morgan}, E.~H., {Remillard}, R.~A., and {Greiner}, J. 1997.
\newblock {RXTE Observations of QPOs in the Black Hole Candidate GRS 1915+105}.
\newblock {\em \apj}, {\bf 482}(June), 993--1010.

\bibitem[\protect\citename{{Motta} {et~al.}, }2009]{mottacutoff}
{Motta}, S., {Belloni}, T., and {Homan}, J. 2009.
\newblock {The evolution of the high-energy cut-off in the X-ray spectrum of GX
  339-4 across a hard-to-soft transition}.
\newblock {\em \mnras}, {\bf 400}(Dec.), 1603--1612.

\bibitem[\protect\citename{{Motta} {et~al.}, }2011]{motta2011}
{Motta}, S., {Mu{\~n}oz-Darias}, T., {Casella}, P., {Belloni}, T., and {Homan},
  J. 2011.
\newblock {Low-frequency oscillations in black holes: a spectral-timing
  approach to the case of GX 339-4}.
\newblock {\em \mnras}, {\bf 418}(Dec.), 2292--2307.

\bibitem[\protect\citename{{Motta} {et~al.}, }2012]{motta2012}
{Motta}, S., {Homan}, J., {Mu{\~n}oz Darias}, T., {Casella}, P., {Belloni},
  T.~M., {Hiemstra}, B., and {M{\'e}ndez}, M. 2012.
\newblock {Discovery of two simultaneous non-harmonically related
  quasi-periodic oscillations in the 2005 outburst of the black hole binary GRO
  J1655-40}.
\newblock {\em \mnras}, {\bf 427}(Nov.), 595--606.

\bibitem[\protect\citename{{Motta}, }2016]{mottaAN}
{Motta}, S.~E. 2016.
\newblock {Quasi periodic oscillations in black hole binaries}.
\newblock {\em Astronomische Nachrichten}, {\bf 337}(May), 398.

\bibitem[\protect\citename{{Motta} {et~al.}, }2014a]{motta2014}
{Motta}, S.~E., {Mu{\~n}oz-Darias}, T., {Sanna}, A., {Fender}, R., {Belloni},
  T., and {Stella}, L. 2014a.
\newblock {Black hole spin measurements through the relativistic precession
  model: XTE J1550-564}.
\newblock {\em \mnras}, {\bf 439}(Mar.), L65--L69.

\bibitem[\protect\citename{{Motta} {et~al.}, }2014b]{motta2014b}
{Motta}, S.~E., {Belloni}, T.~M., {Stella}, L., {Mu{\~n}oz-Darias}, T., and
  {Fender}, R. 2014b.
\newblock {Precise mass and spin measurements for a stellar-mass black hole
  through X-ray timing: the case of GRO J1655-40}.
\newblock {\em \mnras}, {\bf 437}(Jan.), 2554--2565.

\bibitem[\protect\citename{{Mu{\~n}oz-Darias} {et~al.}, }2011]{teoRID}
{Mu{\~n}oz-Darias}, T., {Motta}, S., and {Belloni}, T.~M. 2011.
\newblock {Fast variability as a tracer of accretion regimes in black hole
  transients}.
\newblock {\em \mnras}, {\bf 410}(Jan.), 679--684.

\bibitem[\protect\citename{{Mu{\~n}oz-Darias} {et~al.}, }2013]{Teo2013}
{Mu{\~n}oz-Darias}, T., {Coriat}, M., {Plant}, D.~S., {Ponti}, G., {Fender},
  R.~P., and {Dunn}, R.~J.~H. 2013.
\newblock {Inclination and relativistic effects in the outburst evolution of
  black hole transients}.
\newblock {\em \mnras}, {\bf 432}(June), 1330--1337.

\bibitem[\protect\citename{{Mu{\~n}oz-Darias} {et~al.}, }2014]{Teo2014}
{Mu{\~n}oz-Darias}, T., {Fender}, R.~P., {Motta}, S.~E., and {Belloni}, T.~M.
  2014.
\newblock {Black hole-like hysteresis and accretion states in neutron star
  low-mass X-ray binaries}.
\newblock {\em \mnras}, {\bf 443}(Oct.), 3270--3283.

\bibitem[\protect\citename{{Narayan} and {Heyl}, }2002]{narayan2002}
{Narayan}, R., and {Heyl}, J.~S. 2002.
\newblock {On the Lack of Type I X-Ray Bursts in Black Hole X-Ray Binaries:
  Evidence for the Event Horizon?}
\newblock {\em \apjl}, {\bf 574}(Aug.), L139--L142.

\bibitem[\protect\citename{{Narayan} {et~al.}, }1996]{adaf}
{Narayan}, R., {McClintock}, J.~E., and {Yi}, I. 1996.
\newblock {A New Model for Black Hole Soft X-Ray Transients in Quiescence}.
\newblock {\em \apj}, {\bf 457}(Feb.), 821.

\bibitem[\protect\citename{{Nespoli} {et~al.}, }2003]{nespoli2003}
{Nespoli}, E., {Belloni}, T., {Homan}, J., {Miller}, J.~M., {Lewin}, W.~H.~G.,
  {M{\'e}ndez}, M., and {van der Klis}, M. 2003.
\newblock {A transient variable 6 Hz QPO from GX 339-4}.
\newblock {\em \aap}, {\bf 412}(Dec.), 235--240.

\bibitem[\protect\citename{{Nowak}, }2000]{nowak2000}
{Nowak}, M.~A. 2000.
\newblock {Are there three peaks in the power spectra of GX 339-4 and Cyg X-1?}
\newblock {\em \mnras}, {\bf 318}(Oct.), 361--367.

\bibitem[\protect\citename{{Olive} {et~al.}, }1998]{olive1998}
{Olive}, J.~F., {Barret}, D., {Boirin}, L., {Grindlay}, J.~E., {Swank}, J.~H.,
  and {Smale}, A.~P. 1998.
\newblock {RXTE observation of the X-ray burster 1E 1724-3045. I. Timing study
  of the persistent X-ray emission with the PCA}.
\newblock {\em \aap}, {\bf 333}(May), 942--951.

\bibitem[\protect\citename{{Petrucci}, }2008]{petrucci2008}
{Petrucci}, P.~O. 2008.
\newblock {Models of comptonization.}
\newblock {\em \memsai}, {\bf 79}, 118.

\bibitem[\protect\citename{{Petrucci} {et~al.}, }2008]{petrucci2008b}
{Petrucci}, P.-O., {Ferreira}, J., {Henri}, G., and {Pelletier}, G. 2008.
\newblock {The role of the disc magnetization on the hysteresis behaviour of
  X-ray binaries}.
\newblock {\em \mnras}, {\bf 385}(Mar.), L88--L92.

\bibitem[\protect\citename{{Plotkin} {et~al.}, }2013]{plotkin}
{Plotkin}, R.~M., {Gallo}, E., and {Jonker}, P.~G. 2013.
\newblock {The X-Ray Spectral Evolution of Galactic Black Hole X-Ray Binaries
  toward Quiescence}.
\newblock {\em \apj}, {\bf 773}(Aug.), 59.

\bibitem[\protect\citename{{Plotkin} {et~al.}, }2015]{plotkin133}
{Plotkin}, R.~M., {Gallo}, E., {Markoff}, S., {Homan}, J., {Jonker}, P.~G.,
  {Miller-Jones}, J.~C.~A., {Russell}, D.~M., and {Drappeau}, S. 2015.
\newblock {Constraints on relativistic jets in quiescent black hole X-ray
  binaries from broad-band spectral modelling}.
\newblock {\em \mnras}, {\bf 446}(Feb.), 4098--4111.

\bibitem[\protect\citename{{Pottschmidt} {et~al.}, }2003]{pottschmidt2003}
{Pottschmidt}, K., {Wilms}, J., {Nowak}, M.~A., {Pooley}, G.~G., {Gleissner},
  T., {Heindl}, W.~A., {Smith}, D.~M., {Remillard}, R., and {Staubert}, R.
  2003.
\newblock {Long term variability of Cygnus X-1. I. X-ray spectral-temporal
  correlations in the hard state}.
\newblock {\em \aap}, {\bf 407}(Sept.), 1039--1058.

\bibitem[\protect\citename{{Poutanen} {et~al.}, }1997]{poutanen}
{Poutanen}, J., {Krolik}, J.~H., and {Ryde}, F. 1997.
\newblock {The nature of spectral transitions in accreting black holes - The
  case of CYG X-1}.
\newblock {\em \mnras}, {\bf 292}(Nov.), L21--L25.

\bibitem[\protect\citename{{Psaltis}, }2008]{psaltis2008}
{Psaltis}, D. 2008.
\newblock {Probes and Tests of Strong-Field Gravity with Observations in the
  Electromagnetic Spectrum}.
\newblock {\em Living Reviews in Relativity}, {\bf 11}(Dec.), 9.

\bibitem[\protect\citename{{Psaltis} {et~al.}, }1999a]{pbk}
{Psaltis}, D., {Belloni}, T., and {van der Klis}, M. 1999a.
\newblock {Correlations in Quasi-periodic Oscillation and Noise Frequencies
  among Neutron Star and Black Hole X-Ray Binaries}.
\newblock {\em \apj}, {\bf 520}(July), 262--270.

\bibitem[\protect\citename{{Psaltis} {et~al.}, }1999b]{psaltisHBO}
{Psaltis}, D., {Wijnands}, R., {Homan}, J., {Jonker}, P.~G., {van der Klis},
  M., {Miller}, M.~C., {Lamb}, F.~K., {Kuulkers}, E., {van Paradijs}, J., and
  {Lewin}, W.~H.~G. 1999b.
\newblock {On the Magnetospheric Beat-Frequency and Lense-Thirring
  Interpretations of the Horizontal-Branch Oscillation in the Z Sources}.
\newblock {\em \apj}, {\bf 520}(Aug.), 763--775.

\bibitem[\protect\citename{{Rao} {et~al.}, }2010]{rao2010}
{Rao}, F., {Belloni}, T., {Stella}, L., {Zhang}, S.~N., and {Li}, T. 2010.
\newblock {Low-frequency Oscillations in XTE J1550-564}.
\newblock {\em \apj}, {\bf 714}(May), 1065--1071.

\bibitem[\protect\citename{{Ratti} {et~al.}, }2012]{ratti}
{Ratti}, E.~M., {Belloni}, T.~M., and {Motta}, S.~E. 2012.
\newblock {On the harmonics of the low-frequency quasi-periodic oscillation in
  GRS 1915+105}.
\newblock {\em \mnras}, {\bf 423}(June), 694--701.

\bibitem[\protect\citename{{Reid} {et~al.}, }2014]{reid2014}
{Reid}, M.~J., {McClintock}, J.~E., {Steiner}, J.~F., {Steeghs}, D.,
  {Remillard}, R.~A., {Dhawan}, V., and {Narayan}, R. 2014.
\newblock {A Parallax Distance to the Microquasar GRS 1915+105 and a Revised
  Estimate of its Black Hole Mass}.
\newblock {\em \apj}, {\bf 796}(Nov.), 2.

\bibitem[\protect\citename{{Remillard} {et~al.}, }1999a]{remillard1999b}
{Remillard}, R.~A., {Morgan}, E.~H., {McClintock}, J.~E., {Bailyn}, C.~D., and
  {Orosz}, J.~A. 1999a.
\newblock {RXTE Observations of 0.1-300 HZ Quasi-periodic Oscillationsin the
  Microquasar GRO J1655-40}.
\newblock {\em \apj}, {\bf 522}(Sept.), 397--412.

\bibitem[\protect\citename{{Remillard} {et~al.}, }1999b]{remillard1999a}
{Remillard}, R.~A., {McClintock}, J.~E., {Sobczak}, G.~J., {Bailyn}, C.~D.,
  {Orosz}, J.~A., {Morgan}, E.~H., and {Levine}, A.~M. 1999b.
\newblock {X-Ray Nova XTE J1550-564: Discovery of a Quasi-periodic Oscillation
  near 185 HZ}.
\newblock {\em \apjl}, {\bf 517}(June), L127--L130.

\bibitem[\protect\citename{{Remillard} {et~al.}, }2002]{remillard2002}
{Remillard}, R.~A., {Muno}, M.~P., {McClintock}, J.~E., and {Orosz}, J.~A.
  2002.
\newblock {Evidence for Harmonic Relationships in the High-Frequency
  Quasi-periodic Oscillations of XTE J1550-564 and GRO J1655-40}.
\newblock {\em \apj}, {\bf 580}(Dec.), 1030--1042.

\bibitem[\protect\citename{{Remillard} {et~al.}, }2006a]{Remillard2006}
{Remillard}, R.~A., {Lin}, D., {ASM Team at MIT}, and {NASA/GSFC}. 2006a.
\newblock {New X-ray Transient, XTE J1701-462}.
\newblock {\em The Astronomer's Telegram}, {\bf 696}(Jan.).

\bibitem[\protect\citename{{Remillard} {et~al.}, }2006b]{Remillard2006c}
{Remillard}, R.~A., {Lin}, D., {Cooper}, R.~L., and {Narayan}, R. 2006b.
\newblock {The Rates of Type I X-Ray Bursts from Transients Observed with RXTE:
  Evidence for Black Hole Event Horizons}.
\newblock {\em \apj}, {\bf 646}(July), 407--419.

\bibitem[\protect\citename{{Remillard} {et~al.}, }2006c]{Remillard2006b}
{Remillard}, R.~A., {McClintock}, J.~E., {Orosz}, J.~A., and {Levine}, A.~M.
  2006c.
\newblock {The X-Ray Outburst of H1743-322 in 2003: High-Frequency QPOs with a
  3:2 Frequency Ratio}.
\newblock {\em \apj}, {\bf 637}(Feb.), 1002--1009.

\bibitem[\protect\citename{{Sanna} {et~al.}, }2012]{sanna2012}
{Sanna}, A., {M{\'e}ndez}, M., {Belloni}, T., and {Altamirano}, D. 2012.
\newblock {The time derivative of the kilohertz quasi-periodic oscillations in
  4U 1636-53}.
\newblock {\em \mnras}, {\bf 424}(Aug.), 2936--2942.

\bibitem[\protect\citename{{Sanna} {et~al.}, }2014]{sanna2014}
{Sanna}, A., {M{\'e}ndez}, M., {Altamirano}, D., {Belloni}, T., {Hiemstra}, B.,
  and {Linares}, M. 2014.
\newblock {Broad iron emission line and kilohertz quasi-periodic oscillations
  in the neutron star system 4U 1636-53}.
\newblock {\em \mnras}, {\bf 440}(June), 3275--3284.

\bibitem[\protect\citename{{Shakura} and {Sunyaev}, }1973]{ss}
{Shakura}, N.~I., and {Sunyaev}, R.~A. 1973.
\newblock {Black holes in binary systems. Observational appearance.}
\newblock {\em \aap}, {\bf 24}, 337--355.

\bibitem[\protect\citename{{Sobolewska} {et~al.}, }2011]{sobolewska}
{Sobolewska}, M.~A., {Papadakis}, I.~E., {Done}, C., and {Malzac}, J. 2011.
\newblock {Evidence for a change in the X-ray radiation mechanism in the hard
  state of Galactic black holes}.
\newblock {\em \mnras}, {\bf 417}(Oct.), 280--288.

\bibitem[\protect\citename{{Steiner} {et~al.}, }2016]{steiner2016}
{Steiner}, J.~F., {Remillard}, R.~A., {Garc{\'{\i}}a}, J.~A., and {McClintock},
  J.~E. 2016.
\newblock {Stronger Reflection from Black Hole Accretion Disks in Soft X-Ray
  States}.
\newblock {\em \apjl}, {\bf 829}(Oct.), L22.

\bibitem[\protect\citename{{Stella} and {Vietri}, }1998]{StellaVietri1998}
{Stella}, L., and {Vietri}, M. 1998.
\newblock {Lense-Thirring Precession and Quasi-periodic Oscillations in
  Low-Mass X-Ray Binaries}.
\newblock {\em \apjl}, {\bf 492}(Jan.), L59--L62.

\bibitem[\protect\citename{{Stella} and {Vietri}, }1999]{StellaVietri1999}
{Stella}, L., and {Vietri}, M. 1999.
\newblock {kHz Quasiperiodic Oscillations in Low-Mass X-Ray Binaries as Probes
  of General Relativity in the Strong-Field Regime}.
\newblock {\em Physical Review Letters}, {\bf 82}(Jan.), 17--20.

\bibitem[\protect\citename{{Stella} {et~al.}, }1999]{StellaVietriMorsink1999}
{Stella}, L., {Vietri}, M., and {Morsink}, S.~M. 1999.
\newblock {Correlations in the Quasi-periodic Oscillation Frequencies of
  Low-Mass X-Ray Binaries and the Relativistic Precession Model}.
\newblock {\em \apjl}, {\bf 524}(Oct.), L63--L66.

\bibitem[\protect\citename{{Stiele} {et~al.}, }2011]{stiele}
{Stiele}, H., {Motta}, S., {Mu{\~n}oz-Darias}, T., and {Belloni}, T.~M. 2011.
\newblock {Spectral properties of transitions between soft and hard states in
  GX 339-4}.
\newblock {\em \mnras}, {\bf 418}(Dec.), 1746--1752.

\bibitem[\protect\citename{{Strohmayera}, }2001]{strohmayer2001a}
{Strohmayera}, T.~E. 2001.
\newblock {Discovery of a 450 HZ Quasi-periodic Oscillation from the
  Microquasar GRO J1655-40 with the Rossi X-Ray Timing Explorer}.
\newblock {\em \apjl}, {\bf 552}(May), L49--L53.

\bibitem[\protect\citename{{Strohmayerb}, }2001]{strohmayer2001b}
{Strohmayerb}, T.~E. 2001.
\newblock {Discovery of a Second High-Frequency Quasi-periodic Oscillation from
  the Microquasar GRS 1915+105}.
\newblock {\em \apjl}, {\bf 554}(June), L169--L172.

\bibitem[\protect\citename{{Sunyaev} and {Revnivtsev}, }2000]{sunyaev}
{Sunyaev}, R., and {Revnivtsev}, M. 2000.
\newblock {Fourier power spectra at high frequencies: a way to distinguish a
  neutron star from a black hole}.
\newblock {\em \aap}, {\bf 358}(June), 617--623.

\bibitem[\protect\citename{{Takizawa} {et~al.}, }1997]{takizawa}
{Takizawa}, M., {Dotani}, T., {Mitsuda}, K., {Matsuba}, E., {Ogawa}, M.,
  {Aoki}, T., {Asai}, K., {Ebisawa}, K., {Makishima}, K., {Miyamoto}, S.,
  {Iga}, S., {Vaughan}, B., {Rutledge}, R.~E., and {Lewin}, W.~H.~G. 1997.
\newblock {Spectral and Temporal Variability in the X-Ray Flux of GS 1124-683,
  Nova Muscae 1991}.
\newblock {\em \apj}, {\bf 489}(Nov.), 272--283.

\bibitem[\protect\citename{{Tananbaum} {et~al.}, }1972]{tananbaum1972}
{Tananbaum}, H., {Gursky}, H., {Kellogg}, E., {Giacconi}, R., and {Jones}, C.
  1972.
\newblock {Observation of a Correlated X-Ray Transition in Cygnus X-1}.
\newblock {\em \apjl}, {\bf 177}(Oct.), L5.

\bibitem[\protect\citename{{Titarchuk} and {Fiorito}, }2004]{titarchukfiorito}
{Titarchuk}, L., and {Fiorito}, R. 2004.
\newblock {Spectral Index and Quasi-Periodic Oscillation Frequency Correlation
  in Black Hole Sources: Observational Evidence of Two Phases and Phase
  Transition in Black Holes}.
\newblock {\em \apj}, {\bf 612}(Sept.), 988--999.

\bibitem[\protect\citename{{Turolla} {et~al.}, }2002]{turolla2002}
{Turolla}, R., {Zane}, S., and {Titarchuk}, L. 2002.
\newblock {Power-Law Tails from Dynamical Comptonization in Converging Flows}.
\newblock {\em \apj}, {\bf 576}(Sept.), 349--356.

\bibitem[\protect\citename{{Uttley} {et~al.}, }2011]{uttley2011}
{Uttley}, P., {Wilkinson}, T., {Cassatella}, P., {Wilms}, J., {Pottschmidt},
  K., {Hanke}, M., and {B{\"o}ck}, M. 2011.
\newblock {The causal connection between disc and power-law variability in hard
  state black hole X-ray binaries}.
\newblock {\em \mnras}, {\bf 414}(June), L60--L64.

\bibitem[\protect\citename{{van der Klis}, }1994]{Klis:1994}
{van der Klis}, M. 1994.
\newblock {Similarities in neutron star and black hole accretion}.
\newblock {\em \apjs}, {\bf 92}(June), 511--519.

\bibitem[\protect\citename{{van der Klis}, }2001]{Klis:2001}
{van der Klis}, M. 2001.
\newblock {A Possible Explanation for the ``Parallel Tracks'' Phenomenon in
  Low-Mass X-Ray Binaries}.
\newblock {\em \apj}, {\bf 561}(Nov.), 943--949.

\bibitem[\protect\citename{{van der Klis}, }2006]{Klis:2006}
{van der Klis}, M. 2006 (Apr.).
\newblock {Rapid X-ray Variability}.
\newblock {Pages  39--112 of:} {Lewin}, W.~H.~G., and {van der Klis}, M. (eds),
  {\em Compact stellar X-ray sources}.

\bibitem[\protect\citename{{van Straaten} {et~al.}, }2005]{vanstraaten2005}
{van Straaten}, S., {van der Klis}, M., and {Wijnands}, R. 2005.
\newblock {Relations Between Timing Features and Colors in Accreting
  Millisecond Pulsars}.
\newblock {\em \apj}, {\bf 619}(Jan.), 455--482.

\bibitem[\protect\citename{{Warner} {et~al.}, }2003]{Warner03}
{Warner}, B., {Woudt}, P.~A., and {Pretorius}, M.~L. 2003.
\newblock {Dwarf nova oscillations and quasi-periodic oscillations in
  cataclysmic variables - III. A new kind of dwarf nova oscillation, and
  further examples of the similarities to X-ray binaries}.
\newblock {\em \mnras}, {\bf 344}(Oct.), 1193--1209.

\bibitem[\protect\citename{{White} {et~al.}, }1988]{White1988}
{White}, N.~E., {Stella}, L., and {Parmar}, A.~N. 1988.
\newblock {The X-ray spectral properties of accretion discs in X-ray binaries}.
\newblock {\em \apj}, {\bf 324}(Jan.), 363--378.

\bibitem[\protect\citename{{Wijnands} and {van der Klis}, }1999]{WK}
{Wijnands}, R., and {van der Klis}, M. 1999.
\newblock {The Broadband Power Spectra of X-Ray Binaries}.
\newblock {\em \apj}, {\bf 514}(Apr.), 939--944.

\bibitem[\protect\citename{{Wijnands} {et~al.}, }2003]{Wijnands2003}
{Wijnands}, R., {van der Klis}, M., {Homan}, J., {Chakrabarty}, D.,
  {Markwardt}, C.~B., and {Morgan}, E.~H. 2003.
\newblock {Quasi-periodic X-ray brightness fluctuations in an accreting
  millisecond pulsar}.
\newblock {\em \nat}, {\bf 424}(July), 44--47.

\bibitem[\protect\citename{{Wijnands} {et~al.}, }2015]{Wijnands2015}
{Wijnands}, R., {Degenaar}, N., {Armas Padilla}, M., {Altamirano}, D.,
  {Cavecchi}, Y., {Linares}, M., {Bahramian}, A., and {Heinke}, C.~O. 2015.
\newblock {Low-level accretion in neutron star X-ray binaries}.
\newblock {\em \mnras}, {\bf 454}(Dec.), 1371--1386.

\bibitem[\protect\citename{{Wilkinson} and {Uttley}, }2009]{Wilkinson09}
{Wilkinson}, T., and {Uttley}, P. 2009.
\newblock {Accretion disc variability in the hard state of black hole X-ray
  binaries}.
\newblock {\em \mnras}, {\bf 397}(Aug.), 666--676.

\bibitem[\protect\citename{{Wilms} {et~al.}, }2006]{wilms2006}
{Wilms}, J., {Nowak}, M.~A., {Pottschmidt}, K., {Pooley}, G.~G., and {Fritz},
  S. 2006.
\newblock {Long term variability of Cygnus X-1. IV. Spectral evolution
  1999-2004}.
\newblock {\em \aap}, {\bf 447}(Feb.), 245--261.

\bibitem[\protect\citename{{Zdziarski} {et~al.}, }2012]{zdz2012}
{Zdziarski}, A.~A., {Lubi{\'n}ski}, P., and {Sikora}, M. 2012.
\newblock {The MeV spectral tail in Cyg X-1 and optically thin emission of
  jets}.
\newblock {\em \mnras}, {\bf 423}(June), 663--675.

\end{thebibliography}

\end{document}